\documentclass[11pt, a4paper, oneside]{Thesis} 

\graphicspath{{Pictures/}} 

\usepackage[square,sort,compress,numbers]{natbib}
\usepackage{cite}
\usepackage{amsmath,latexsym} 
\usepackage{float}

\title{\ttitle} 
\DeclareUnicodeCharacter{2212}{-}
\DeclareUnicodeCharacter{2212}{+}
\begin{document}

\setstretch{1.3} 

\fancyhead{} 
\rhead{\thepage} 
\lhead{} 

%

\thesistitle{Accelerated Expansion of the Universe in Non-minimally Coupled Gravity}
\documenttype{Thesis}
\supervisor{Prof. Pradyumn Kumar Sahoo}
\supervisorposition{Professor}
\supervisorinstitute{BITS-Pilani, Hyderabad Campus}
\examiner{}
\degree{Ph.D. Research Scholar}
\coursecode{DOCTOR OF PHILOSOPHY}
\coursename{Thesis}
\authors{\textbf{SANJAY MANDAL}}
\IDNumber{2018PHXF0444H}
\addresses{}
\subject{}
\keywords{}
\university{\texorpdfstring{\href{http://www.bits-pilani.ac.in/} 
                {Birla Institute of Technology and Science, Pilani}} 
                {Birla Institute of Technology and Science, Pilani}}
\UNIVERSITY{\texorpdfstring{\href{http://www.bits-pilani.ac.in/} 
                {BIRLA INSTITUTE OF TECHNOLOGY AND SCIENCE, PILANI}} 
                {BIRLA INSTITUTE OF TECHNOLOGY AND SCIENCE, PILANI}}



\department{\texorpdfstring{\href{http://www.bits-pilani.ac.in/pilani/Mathematics/Mathematics} 
                {Mathematics}} 
                {Mathematics}}
\DEPARTMENT{\texorpdfstring{\href{http://www.bits-pilani.ac.in/pilani/Mathematics/Mathematics} 
                {Mathematics}} 
                {Mathematics}}
\group{\texorpdfstring{\href{Research Group Web Site URL Here (include http://)}
                {Research Group Name}} 
                {Research Group Name}}
\GROUP{\texorpdfstring{\href{Research Group Web Site URL Here (include http://)}
                {RESEARCH GROUP NAME (IN BLOCK CAPITALS)}}
                {RESEARCH GROUP NAME (IN BLOCK CAPITALS)}}
\faculty{\texorpdfstring{\href{Faculty Web Site URL Here (include http://)}
                {Faculty Name}}
                {Faculty Name}}
\FACULTY{\texorpdfstring{\href{Faculty Web Site URL Here (include http://)}
                {FACULTY NAME (IN BLOCK CAPITALS)}}
                {FACULTY NAME (IN BLOCK CAPITALS)}}

\maketitle

\clearpage
\setstretch{1.3} 

\pagestyle{empty} 
\pagenumbering{gobble}


\addtocontents{toc}{\vspace{2em}} 

\frontmatter 
\Certificate
\Dedicatory{\bf \begin{LARGE}
Dedicated to
\end{LARGE} 
\\
\vspace{3cm}
 Maa, Baba, Dada \\}
 \Declaration
\begin{abstract}
The accelerated expansion of the universe is the most debatable cosmological scenario in the last two decades. Looking for the proper explanation of this scenario, researchers have presented a lot of proposals to discuss it. However, somehow, most of the proposal seems incomplete and needs to improve for a better representation of the universe. Therefore, this thesis aims to investigate some problems searching for an accurate explanation of the accelerated scenario of the universe.

Before discussing about the investigated problems, the preliminaries are discussed in Chapter-\ref{Chapter1}. Chapter-\ref{Chapter1} begins by discussing the evolution of the understanding of the universe from our ancestors to the modern days. However, the proper geometrical representation of the universe started with Einstein's general theory of relativity. Later, this chapter discusses the fundamental mathematical frameworks, their applications to cosmology, and cosmological observations, which help to validate the cosmological models. The fundamental theory of gravity, like general relativity, fails to address some issues like fine-tuning, flatness issues of the universe, and it seems incomplete. Therefore, its modifications and generalization are more successful in addressing this issue. Hence, this chapter concludes by over-viewing the modified theories of gravity.

In Chapters-\ref{Chapter3}, \ref{Chapter4}, and \ref{Chapter5}; the studies are based on minimal coupling between matter and geometry. Chapter-\ref{Chapter3} aims to discuss the accelerated expansion of the universe in teleparallel gravity. A well-known deceleration parameter is considered, and its free parameters constraints using observational data and the cosmological model's energy conditions are discussed. Chapter-\ref{Chapter4} extends the previous analysis to put bounds on the cosmological model's parameter using energy conditions. For this analysis, the observational values of cosmographic parameters for the accelerated expansion of the universe are used. Further, Chapter-\ref{Chapter5} aims to constraint the Lagrangian function in the action for the accelerated expansion of the universe. For this study, the parametrization technique and observational datasets are used.

In Chapters-\ref{Chapter6} and \ref{Chapter2}; the studies are based on non-minimal coupling between matter and geometry. Chapter-\ref{Chapter6} aims to study the accelerated expansion of the universe by focusing on the energy conditions. Using the profile of the equation of state parameter ($\omega$), this chapter compares the constructed cosmological models with the $\Lambda$CDM model. It is well-known that $\omega$ plays a significant role in describing the universe's various phases of evolution. Then, Chapter-\ref{Chapter2} extends the above analysis focusing on constraint $\omega$ for the current accelerated expansion of the universe. Different observational datasets were used for this study and successfully constraint $\omega$. Finally, Chapter-\ref{Chapter7} concludes by gathering all the outcomes of our studies. 


\end{abstract} 

\begin{acknowledgements}

It is a genuine pleasure to express my deep sense of thanks and gratitude to my mentor and guide, \textbf{Prof. Pradyumn Kumar Sahoo} (P. K. Sahoo), Professor, Department of Mathematics, BITS-Pilani, Hyderabad Campus, Hyderabad, Telangana. His dedication and keen interest, above all, his overwhelming attitude to help his students had been solely and mainly responsible for completing my work. His timely advice, meticulous scrutiny, scholarly advice, and scientific approach have helped me to a very great to accomplish this task.

I sincerely thank my Doctoral Advisory Committee (DAC) members, \textbf{Prof. Bivudutta Mishra}, \textbf{Dr. Nirman Ganguly}, and \textbf{Prof. Suresh Kumar}, for their valuable suggestions and constant encouragement to improve my research works.

It is my privilege to thank HoD, the DRC convener, faculty members, my colleagues, and the Department of Mathematics staff for supporting this amazing journey of my Ph.D. career.

I owe a deep sense of gratitude to all my co-authors for their valuable suggestions, discussions, encouragement, and help in working out my research works.

I gratefully acknowledge BITS-Pilani, Hyderabad Campus, for providing me with the necessary facilities, and the Department of Science and Technology (DST), Government of India, Delhi, for providing Inspire Fellowship (File No. DST/INSPIRE Fellowship/2018/IF180676) to carry out my research works.

\vspace{1.2 cm}
Sanjay Mandal,\\
ID: 2018PHXF0444H.

\end{acknowledgements}



\lhead{\emph{Contents}} 
\tableofcontents 
\addtocontents{toc}{\vspace{1em}}
\lhead{\emph{List of Tables}}
\listoftables 
\addtocontents{toc}{\vspace{1em}}
\lhead{\emph{List of Figures}}
\listoffigures 
\addtocontents{toc}{\vspace{1em}}




\lhead{\emph{List of symbols}}
\listofsymbols{ll}{

$g^{ij}:$ \,\,\,\,\,\,\, Lorentzian Metric\\
$g:$ \,\,\,\,\,\,\,\,\,\, Determinant of $g^{ij}$\\
$\Gamma^{\lambda}_{ij}:$ \,\,\,\,\, General Affine Connection\\
$\lbrace^{\lambda}_{ij}\rbrace $\,\,\,\,\,\,\,\, Levi-Civita connection\\
$\nabla_{i}$: \,\,\,\,\,\, Co-variant derivative w.r.t. Levi- Civita connection\\
$(ij):$ \,\,\,\, Symmetrization over the indices $i$ and $j$\\
$[ij]:$ \,\,\,\,\, Anti-symmetrization over the indices $i$ and $j$\\
$R^{\lambda}_{\sigma ij}:$ \,\,\, Riemann tensor \\
$R_{ij}:$\,\,\,\,\,\,\, Ricci tensor \\
$R:$ \,\,\,\,\, \,\,\,  Ricci scalar \\
$S_{M}:$ \,\,\,\,\, Matter action\\
$T_{ij}:$ \,\,\,\,\,\, Stress-energy tensor\\
$ S^{\mu \nu}_{\gamma}: $ \,\,\,\, Superpotential tensor\\
$Q_{\gamma\mu\nu}: $\,\,\, Non-metricity tensor\\
GR:\,\,\,\,\, \,\, General relativity\\
$\Lambda$CDM: $\Lambda$ cold dark matter\\
SCDM: Standard cold dark matter\\
ECs:\,\,\,\,\,  Energy Conditions\\
EoS:\,\,\,\,\,    Equation of state parameter\\
SNIa:\,\,\,   Type Ia supernovae\\
CMB: \,\,Cosmic microwave background\\
BAO: \,\,  Baryon acoustic oscillations\\
DE:\,\,\,\,\, \,  Dark energy\\
DM:\,\,\,\,\, \,   Dark matter\\
HDE:\,\,\,\,\,    Holographic dark energy\\
CG: \,\,\,\,\, \,\,  Chaplygin gas\\
MCMC:\,   Markov chain monte carlo}

\addtocontents{toc}{\vspace{2em}}

%
%


\clearpage 





\mainmatter 

\pagestyle{fancy} 


\chapter{Preliminaries in a Nutshell} 
\label{Chapter1}

\lhead{Chapter 1. \emph{Preliminaries in a Nutshell}} 

\clearpage
\pagebreak
This thesis titled {\bf Accelerated Expansion of the Universe in Non-minimally Coupled Gravity} has been focused on investigating the accelerated expansion of the universe. Before going to the investigated problems, the present chapter discusses the history, mathematical notations, basic elements, cosmological applications, fundamental theories of gravity, and cosmological observations.

\section{Historical Overviews}
The cosmic evolution of the universe always has been mysterious to everyone, and it challenges us to understand its nature. If someone looks back at history, they can find from our ancestors to the current generation, always trying to understand the universe. Currently, scientists are in search of the exact theory which will be able to explain the evolution process accurately. Starting from the day of Brahma (the ideas are described in ancient Indian mythological book, like the \textit{Bhagavatam}), where the time scale is characterized in terms of a cycle of \textit{yugas} such as \textit{Satyayuga, Tretayuga, Dwaparayuga}, and \textit{Kaliyuga}. According to this belief, the universe is going through the \textit{Kaliyuga}. The ancient Egypt mythologies were different. They believed that the Sun God \textit{Ra} was supposed to travel on a boat called \textit{Manjeet} in twelve hours of daylight. At the end of the day, \textit{Ra} died and traveled in another boat called \textit{Mesektet-boat} for twelve hours of nights and was reborn in the morning in the east with a new Ra. In the Nordic civilization of northern Europe, the Norse world tree concept is a world tree in which the whole universe on its roots and branches. They characterized the tree into three levels which include nine worlds. Chinese mythology is very rich and diverse. First, there was darkness and chaos everywhere. Then from the darkness emerged a cosmic egg. Inside that egg, a giant name \textit{Pangu} had been sleeping for billions of years. When he grew and woke up, he came out breaking the egg. He grew for 18000 years at the rate of 3 meters per day and felt tired. Then he slept and never woke up. When \textit{Pangu} went into eternal sleep, his body became various parts of the universe. Moreover, the theoretical development toward understanding the universe was started by a great thinker and mystic named Pythagoras. His theorem plays a key role in Euclid's geometry, especially where the measurements of distance are involved. After that, there are several concepts proposed to understand the evolution process, such as the central fire concept, epicycles, and the geocentric theory. There are few great minds like Aristarchus (c. 310-230 BC), Nicolaus Copernicus (1473-1543), Galileo Galilei (1564-1642), Tycho Brahe (1564-1601), Johannes Kepler (1571-1630), Isaac Newton, and Albert Einstein (1879-1955), who presented some great ideas which help us to understand our Universe from generations. The recent developments in the observations will tell us that the ancient ideas or concepts make no sense. However, they represent the most basic human aspiration to know the answer during one's lifetime.

Albert Einstein's General Theory of Relativity (GR) is one of the human mind's greatest intellectual achievements ever conceived. The fundamental supposition of this theory is that the gravitational field can be deciphered geometrically. Moreover, it is directly connected with the significant variation of the spacetime metric $g_{\mu\nu}$. Geometrically, the metric tensor provides the infinitesimal distance between two adjoining points of the spacetime continuum. Therefore, in GR, the gravitational field is fully determined by the quantities that describe the intrinsic geometrical properties and the structure of spacetime. This important idea has the fundamental implication that physical phenomena and processes locally induce the spacetime geometry itself and that space and time are not a prior determined absolute concept. For an arbitrary gravitational field, which generally varies in both space and time, the metric of the four-dimensional spacetime is non-Euclidean. Therefore, its geometric properties cannot be described any longer by the simple and well-known results of Euclidean geometry, which is constructed based on Euclid's fifth postulate of the parallels. That dictates that through an arbitrary point, one can construct one and only one parallel to a given straight line.

In 1917, Einstein proposed the first-ever relativistic cosmological model for a static, homogeneous, and isotropic universe with spherical geometry \cite{einstein}. This model raises many interesting questions, especially the gravitational instability that causes the acceleration of the model. Later, Einstein modified his equation to overcome this by introducing the cosmological constant. But, somehow, it also failed in the small-scale perturbations. Willem de Sitter (1872-1934) was the first astronomer who studied the geometrical and physical properties of the Einstein's static universe \cite{desitter, desitter2}. In his study, he found a new solution to Einstein's field equation for the vacuum universe by imposing vanishing energy density and pressure. In the early 1920s, Alexander Fridmann (1889-1925) realized that Einstein's field equations have non-static solutions that could describe the expansion of the universe and can be expressed as a function of time \cite{friedman}. His study showed that our universe started from a single event, and matter, space, and time appeared at once due to the initial explosion \cite{lemaitre}. In 1929, Edwin Hubble (1885-1972) measured the distance between the galaxies using a 100-inch telescope known as the Hubble telescope and stated that all the galaxies recede from us. Moreover, he also presented the expansion law of the universe, which is given as $v=H_0 d$, where $v$ is the velocity of the galaxy, $d$ is the distance from the observed galaxy to earth, and $H_0$ is the Hubble constant in $km/s/Mpc$ units. Currently, the Hubble constant has become a blunder in observational cosmology. Fred Hoyle (1915-2001) strongly supported the alternative theory to cosmological theory called the steady-state theory, which is later known as the ``big bang" theory. In 1964, Arno Penzias and Robert Wilson accidentally detected an isotropic cosmological microwave background, the first major confirmation of the big bang theory \cite{penzias}. Modern cosmology archived another milestone in the year 1992 by discovering anisotropies in the Cosmic Microwave Background radiation (CMB) at the level of $10^{-5}$K, which was detected by the COBE satellite team. The COBE research team initiated another  Wilkinson Microwave Anisotropy Probe (WAMP) satellite experiment. The WAMP team provided details of the full-sky map of the electromagnetic radiation from 379,000 years after the big bang.

In 1998-1999, two research groups led by Riess and Schmidt \cite{riess/1998}, and by Perlmutter \cite{perlmutter/1999} discovered the accelerated expansion of the universe; this discovery astonished the cosmologists and relativists. Because they believed that the attractive force of gravity only characterizes the expansion dynamics of the universe, that is why the universe would expect to be rapidly decelerating. In the last two decades, many astronomical observations and theoretical studies have been performed to study this unexpected scenario of the universe. Moreover, many observations, including Planck satellite data \cite{planck/2015} suggest that our universe contains only $4-5\%$ of ordinary matter (such as baryons, electrons, and other elementary particles). Rest $95\%$ of matter-energy of the universe is distributed in two unknown components known as dark matter ($~25\%$) and dark energy ($~70\%$), respectively. These outcomes lead to another cosmological paradigm called $\Lambda$CDM ($\Lambda$ Cold Dark Matter), which presumes cold dark matter as a major matter component of the universe. After that, the cosmological constant again come to picture. And the cosmological constant in Einstein's equation able to describe the late-time dynamics of the universe. To shed more light on this cosmological accelerated scenario, this thesis will discuss the late-time cosmology of the universe. Before discussing late-time cosmology, let us start with some basic mathematical foundations.

\section{Gravitation}

The first step in constructing a theory of gravitation is to set up a mathematical framework that will allow us to verify the laws of physics. As it is well known, the scalars, vectors, tensors, and mathematical quantities help us to test the properties of physical objects and dynamical systems. These mathematical quantities will be used to construct the framework. Moreover, when we begin to study the gravitational fields, in order to verify the physical laws, we need an arbitrary frame of reference. We need to develop a four-dimensional geometry in an arbitrary system. Hence, this section contains some basic mathematical quantities developed using vectors, tensors, and their differential operators.

\subsection{The Metric Tensor}
We define a differential manifold as a Riemannian space with the basic property that each point of that manifold can be represented as tensor \cite{dub}
\begin{equation}\label{1}
g_{\mu\nu}(x)=g_{\mu\nu}(x^1,x^2,x^3,...,x^n).
\end{equation}
This is a symmetric, twice covariant, and non-degenerate tensor, and we call it a metric tensor. So 
\begin{equation}\label{2}
g_{\mu\nu}=g_{\nu\mu},\,\,g=\text{det}|g_{\mu\nu}|.
\end{equation}
The fundamental properties of the function $g_{\mu\nu}$ should be continuous and has a continuous derivatives with respect to all coordinates $(x^1,x^2,x^3,…,x^n)$. This metric helps us to construct a invariant second-order differential form in a Riemannian space, defined as
\begin{equation}\label{3}
ds^2=g_{\mu\nu}dx^{\mu}dx^{\nu}.
\end{equation}
This quantity is called interval \cite{lan}. It is well known in linear algebra that a square matrix can be reduced to a diagonal form by determining the eigenvalues of the matrix. Similarly, one can reduce the matrix $g_{\mu\nu}$ to the diagonal form by considering a proper coordinate transformation. In that case, the diagonal elements of $g_{\mu\nu}$ may have different signs, and the difference between the positive and negative numbers is called the Riemannian metric's signature. Researchers generally use two types of signature conventions to explore the fate of the universe, such as (+,-,-,-) or (-,+,+,+). Throughout the thesis, we will adopt the latter signature for the metric. The differential form in equation \eqref{3} can have any sign in Riemannian space. Based on its invariant property, this interval can be characterized in the following types such as spacelike, timelike, or null-like.

\subsection{General Affine Connection}
The fundamental nature of the gravitational fields can be explored by the basic properties of the dynamics of the physical objects. This can be done by analysing the geodesics nature/ line of motion of the test particle; for this, the affine connection plays an important role. In differential geometry, the general affine connection can be written as,

\begin{equation}\label{4}
{\Gamma^{\lambda}}_{\mu\nu}=\left\lbrace{^{\lambda}}_{\mu\nu}\right\rbrace+{K^{\lambda}}_{\mu\nu}+{L^{\lambda}}_{\mu\nu},
\end{equation}
where the Levi-Civita connection/Christoffel symbol of the metric is
\begin{equation}\label{5}
\left\lbrace{^{\lambda}}_{\mu\nu}\right\rbrace\equiv\frac{1}{2}g^{\lambda \beta}\left(\partial_{\mu}g_{\beta \nu}+\partial_{\nu}g_{\beta \mu}-\partial_{\beta}g_{\mu \nu}\right),
\end{equation}
and the contortion is
\begin{equation}\label{6}
{K^{\lambda}}_{\mu\nu}\equiv\frac{1}{2}{T^{\lambda}}_{\mu\nu}+T_{(\mu}{}^{\lambda}{}_{\nu)},
\end{equation}
with the torsion tensor ${T^{\lambda}}_{\mu\nu}\equiv 2{\Gamma^{\lambda}}_{[\mu\nu]}$. The disformation $L^{\lambda}_{\mu\nu}$ in terms of the non-metricity tensor can be read as
\begin{equation}\label{7}
{L^{\lambda}}_{\mu\nu}\equiv \frac{1}{2} g^{\lambda\beta}\left(-Q_{\mu\beta\nu}-Q_{\nu \beta\mu}+Q_{\beta\mu\nu}\right).
\end{equation}
Here, the non-metricity tensor $Q_{\gamma \mu\nu}$ is defined as the covariant derivative of the metric tensor with respect to the Weyl-Cartan connection ${\Gamma^{\lambda}}_{\mu\nu}$, $Q_{\gamma \mu\nu}\equiv \nabla_{\gamma}g_{\mu\nu}$, and it can be written as \cite{hehl/1976};
\begin{equation}\label{8}
Q_{\gamma \mu\nu}=-\frac{\partial g_{\mu\nu}}{\partial x^{\gamma}}+g_{\nu \lambda}{\Gamma^{\lambda}}_{\mu\gamma}+g_{\lambda\mu}{\Gamma^{\lambda}}_{\nu\gamma}.
\end{equation}
Notice that the non-torsional part of connection $\Gamma^{\alpha}_{\mu\nu}$ is the Levi-Civita connection, whereas contorsion and disformation tensors have torsional transformation properties under the change of coordinates. As we have put together the necessary geometrical objects, we can now characterize a geometry of spacetime as follows:
\begin{itemize}
\item \textit{Metric}: the connection is metric-compatible, which indicates that $Q_{\alpha \mu\nu}(\Gamma, g)=0$. The length of vectors is conserved in metric spaces; hence non-metricity gauges how much their length changes when we parallel transport them.
\item \textit{Torsionless}: $T^{\alpha}_{\mu\nu}(\Gamma)=0$ and the connection is symmetric. The non-closure of the parallelogram generated when two infinitesimal vectors are parallel carried along one other is measured by torsion. As a result, it is commonly assumed that parallelograms do not close when torsion is present.
\item \textit{Flat}: $R^{\alpha}_{\beta\mu\nu}=0$ and the connection is not curved. Curvature is the rotation that a vector undergoes as it travels parallel along a closed curve. This creates a barrier to comparing vectors defined at various places in spacetime. However, in flat spaces, vectors do not rotate as they are conveyed, giving a stronger sense of parallelism at a distance. This is why theories developed in these environments are known be teleparallel.
\end{itemize}
Einstein's general relativity is formulated on a metric and torsionless spacetime. Also, it attributes gravity to the curvature. However, it is natural to wonder, as Einstein did later, gravity may be attributed to the other qualities that spacetime can have, such as torsion and non-metricity. So far, these three theories of gravity equivalently described GR and knocking into shape a geometrical trinity of gravity. The usual formulation of GR, for example, assumes a Levi-Civita connection, which requires vanishing both torsion and non-metricity. However, its teleparallel equivalent (TEGR) assumes a Weitzenb\"ock connection which entails zero curvature, and non-metricity \cite{maluf}. The Weitzenb\"ock condition of the vanishing sum of the curvature and torsion scalar was studied in a gravitational model with Weyl-Cartan spacetime \cite{Haghani}. Another similar formulation of GR, known as the symmetric teleparallel equivalent of GR, is a relatively unmapped field (STEGR). The gravitational interaction is described by the nonmetricity tensor $Q$, which takes into account vanishing curvature and torsion. The STEGR was first presented in a brief paper \cite{nester}, in which the authors emphasize that the formulation brings a new perspective to GR and that the gravitational interaction effects, via nonmetricity, have a character similar to the Newtonian force and are derived from a potential, namely the metric. The formulation, on the other hand, is geometric and covariant. Therefore, to represent the same physical interpretation, GR can be described by Einstein-Hibert action as $\int \sqrt{g}\,R d^4x$, the action of teleparallel equivalent $\int \sqrt{g}\,\mathcal{T} d^4x$ \cite{olmo}and the coincident GR $\int \sqrt{g}\,Q d^4x$ \cite{heisenberg, heisenberg1}. The equivalent descriptions to GR by curvature, torsion, and non-metricity provide the starting point for modified theories of gravity once the respective scalar is replaced by the arbitrary functions. These three fundamental theories are called \textit{`Geometrical Trinity of Gravity'}. We will discuss these three geometries in the following subsections.

\subsection{General Relativity}
The amazing fact about GR is that spacetime is not only curved but also dynamic. In other words, not only matter motion is affected by curved space, but also matter creates curvature in space. The matter source of spacetime can be described by Einstein equation,
\begin{equation}\label{9}
G_{\mu\nu}\equiv R_{\mu\nu}-\frac{1}{2}g_{\mu\nu}R=8\pi G T_{\mu\nu}.
\end{equation}
Here, $G_{\mu\nu}$, $R_{\mu\nu}$, and $R$ represents the Einstein tensor, Riemannian  tensor, and Ricci scalar, respectively. These tensors can be written in terms of metric tensor, Christoffel symbols and their derivatives as
\begin{equation}\label{10}
R^{\rho}_{\sigma\mu\nu}=\delta_{\mu}\left\lbrace^{\rho}_{\nu\sigma}\right\rbrace-\delta_{\nu}\left\lbrace^{\rho}_{\mu\sigma}\right\rbrace+\left\lbrace^{\rho}_{\mu\lambda}\right\rbrace \left\lbrace^{\lambda}_{\nu\sigma}\right\rbrace-\left\lbrace^{\rho}_{\nu\lambda}\right\rbrace \left\lbrace^{\lambda}_{\mu\sigma}\right\rbrace,
\end{equation}
\begin{equation}\label{11}
R_{\mu\nu}=R^{\lambda}_{\mu\lambda\nu},\,\,\,\,\, R=g^{\mu\nu}R_{\mu\nu}.
\end{equation}
Also, $T_{\mu\nu}$ describes the energy-momentum component of the universe. We shall discuss about $T_{\mu\nu}$ in the upcoming section.

It is worthy of mentioning here that the Einstein equation can be derived using variation principle from varying the Einstein-Hilbert action
\begin{equation}\label{12}
S=\frac{M_P^2}{2}\int d^4x\,\sqrt{-g}\, R+\int d^4x\, \sqrt{-g}\, L_m,
\end{equation}
where $M_P\equiv 1/\sqrt{8\pi G}$ is called reduced Planck mass. In the above action, the first term is the gravitational part and the second term is the matter part. 

Einstein's GR formulation is based on some fundamental concepts such as

\begin{itemize}
\item General co-variance: It expresses the relativity principle that laws of physics take the same form in all coordinate systems.
\item Equivalence principle: It defines the equality of gravitational and inertial mass.
\item Spacetime curvature: It provides the mass by which gravitation controls the dynamics.
\item Levi-Civita connection: It formulates without presence of the torsion and non-metricity.
\end{itemize}
\subsection{Teleparallel Equivalent to GR}
The vierbein fields, $e_{\mu}(x^i)$, act as a dynamical variable for the teleparallel gravity. As usual, $x^i$ is used to run over the spacetime coordinates, and $\mu$ denotes the tangent spacetime coordinates. At each point of the manifold, the vierbein fields form an orthonormal basis for the tangent space, which is presented by the line element of four-dimensional Minkowski spacetime i.e., $e_{\mu}e_{\nu}=\eta_{\mu\nu}=$diag$(-1,+1,+1,+1)$. In the vector component, the vierbein fields can be expressed as $e^i_{\mu}\partial_i$, and the metric tensor can be written as
\begin{equation}\label{13}
g_{\mu\nu}=\eta_{ij}e^i_{\mu}(x)e^j_{\nu}(x).
\end{equation}
Moreover, the vierbein basis follow the general relation $e^i_{\mu}e^{\mu}_j=\delta^i_j$ and $e^i_{\mu}e^{\nu}_i=\delta^{\nu}_{\mu}$.
In teleparallel gravity, the curvatureless Weitzenb$\ddot{o}$ck connection \cite{r/1923} defined as
\begin{equation}\label{14}
\hat{\Gamma}^{\gamma}_{\mu\nu}\equiv e^{\gamma}_i\partial_{\nu}e^i_{\mu}\equiv -e^i_{\mu}\partial_{\nu}e^{\gamma}_i.
\end{equation}
Using Weitzenb$\ddot{o}$ck connection one can write the non-zero torsion tensor as
\begin{equation}\label{15}
T^{\gamma}_{\mu \nu}\equiv \hat{\Gamma}^{\gamma}_{\mu\nu}-\hat{\Gamma}^{\gamma}_{\nu\mu} \equiv e^{\gamma}_i(\partial_{\mu} e^i_{\nu}-\partial_{\nu} e^i_{\mu}).
\end{equation}
The contracted form of the above torsion tensor can be written as follows: \cite{Maluf/1994,Hayashi/1979,Arcos/2004}
\begin{equation}\label{16}
\mathcal{T}\equiv S^{\mu \nu}_{\gamma}T^{\gamma}_{\mu \nu}\equiv \frac{1}{4}T^{\gamma \mu \nu}T_{\gamma \mu \nu}+\frac{1}{2}T^{\gamma \mu \nu}T_{\nu \mu \gamma}-T^{\gamma}_{\gamma \mu}T^{\nu \mu}_{\nu},
\end{equation}
where
\begin{equation}\label{17}
S^{\mu \nu}_{\gamma}=\frac{1}{2}(K^{\mu \nu}_{\gamma}+\delta^{\mu}_{\gamma}T^{\alpha \nu}_{\alpha}-\delta^{\nu}_{\gamma}T^{\alpha \mu}_{\alpha}),
\end{equation}
represents the superpotential tensor. The
difference between the Levi-Civita and Weitzenb$\ddot{o}$ck connections is the contortion tensor which is defined as
\begin{equation}\label{18}
K^{\mu \nu}_{\gamma}=-\frac{1}{2}(T^{\mu \nu}_{\gamma}-T^{\nu \mu}_{\gamma}-T^{\mu \nu}_{\gamma}).
\end{equation}
The action for  teleparallel equivalent to general relativity reads
\begin{equation}\label{19}
S=-\frac{M_P^2}{2}\int  d^4x\, e\, \mathcal{T} +\int d^4x\, e\, L_m,
\end{equation}
where $e=\sqrt{-g}$. Note that, the flat and teleparallel connections are employed in the transition from Einstein-Hilbert action to TEGR action. Other than this, there in no change in the matter action.
\subsection{Symmetric Teleparallel Equivalent to GR}
Symmetric teleparallel equivalent to GR is formulated by considering flat, vanishing torsion in the general connection, and the nonmetricity tensors. In this subsection, we shall discuss the non-metricity tensors.

The nonmetricity tensor and its traces are such that
\begin{equation}
\label{20}
Q_{\gamma\mu\nu}=\nabla_{\gamma}g_{\mu\nu}\,,
\end{equation}
\begin{equation}
\label{21}
Q_{\gamma}={{Q_{\gamma}}^{\mu}}_{\mu}\,, \qquad \widetilde{Q}_{\gamma}={Q^{\mu}}_{\gamma\mu}\,.
\end{equation}
Moreover, the superpotential as a function of nonmetricity tensor is given by
\begin{equation}
\label{22}
4{P^{\gamma}}_{\mu\nu}=-{Q^{\gamma}}_{\mu\nu}+2Q_{({\mu^{^{\gamma}}}{\nu})}-Q^{\gamma}g_{\mu\nu}-\widetilde{Q}^{\gamma}g_{\mu\nu}-\delta^{\gamma}_{{(\gamma^{^{Q}}}\nu)}\,,
\end{equation}
where the trace of nonmetricity tensor \cite{Jimenez/2018} reads
\begin{equation}
\label{23}
Q=-Q_{\gamma\mu\nu}P^{\gamma\mu\nu}\,.
\end{equation}
The action for symmetric teleparallel equivalent to general relativity is the following
\begin{equation}\label{24}
S=-\frac{M_P^2}{2}\int d^4x\, \sqrt{-g}\, Q +\int d^4x\, \sqrt{-g}\, L_m.
\end{equation}
Note that, the flat and non-metricity tensors are employed in the transition from Einstein-Hilbert action to STGR action with no change in the matter action.
\section{Matter Components}
The energy-momentum tensor can describe the matter source and its energy, fluxes, and momentum in spacetime. It is a second-rank tensor quantity and represents the amount of energy, momentum, and fluxes in a unit volume. With its help, one can discuss various laws of physics for the matter source. Mathematically, it is a $4\times 4$ matrix denoted by $T^{\mu\nu}$ and its elements provides the following information:
\begin{itemize}
\item $T^{00}$= energy density
\item $T^{\mu 0}=c\times$ density of $\mu$th component of momentum ($\nu=1,\,2,\,3$)
\item $T^{\nu 0}=c^{-1}\times$ energy flux in the $\nu$th direction ($\nu=1,\,2,\,3$)
\item $T^{\mu\nu}=$ flux in the $\mu$th direction of $\nu$th component of momentum ($\mu,\,\nu=1,\,2,\,3$).
\end{itemize}

The energy-momentum tensor is locally conserved, i.e., $\nabla_{\mu}T^{\mu\nu}=0$. One can understand this in the following aspect. Einstein equation for any matter component satisfies $\nabla_{\mu}G^{\mu\nu}=0$, which indicates that the energy-momentum is conserved locally. Also, one can derive $\nabla_{\mu}T^{\mu\nu}=0$ directly from the Einstein's equation.

In this thesis, we will work on perfect fluid, that viscosity and heat conduction can be neglected. For perfect fluid matter, the energy-momentum tensor in the rest frame can be written as
\begin{equation}\label{25}
T^{\mu}_{\nu}=\text{diag}(-\rho, p,p,p),
\end{equation}
where $\rho$ is the energy density and $p$ is the pressure along all directions in the spacetime. One can write the energy-momentum tensor in an arbitrary coordinate to perform coordinate transformation with the fluid velocity $u^{\mu}(x)$ as
\begin{equation}\label{26}
T_{\mu\nu}=(\rho+p)u_{\mu}u_{\nu}+g_{\mu\nu}p.
\end{equation}
\subsection{Energy Conditions}
It is well known that energy conditions (ECs) play a vital role in cosmology. Before going to any particular cosmological scenario, let us discuss the energy conditions in general. Suppose an observer is moving with a velocity $u^{\mu}$, measured the energy-momentum density $T_{\mu\nu}u^{\mu}$. Further the energy density measured $T_{\mu\nu}u^{\mu}u^{\nu}$.

The null energy condition (NEC) states that
\begin{equation}\label{27}
T_{\mu\nu}u^{\mu}u^{\nu}\geq 0,
\end{equation}
for every null vector field $u^{\mu}$. For an ideal fluid description, the NEC reduces to $\rho+p\geq 0$. NEC plays an important role in cosmology. It helps us to understand whether the universe undergoes inflation or super-inflation; or whether the universe will develop a singularity or a bounce solution. The NEC is also a limit to cosmology; violating this may break many physical laws. NEC is very weak. One can integrate NEC with weak energy conditions.

The weak energy condition (WEC) states that
 \begin{equation}\label{28}
T_{\mu\nu}u^{\mu}u^{\nu}\geq 0,
\end{equation}
for every time-like vector $u^{\mu}$(thus for an observer). For an ideal fluid content, the WEC reduces to $\rho+p\geq 0$, and the energy density should always be non-negative, i.e., $\rho\geq 0$. The latter condition of WEC makes it strong. Because it rules out one of the interesting geometries in string theory, the AdS spacetime, which is believed to be one of the best-understood examples of quantum gravity.

The generalization of WEC is the dominant energy condition (DEC), which states that it holds WEC and the second condition of WEC restricts $\rho\geq |p|$.

The strong energy condition (SEC) states that
\begin{equation}\label{29}
\left(T_{\mu\nu}-\frac{1}{2}g_{\mu\nu}T\right)u^{\mu}u^{\nu}\geq 0,
\end{equation}
for every time-like vector $u^{\mu}$. In the ideal fluid distribution, the SEC reads $\rho+3p\geq 0$. The SEC can be derived from Einstein's equation, and it can be written as $R_{\mu\nu}u^{\mu}u^{\nu}\geq 0$. The SEC states that gravity is attractive. On the other hand, currently, the universe is going through the accelerated expansion phase, which was confirmed through the CMB observations. Although the dark energy observations contradict the SEC, however SEC is itself a strong statement.

\section{Cosmology}
Our universe is extremely large and complicated. It is very difficult to understand its nature and cosmic dynamics. To formulate a theory of cosmology, we need to set up some levels of approximations. It is well-known that GR is the fundamental theory to explore the structure and evolution of the universe in large-scale structure, which is based on cosmological principle. The cosmological principle first introduced by Einstein states that our universe is homogeneous and isotropic. Homogeneous means our universe evolves and looks the same at all locations, and isotropic means our universe looks the same in all spatial directions. Initially, there were not enough experimental data to test the cosmological principle, but later it was well tested by the observable universe. The deviation from the cosmological principle can be treated as perturbations. The application of the cosmological principle started from the Friedmann equations. There are two ways to derive the fundamental setup for the Friedmann equation: Newtonian cosmology and Relativistic cosmology.

\subsection{Newtonian Cosmology}
Consider a homogeneous sphere with mass $M$ and  that a sphere is expanding or contracting with radius $r(t)$. A test particle with infinitesimal mass $m$ is placed at the surface of a sphere. According to Newton's law of gravity, the gravitational force $F$ experienced by the test particle is,
\begin{equation}\label{30}
F=-\frac{G M m}{r^2},
\end{equation}
where $M$ and $m$ are the masses of the two objects respectively. Further, Newton's second law of motion says that the force and acceleration are related by the following relation
\begin{equation}\label{31}
F=m \ddot{r}.
\end{equation}
Now comparing the above two equations, we get 
\begin{equation}\label{32}
\ddot{r}=-\frac{G M}{r^2}.
\end{equation}
Consider $r$ as a function of time $t$ and replace by the scale factor $a(t)$. So, we can write $r(t)$ as
\begin{equation}\label{33}
r(t)= a(t)x_r,
\end{equation}
where $x_r$ is the comoving radius of a sphere.
 Multiplying $da/dt$ at both sides of the equation \eqref{32} and integrating it, gives
\begin{equation}\label{34}
\left(\frac{\dot{a}}{a}\right)^2=\frac{8\pi G}{3c^2}\rho(t)-\frac{\kappa c^2}{R_0a^2}.
\end{equation}
This is known as Friedmann equation. Here $\kappa$ can be $+1, 0, -1$ for closed, flat and open universe, respectively. Now, we have one equation with unknown as $a(t)$ and $\rho(t)$. So, to study the unknown quantities, we need another equation. For that, we have taken
the first law of thermodynamics in consideration:
\begin{equation}\label{35}
dQ=dE+p dV,
\end{equation}
where $dQ$ represents the change in heat flow, $dE$ represents the change in internal energy, $p$ is the pressure, and $dV$ is the change in volume. As we are studying the homogeneous and isotropic universe, so $dQ=0$. After some algebraic manipulation, one can find the following equation
\begin{equation}\label{36}
\dot{\rho}+3 \frac{\dot{a}}{a}(\rho+p)=0.
\end{equation}
This equation is known as the fluid equation/energy conservation equation. Combining the first Friedmann equation and fluid equation, one can find the acceleration equation as,
\begin{equation}\label{37}
\frac{\ddot{a}}{a}=-\frac{4\pi G}{3c^2}(\rho+3p).
\end{equation}
If the matter components are filled with ordinary baryonic matter, then the energy density $\rho$ and pressure $p$ become positive, resulting in the decreasing expansion of the system. The expansion rate will go up if energy density $\rho>0$ and
\begin{equation}\label{38}
p<-\frac{1}{3}\rho.
\end{equation}
One can check this from equation \eqref{37}.

\subsection{Relativistic Cosmology}
The first step into relativistic cosmology is finding an appropriate metric of the universe. This task is very much simple as we already assume our universe to be homogeneous and isotropic. In a homogeneous and isotropic geometry, the scalar curvature must have a finite value at all points of the universe. Therefore in the early 1920s, anticipated by Friedmann and Lema\^itre, rigorously shown by Robertson \cite{robert} and Walker \cite{walker}, that the only metric which can describe such a universe in a spherical coordinate system $(r, \theta, \phi)$ is given by
\begin{equation}\label{39}
ds^2=-c^2dt^2+a^2(t)\left[\frac{dr^2}{1-\kappa r^2}+r^2(d\theta^2+sin^2\theta d\phi^2)\right],
\end{equation}
where $a(t)$ is the scale factor, and $\kappa$ is the constant discussed above. This metric is known as Friedmann-Lema\^itre-Robertson-Walker (FLRW) metric. This FLRW metric gives the spacetime interval between two arbitrary events occurring independently in the universe. Hence, it is defined in a comoving coordinate system. For $\kappa=0$, equation \eqref{39} reduces to
\begin{equation}\label{flrw}
ds^2=-c^2dt^2+a^2(t)\left[dr^2+r^2(d\theta^2+sin^2\theta d\phi^2)\right].
\end{equation}
Now, using FLRW metric \eqref{39}, Einstein's gravitational field equation \eqref{9}, and energy-momentum tensor for perfect fluid \eqref{26}, one can find the Friedmann equations:
\begin{equation}\label{41}
3H^2+\frac{3\kappa c^2}{a^2}=8\pi G \rho(t),
\end{equation}
\begin{equation}\label{42}
2\dot{H}+3H^2+\frac{\kappa c^2}{a^2}=-\frac{8 \pi G}{c^2}p
\end{equation}
where `.' represents one time derivative with respect to $t$, $H=\frac{\dot{a}}{a}$ is the Hubble parameter.

In order to close the system of cosmological evolution, we need to specify the equation of state for the fluid distribution, which is given by
\begin{equation}\label{43}
p=\omega \rho,
\end{equation}
where $\omega$ is known as the equation of state parameter.

It is worthy to mention here that, the deceleration parameter is an important cosmological quantity, defined as
\begin{equation}\label{44}
q=-1-\frac{\dot{H}}{H^2}=-\frac{\ddot{a}}{aH^2}.
\end{equation}
Also, one can rewrite the deceleration parameter using Friedmann equation with $\kappa=0$ as
\begin{equation}\label{45}
q=\frac{1}{2}(1+3\omega).
\end{equation}
Now, we have a complete set of differential equations by which one can discuss some exact cosmological models. Let us discuss some widely used fundamental models.

For the assumed equation of state with constant equation of state parameter $\omega$, one can easily integrate the fluid equation to obtain
\begin{equation}\label{46}
\rho(a)=\frac{\rho_0}{a^{3(1+\omega)}},
\end{equation}
where $\rho_0$ is an integration constant. Further, one can find the solution for scale factor as a function of $t$ from the first equation of Friedmann as
\begin{equation}\label{47}
  a(t)\propto \begin{cases}
    t^{2/3(1+\omega)}, & \omega\neq -1,\\
    e^{\lambda t}, & \omega=-1,
  \end{cases}
\end{equation}
where $\lambda$ is an integration constant. For different value of $\omega$, the above general expression of scale factor $a(t)$ and $\rho$ explains different cosmological evolutions. Such as
\begin{itemize}
\item $\omega=1$ represents stiff fluid dominated universe,
\item $\omega=1/3$ represents radiation dominated universe,
\item $\omega=0$ represents matter dominated universe,
\item $\omega=-1$ represents $\Lambda$CDM universe; the first candidate of dark energy.
\end{itemize}
There are some cosmological quantities that help us to understand cosmic dynamics through observational cosmology. Such as the dimensionless energy density, defined as
\begin{equation}\label{48}
\Omega(t)=\frac{\rho(t)}{\rho_c(t)},
\end{equation}
where $\rho_c(t)=\frac{3H^2}{8\pi G}$ is the critical energy density. Moreover, the relation between scale factor $a(t)$ and redshift $z$ plays a key role in observational cosmology, is given by
\begin{equation}\label{49}
\frac{a}{a_0}=\frac{1}{1+z},
\end{equation}
where $a_0$ is the present-day scale factor.

\section{Cosmological Observations}
Since 1929, cosmological observations have played a vital role in studying the expansion history of the universe. The effect of the dark energy probe can be mainly detected via the luminosity distance $d_L(z)$ and the angular diameter distance $D_A(z)$. In this section, we will briefly overview some cosmological observations in the search of cosmic expansion.
\subsection{Type Ia Supernovae}
Type Ia Supernovae (SNIa) is a sub-category of a massive explosion in a large-scale structure that happens due to the explosion of a white dwarf star. A white dwarf star can accumulate mass from its nearby star and approach the Chandrasekhar mass limit, resulting in an explosion \cite{1}. Therefore, SNIa can be used as a standard candle to measure the luminosity distance \cite{2,3,4}. In 1998, Riess et al. \cite{riess/1998} discovered the accelerated expansion of the universe using 16 distant and 34 nearby SNIa from the Hubble telescope observations. In 1999, Perlmutter et al. \cite{perlmutter/1999} confirmed the cosmic acceleration by analyzing 18 nearby supernovae (SN) from the Calan-Tololo sample and 42- high-redshift SN. This discovery has surprised the scientific community since Edwin Hubble's cosmic expansion discovery in 1929. For this ground-breaking discovery Saul Perlmutter, Brian Schmidt, and Adam Riess won the Nobel prize in Physics in 2011. In recent years, surveys on SNIa have drawn more and more attention \cite{7,9,10}. Many research groups have focused on this field, such as the Sloan Digital Sky Survey (SDSS) SN Survey \cite{19}, the Lick Observatory Supernova Search (LOSS) \cite{18}, the Carnegie Supernova Project (CSP) \cite{17}, the Nearby Supernova Factory (NSF) \cite{16}, the Equation of State: SupErNovae trace Cosmic Expansion (ESSENCE) \cite{15}, the Supernova Legacy Survey (SNLS) \cite{1314}, and the Higher-Z Team \cite{1112}, etc.
\subsection{Cosmic Microwave Background }
Cosmic Microwave Background (CMB) is the legacy of the cosmic recombination epoch. It provides abundant information about the early universe. In 1964, Penzias and Wilson \cite{37} firstly detected the CMB radiation, and won the Nobel prize in physics in 1978 for the achievement. Their work strongly supported the Big Bang cosmology of the universe \cite{38}. In 1989, the COBE research group launched the first generation of CMB satellite and discovered CMB anisotropy of the universe \cite{39}. Their discovery helps to explore the dynamics of the universe more precisely. Two lead researchers of the COBE research group, Smooth, and Mather, received the Nobel prize in Physics in 2006. The BOOMERang, TOCO, and maxima experiments \cite{40} were the first to measure the acoustic oscillations in the CMB anisotropy angular spectrum \cite{41,42}. The Wilkison Microwave Anisotropy Probe (WAMP) is the second generation of CMB satellite, launched in 2001, and it measured the CMB spectrum and probing the cosmological parameters with higher accuracy \cite{WAMP}. Recently, its successor, namely, the Planck satellite, was launched in 2009, and the early result was released \cite{Planck/2011}.
\subsection{Baryon Acoustic Oscillations (BAO)}
Baryon Acoustic Oscillations (BAO) refer to the clustering or overdensity of the baryonic matter at certain length scales due to acoustic waves which propagate in the early universe \cite{41,62}. Similar to SNIa, BAO provides a standard candle for the length scales in cosmology, which helps us to explore the expansion history of the universe. The BAO measurement does not require the precise measurement of galaxy magnitude and galaxy image resolved; instead, it requires the three-dimensional position of the galaxy. These observations are less affected by the astronomical uncertainties than the other probes of dark energy. Still, it also suffers some uncertainties, such as the redshift distortion of clustering and the effect of non-gravitational evolution \cite{69}. There are many experiments conducted for BAO measurements, such as the Two-degree-Field Galaxy Redshift Survey (2dFGRS) \cite{72}, SDSS \cite{73}, etc. SDSS is the most successful survey for BAO observation, which continuously released its eighth data in 2011 \cite{75}.
\subsection{Hubble Parameter Measurements}
In 1929, Hubble discovered the linear correlation between the apparent distance to the galaxies $D$ and their recession velocity $v$:
\begin{equation}
v=H_0 D,
\end{equation}
where $H_0$ is the Hubble constant. This discovery opened a new era in modern cosmology and provided a piece of strong evidence that our universe is expanding. Soon after, researchers found that the $H_0$ should be replaced by $H(z)$ as a function of $z$ in Hubble's law. As discussed previously, $H(z)$ describes the expansion history of the universe and plays a crucial role in modern cosmology and observations. In the beginning, Hubble and Humason \cite{136} measured a value of $H_0=500$ km/s/Mpc, which is comparatively very high compared to the present measurements because of the high uncertainties/errors. Since the Hubble Space Telescope (HST) launch, the estimated value of $H_0$ is between $50$ and $100$ km/s/Mpc. For example, Friedmann et al. \cite{138}, obtained $H_0=72\pm 8$ km/s/Mpc, Sandage et al., \cite{139} gave $H_0=62.3\pm 6.3$ km/s/Mpc, Riess et al. \cite{140,141} advocated $H_0=73.8\pm 3.6$ km/s/Mpc with a $4\%$ uncertainties and $H_0=73.8\pm 2.4$ km/s/Mpc with a $3.3\%$ uncertainties. In the future, the goal will be to find $H_0$ with an uncertainty of $1\%$ for the next decade \cite{145}.
\subsection{Other Cosmological Probes}
Several other cosmological observations also help us to understand the dynamics of the universe. For example, Weak lensing (WL) is the slight distortions of the distant galaxies' images due to  the gravitational bending of light by structures in the universe, Galaxy Clusters (CL) are the largest bound objects in the universe. Gamma-Ray Burst (GRB) are the most luminous electromagnetic events happens due to the highly energetic explosions in distant galaxies, X-Ray observations deal with the X-ray coming from celestial objects, cosmic age tests deal with the cosmic age problem of the universe, etc. For more detail on these observations, one can see \cite{perkin}.
\section{Accelerated Expansion}
At the beginning of the 20th century, Albert Einstein proposed GR, which changed our understanding of the universe. It is growing by a prominent number of correct observations and exploring the hidden scenarios of the universe in modern cosmology. Later on, a group of supernovae observations confirmed that our universe is currently going through the accelerated expansion phase \cite{riess/1998}. This causes high negative pressure in the universe produced by the unknown form of energy and matter called dark energy (DE) and dark matter (DM). Finding the properties of the unknown form of energy is a challenging task for researchers in the modern era. In GR, the cosmological constant, $\Lambda$, is the simplest candidate, which explains the vacuum energy \cite{carroll/1992, Sahni/2000}. With the presence of cosmological constant $\Lambda$, the Friedmann equations can be written as
\begin{equation}\label{50}
3H^2+\frac{3\kappa c^2}{a^2}=8\pi G \rho +\Lambda c^2,
\end{equation}
\begin{equation}\label{51}
2\dot{H}+3H^2+\frac{\kappa c^2}{a^2}=-\frac{8 \pi G}{c^2}p+\Lambda c^2.
\end{equation}
In 1917, Einstein introduced the cosmological constant in his field equations to build the first static general relativistic cosmological model. For example, one can simply check that for the case of a flat vacuum universe with $\rho=0=p$, and $\kappa=0$. For this one can find the following solution for the Hubble parameter and scale factor as
\begin{equation}\label{52}
H\equiv H_0=\sqrt{\frac{\Lambda c^2}{3}}=constant,
\end{equation}
\begin{equation}\label{53}
a(t)=a_0exp\left(\sqrt{\frac{\Lambda c^2}{3}}t\right)=a_0exp(H_0t).
\end{equation}
This solution to the Einstein field equations is known as the de Sitter solution, and it plays a fundamental role in modern cosmology. Later, this $\Lambda$ plays a central role for dark energy to describe the late-time acceleration of the universe, and also passes the recent dark energy probes. But, somehow, it fails to overcome some fundamental issues to describe the evolution of the universe. Therefore, researchers feel GR is needed to modify. However, in the search of proper description of gravity, several modified theories of gravity are proposed in the literature, considering GR a fundamental theory of gravity. Modified gravity theories are the generalization of GR, and it violates the Strong Equivalence Principle (SEP) \cite{joyce/2016}. Despite little progress so far in understanding cosmic acceleration \cite{baker/2019}, modified gravity studies are important as they provide reliable, logical alternatives to GR and may ease some of the current problems. Also, modified theories of gravity are well-known for their successful description of the accelerated expansion of the universe. In the last two decades, many works have been carried out in modified gravity theories to explore and understand the evolution process of the universe (see the references \cite{peracaula/2019}).

\section{Modified Theories of Gravity}
GR has attractive features, but several theoretical challenges still motivate us to explore some modifications in GR. For example, GR does not provide us with sufficient ideas to reconcile shortcomings like the initial singularity, fine-tuning, flatness issues,  cosmological constant, and cosmic coincidence problems \cite{Sahni/2000, Padmanabhan/2003}. Also, some theoretical arguments indicate that GR suffers from shortcomings like GR fails to explain the local energy-momentum conservation. This property of GR is not satisfactory as all the fundamental interactions in the universe follow the principle of local conservation of energy-momentum tensor. GR fails to be quantized: the formulation of gravity through quantum field theory is essential for the unification of all fundamental interactions. However, all the attempts to find a consistent quantum Gauge field theory for GR have failed.

Several modified theories are introduced in the literature to overcome all these problems. 
Modified theories of gravity are essential for explaining the late-time cosmic acceleration of the universe.
\begin{itemize}
\item	It gives appropriate unification of early-time inflation and late-time cosmic acceleration.
\item	It serves as a basis for some cosmological models providing a unified explanation of dark matter and dark energy.
\item	Modified gravity helps us to study the transition from non-phantom to phantom phase without introducing any exotic matter.
\item	It also sometimes describes the transition from deceleration to acceleration phase in the evolution of the universe.
\end{itemize}
 However, modified theories of gravity can be categorized into two types based on the coupling between matter and geometry: minimal and non-minimal. Usually, minimal couplings are weak couplings, while non-minimal couplings represent strong couplings. In mathematical point of view, the cosmological models with additional terms refereed to as minimal coupling, for example; $f(R,T)=R+\lambda T$, $f(R,T)=\beta_1 R^\mu+\beta_2 R^\nu+\frac{2k^2}{-1+3\omega}T$ etc. in modified theories of gravity.  At the same time, the cosmological models with product terms are treated as non-minimal coupling, for example; $f(R,T)=R(1+\lambda T)$, $f(R,T,R_{\mu \nu}T^{\mu \nu})+\beta T$ in modified theories of gravity. Some of the minimal and non-minimal coupling theories are discussed in the following subsections, and their applications are discussed in the upcoming chapters.

\subsection{$f(R)$ Gravity and Its Extensions}
One of the simple extension of GR is $f(R)$ gravity, in which the Ricci scalar $ R$ in the standard Einstein-Hilbert action is replaced by an arbitrary function of Ricci scalar $R$, was proposed by Buchdahl in  1970 \cite{bud}. Later, there are several works have been done to explore the gravitational interaction of the universe through cosmological applications (see \cite{fr}).  The action in $f(R)$ gravity reads
\begin{equation}
\label{54}
\mathcal{S}=\frac{1}{2\kappa}\int \sqrt{-g}\,f(R)\,d^4x+\int \mathcal{L}_m
d^4x,
\end{equation}
where $\kappa=8\pi G, \mathcal{L}_m$ is the matter Lagrangian.

By varying this action with respect to a metric we find
\begin{equation}
\label{55}
f_R(R)R_{\mu\nu}-\frac{1}{2}g_{\mu\nu} f(R)-(\nabla_{\mu}\nabla_{\nu}-g_{\mu\nu}\square)f_R(R)=\kappa T_{\mu\nu},
\end{equation}
where $f_R(R)=df(R)/dR$, $\nabla$ represents covariant derivative, $\square$ represents d'Alembert operator, and $T_{\mu\nu}$ is the energy momentum tensor of the matter which is defined by,
\begin{align*}
T_{\mu\nu}=-\frac{2}{\sqrt{-g}}\frac{\delta\left(\sqrt{-g}\,\mathcal{L}_m\right)}{\delta g^{\mu\nu}}.
\end{align*}

Considering the contraction of Eq. \eqref{55}, provides the following relationship
\begin{equation}
\label{56}
R f_R(R)-2f(R)+3\square f_R(R)=T,
\end{equation}
where $R$ is the Ricci scalar, and $T=T^{\mu}_{\mu}$ is the stress of the energy-momentum tensor.

The trace equation \eqref{56} can be used to simplify the field equations and that then can be treated as a constraint equation. In addition, there are more generalization of GR have been presented in the literature based on the non-minimal coupling between matter and geometry. Among those theories, $f(R,T)$ gravity is a well-known generalization of $f(R)$ gravity, proposed by T. Harko et al. \cite{Harko/2011}, and action for this gravity can be written as 
\begin{equation}\label{57}
S=\frac{1}{16\pi}\int d^{4}x\sqrt{-g}f(R,T)+\int d^{4}x\sqrt{-g}L_m.
\end{equation}
Apart from these two, there are other modified theories have been formulated to explore the dynamics of the universe such as $f(G)$ gravity \cite{fg},$f(R,L_m)$ gravity \cite{frlm}, etc.

\subsection{$f(\mathcal{T})$ Gravity and Its Extensions}
Teleparallel gravity is the modification of TEGR, which was proposed by R. Ferraro and F. Fiorini \cite{ft}. The action for teleparallel gravity is represented as 
\begin{equation}\label{58}
S=\frac{1}{16\pi G}\int[\mathcal{T}+f(\mathcal{T})]e\, d^4x,
\end{equation}
where $e=det(e^i_\mu)=\sqrt{-g}$ and $G$ is Newtonian gravitational
constant.
Varying the action $S+L_m$, where $L_m$ represent the matter Lagrangian
yields the field equations as 
\begin{multline}  \label{59}
e^{-1}\partial_{\mu}(ee^{\gamma}_i S^{\mu
\nu}_{\gamma})(1+f_{\mathcal{T}})-(1+f_{\mathcal{T}})e^{\lambda}_i T^{\gamma}_{\mu \lambda}S^{\nu
\mu}_{\gamma} +e^{\gamma}_i S^{\mu \nu}_{\gamma}\partial_{\mu}(\mathcal{T})f_{\mathcal{T}\mathcal{T}}+%
\frac{1}{4}e^{\nu}_i[\mathcal{T}+f(\mathcal{T})]=\frac{k^2}{2} e^{\gamma}_iT^{(M)\nu}_{\gamma},
\end{multline}
where $f_{\mathcal{T}}=df(\mathcal{T})/d\mathcal{T}$, $f_{\mathcal{T}\mathcal{T}}=d^2f(\mathcal{T})/d\mathcal{T}^2$.
 Several extensions of $f(\mathcal{T})$ have been proposed in the literature such as $f(\mathcal{T},B)$ gravity (where $B$ represents the boundary term ) \cite{ftb}, $f(\mathcal{T},T_G)$ gravity (where $T_G$ represents teleparallel equivalent Gauss-Bonnet (TEGB) term) \cite{ftg}, etc.
\subsection{$f(Q)$ Gravity and Its Extensions}
Symmetric teleparallel gravity or $f(Q)$ gravity is the generalization of STEGR theory, proposed in \cite{Jimenez/2018}. For this theory, one can consider the action for matter coupling in $f(Q)$ gravity as \cite{Jimenez/2018}
\begin{equation}
\label{60}
S=\int d^4x \sqrt{-g}\left[\frac{1}{2}f_1(Q)+f_2(Q)L_M\right],
\end{equation}
where $g$ is the determinant of metric, $f_1(Q)$ and $f_2(Q)$ are the arbitrary functions of the non-metricity $Q$, and $L_M$ is the Lagrangian for the matter fields.

To simplify the formulation, let us introduce the following notations
\begin{align}
\label{61}
f=f_1(Q)\,+2 f_2(Q)L_M,\\
F=f_1'(Q)+2 f_2'(Q)L_M,
\end{align}
where primes ($'$) represent the derivatives of functions $f_1(Q),$ and $ f_2(Q)$ with respect to $Q$.

By varying action \eqref{60} with respect to metric tensor, we can write the gravitational field equation:
\begin{equation}
\label{62}
\frac{2}{\sqrt{-g}}\nabla_{\gamma}\left( \sqrt{-g}F {P^{\gamma}}_{\mu\nu}\right)+\frac{1}{2}g_{\mu\nu}f_1 
+F \left(P_{\mu\gamma i}{Q_{\nu}}^{\gamma i}-2Q_{\gamma i \mu}{P^{\gamma i}}_{\nu} \right)=-f_2 T_{\mu\nu}\,.
\end{equation}

One can see that, if we consider $f_2(Q)=1$, then the non-minimal coupling theory reduces to minimal coupling theory of gravity. For this scenario the action reads
\begin{equation}
\label{63}
\mathcal{S}=\int \frac{1}{2}\,f(Q)\,\sqrt{-g}\,d^4x+\int \mathcal{L}_m\,\sqrt{-g}\,d^4x\,.
\end{equation}
By varying action \eqref{63} with respect to metric tensor, we can write the gravitational field equation, which is given by
\begin{equation}\label{64}
\frac{2}{\sqrt{-g}}\nabla_{\gamma}\left( \sqrt{-g}f_Q {P^{\gamma}}_{\mu\nu}\right)+\frac{1}{2}g_{\mu\nu}f
+f_Q\left(P_{\mu\gamma i}{Q_{\nu}}^{\gamma i}-2Q_{\gamma i \mu}{P^{\gamma i}}_{\nu} \right)=-T_{\mu\nu}\,,
\end{equation}
where $f_Q=\frac{df}{dQ}$. Besides, we can also take the variation of \eqref{1} with respect to the connection, yielding to 
\begin{equation}\label{65}
\nabla_{\mu}\nabla_{\gamma}\left( \sqrt{-g}f_Q {P^{\gamma}}_{\mu\nu}\right)=0\,.
\end{equation}
$f(Q,T)$ gravity is the generalization of $f(Q)$ gravity, proposed by Xu et al., \cite{fqt}, where in the Einstein-Hilbert action the Lagrangian $f(Q)$ replaced by an arbitrary function of $Q$ and the trace of energy-momentum tensor $T$ as $f(Q,T)$.

\section{Conclusion}

In this chapter,  we have discussed the evolution of the understanding of the universe from our ancestors to the modern days. But, the proper geometrical representation of the universe started with Einstein's general theory of relativity. Later, in this chapter, we have presented the fundamental mathematical frameworks, their applications to cosmology, and cosmological observations, which help to validate the cosmological models. The fundamental theory of gravity, like general relativity, fails to address some fundamental issues of the universe, and it seems incomplete. Therefore, its modifications and generalization are more successful in addressing this issue, and we have concluded this chapter by over-viewing the modified theories of gravity. In the upcoming chapters, some problems are investigated applying the above-discussed modified theories of gravity.




\chapter{Accelerating Universe in Hybrid and Logarithmic Teleparallel Gravity} 

\label{Chapter3} 

\lhead{Chapter 2. \emph{Accelerating Universe in Hybrid and Logarithmic Teleparallel Gravity}} 


\vspace{10 cm}
* The work, in this chapter, is covered by the following publication:

\textit{Accelerating Universe in Hybrid and Logarithmic Teleparallel Gravity}, Physics of the Dark Universe, \textbf{28}, 100551 (2020).
\clearpage
This chapter aims to discuss the accelerated expansion of the universe in the background of hybrid and logarithmic teleparallel gravity. For this purpose, a well-known deceleration parameter is considered and constrained its free parameter using observational measurements. Next, this chapter presents a few geometric diagnostics of this parametrization to understand the nature of dark energy and its deviation from the $\Lambda$CDM cosmology. Finally, the energy conditions to check the consistency of the parameter spaces for both the teleparallel gravity models are studied. The outcomes show that SEC violates both models, which is an essential recipe for obtaining an accelerating universe.

\section{Introduction} 

Teleparallel gravity is a well-established and well-motivated modified
theory of gravity inspired from $f(R)$ gravity \cite{cosmo7to9} (See \cite%
{reviewft} for a review on teleparallel gravity). In teleparallel gravity,
the Ricci scalar $R$ of the underlying geometry in the action is replaced by
an arbitrary functional form of torsion scalar $\mathcal{T}$. Thus, in teleparallel
gravity, instead of using the torsionless Levi-Civita connection (which is
usually assumed in GR), the curvatureless Weitzenb\"{o}ck connection is
employed in which the corresponding dynamical fields are the four linearly
independent verbeins, and $\mathcal{T}$ is related to the antisymmetric connection
following from the non-holonomic basis \cite{cosmoft, cosmo10to12}.\newline
Linear $f(\mathcal{T})$ gravity models are the teleparallel equivalent of GR (TEGR) 
\cite{cosmo14}. Nonetheless, $f(\mathcal{T})$ gravity differ significantly from $f(R)$
gravity in the fact that the field equations in $f(\mathcal{T})$ gravity are always at
second-order compared to the usual fourth-order in $f(R)$ gravity. This owes
to the fact that the torsion scalar contains only the first derivatives of
the vierbeins and thus makes cosmology in $f(\mathcal{T})$ gravity much simpler.
However, despite being a second-order theory, very few exact solutions to
the field equations have been reported in the literature. Power-law solutions in
FLRW spacetime have been reported in \cite{cosmo19to20}, while for
anisotropic spacetimes in \cite{cosmo21}. Solutions for Bianchi I spacetime
and static spherically spacetimes can be found in \cite{cosmoft} and \cite%
{cosmo22to23} respectively.

Since cosmology in $f(\mathcal{T})$ gravity is much simpler compared to other modified
gravity theories, it has been employed to model inflation \cite{inflation},
late time acceleration \cite{late} and big bounce \cite{bounce}. The
instability epochs of self-gravitating objects coupled with anisotropic
radiative matter content and the instability of cylindrical compact object
in $f(\mathcal{T})$ gravity have been discussed in Ref. \cite{bhatti/2017,bhatti/2017a}.

The chapter is organized as follows: Section \ref{IIa} presents an
overview of $f(\mathcal{T})$ gravity. Section \ref{IIIa} describes the kinematic
variables obtained from a parametrization of the deceleration parameter used to
obtain the exact solutions of the field equations. Section \ref{IVa}
present the hybrid and logarithmic teleparallel gravity models and obtain
the expressions of pressure, density, and EoS parameter. Section \ref{Va} presents some geometric diagnostics of the parametrization of the deceleration
parameter. Section \ref{VIa} studies the energy conditions for both the
teleparallel gravity models. In section \ref{VIIa}, we obtain some
observational bounds on the free parameters of the parametrization by
performing a chi-square test using Hubble datasets with $57$ datapoints,
Supernovae datasets consisting of $580$ data points from Union$2.1$
compilation datasets and BAO datasets. Finally, Section \ref{VIIIa} presents this chapter's results and conclusions.

\section{Field Equations in $f(\mathcal{T})$ Gravity}\label{IIa}


For a flat FLRW universe with 
the scale factor $a(t)$, gives 
\begin{equation}  \label{a1`}
e^{i}_{\mu}=diag(1,a,a,a).
\end{equation}
Employing \eqref{flrw} into the field equation \eqref{59}, the modified
Friedman equations read 
\begin{equation}  \label{a2}
H^{2}=\frac{8 \pi G}{3}\rho -\frac{f}{6} + \frac{\mathcal{T} f_{\mathcal{T}}}{3},
\end{equation}
\begin{equation}  \label{a3}
\dot{H}=-\left[\frac{4 \pi G (\rho+p)}{1+f_{\mathcal{T}}+2 \mathcal{T} f_{\mathcal{T}\mathcal{T}}} \right] ,
\end{equation}
where $\rho$ and $p$ be the energy density and
pressure of the matter content and $\mathcal{T}=-6H^{2}$. From equations \eqref{a2}
and \eqref{a3}, we obtain the expressions of density $\rho$, pressure $p$
and EoS parameter $\omega$, respectively as

\begin{equation}  \label{a4}
\rho=3H^2 +\frac{f}{2} +6H^2f_\mathcal{T}
\end{equation}
\begin{equation}  \label{a5}
p=-2\dot{H}(1 + f_\mathcal{T} -12H^2f_{\mathcal{T}\mathcal{T}}) - (3H^2 +\frac{f}{2}+6H^2f_\mathcal{T}),
\end{equation}
\begin{equation}  \label{a6}
\omega=\frac{p}{\rho}=-1-\frac{2\dot{H}(1 + f_\mathcal{T} -12H^2f_{\mathcal{T}\mathcal{T}})}{(3H^2 +\frac{f%
}{2}+6H^2f_\mathcal{T})},
\end{equation}
where we set $8\pi G=1$. Furthermore, the continuity equation reads 
\begin{equation}\label{a7}
\dot{\rho}+3H(1+\omega)\rho=0.
\end{equation}

\section{Kinematic Variables}\label{IIIa}

The system of field equations described above has only two independent
equations with four unknowns. To solve the system completely and in order to
study the temporal evolution of energy density, pressure, and EoS parameter,
we need two more constraint equations (extra conditions). In literature,
there are several arguments for choosing these equations (see \cite{pacif} for
details). The method is well known as the model-independent way approach to
study cosmological models that generally considers a parametrization of any
kinematic variables such as Hubble parameter, deceleration parameter, jerk
parameter, EoS parameter and provide the necessary supplementary equation 
\cite{para}. Bearing that in mind, this chapter shall work with a parametrization of
the deceleration parameter proposed in \cite{banerjee} as 
\begin{equation}
q=-1+\alpha \left[ -1+\frac{1}{1+\left( \frac{1}{1+z}\right) ^{\alpha }}%
\right] ,  \label{a8}
\end{equation}%
where $\alpha $ shall be constrained from a chi-square test using any
observational datasets (ref. section \ref{VIIa}). The motivation use this
parametrization is driven by the fact that equation \eqref{a8} allows a
signature flipping for $-1>\alpha >-2$. Additionally note that,

\begin{itemize}
\item $\alpha=-2$ corresponds to a decelerated universe at $z=0$.

\item $\alpha<-2$ corresponds to an accelerated universe in the future
(i.e., $z<0$).

\item $\alpha\geq-1$ corresponds to an externally accelerating universe.
\end{itemize}

The expression of Hubble parameter for the parametrization \eqref{a8} reads 
\begin{equation}  \label{a9}
H=\beta \left[1+\left(\frac{1}{1+z}\right)^ {\alpha}\right]
\end{equation}
where $\beta$ is the integration constant. To obtain \eqref{a9}, the following
relation is used
\begin{equation}\label{a10}
\frac{H (z)}{H_{0}}=\exp\left[\int^{z}_{0}\frac{1+ q(z^{\prime })}{%
1+z^{\prime }}dz^{\prime }\right]
\end{equation}

\begin{figure}[H]
\centering
\includegraphics[width=8.5 cm]{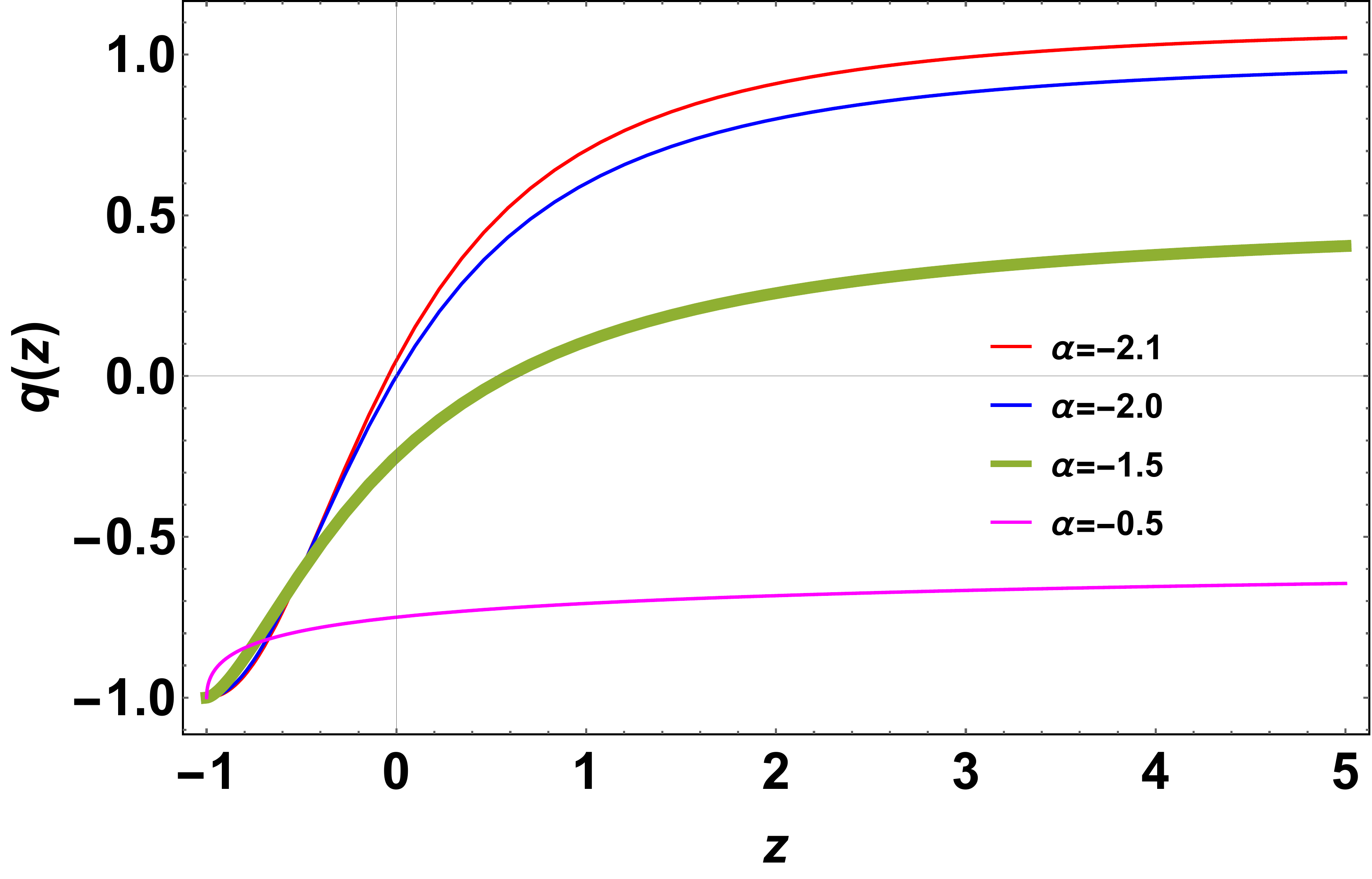}
\caption{Deceleration parameter ($q$) as a function of redshift for different values of model parameter $\protect\alpha $ showing diverge
evolutionary dynamics.}
\label{f1a}
\end{figure}

Higher order derivatives of deceleration parameter such as jerk ($j$), snap (%
$s$) and lerk ($l$) parameters provide important information about the
evolution of the universe. They are represented in \cite{olive21} as,
\begin{align*}
j(z)=(1+z)\frac{dq}{dz}+q(1+2q),
\end{align*}
\begin{align*}
s(z)=-(1+z)\frac{dj}{dz}-j(2+3q),
\end{align*}
\begin{align*}
l(z)=-(1+z)\frac{ds}{dz}-s(3+4q)
\end{align*}
The jerk parameter represents the evolution of the deceleration parameter. Since 
$q$ can be constrained from observations, the jerk parameter is used to predict
the future. Additionally, the jerk parameter, along with higher derivatives
such as snap and lerk parameters, provide useful insights into the emergence
of sudden future singularities \cite{olive21}.

\begin{figure}[H]
\centering
\includegraphics[width=8.5 cm]{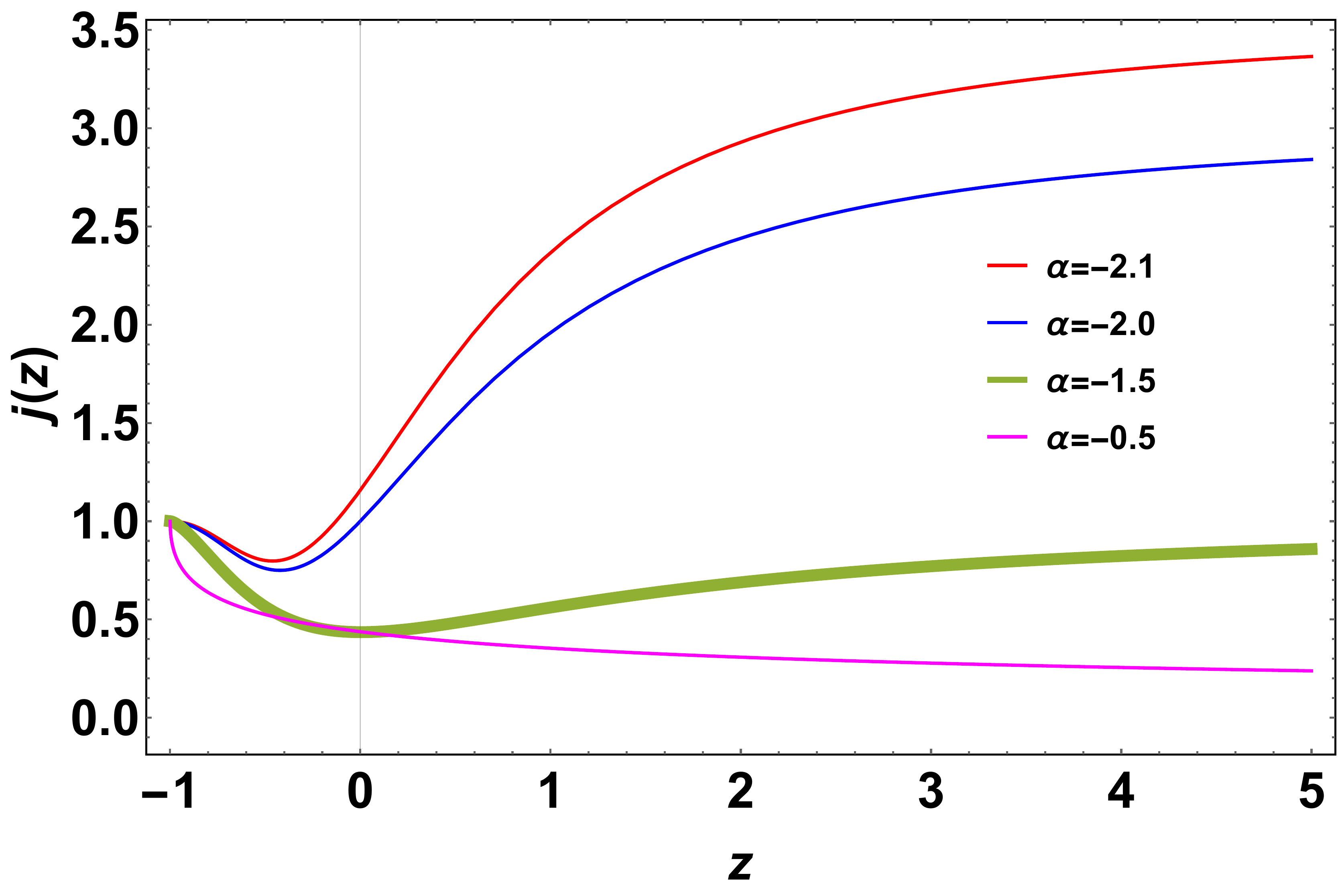}
\caption{Jerk parameter as a function of redshift.}
\label{f2a}
\end{figure}

\begin{figure}[H]
\centering
\includegraphics[width=8.5 cm]{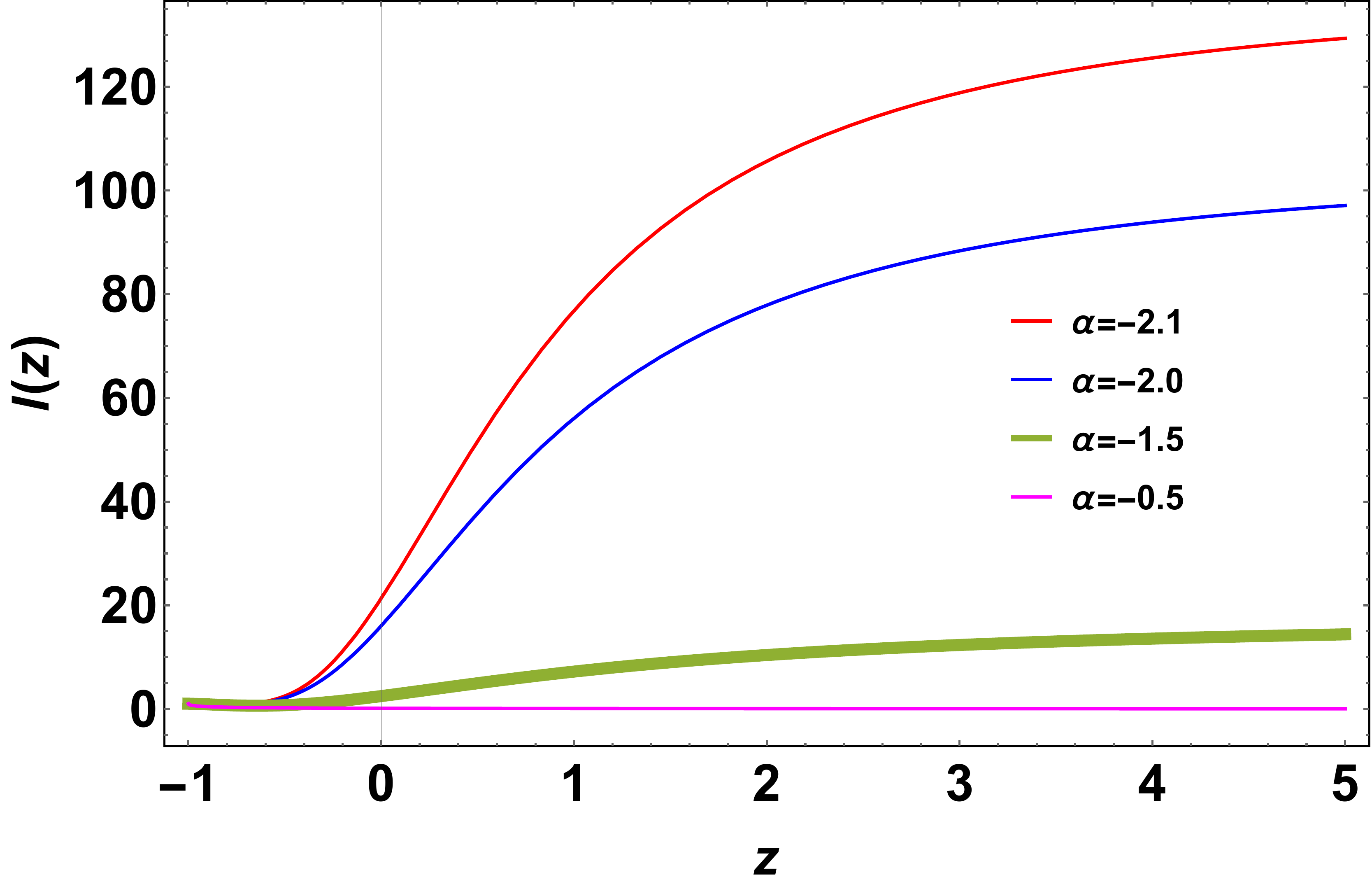}
\caption{Lerk parameter as a function of redshift.}
\label{f4a}
\end{figure}

From Figs. \ref{f2a} and \ref{f4a}, the jerk and lerk parameters are
observed to have decreasing behaviors. Also, as the value of $\alpha$
decreases, the parameters assume higher values at redshift $z=0$. Both of
these parameters are positive, which represents an accelerated expansion. The
snap parameter is negative for all $\alpha$ which also denotes an accelerated
expansion. Interestingly, the jerk parameter does not attain unity at $z=0$
which clearly does not coincide with $\Lambda$CDM model. Interestingly, this
implies that the late time acceleration can be caused due to modifications
of gravity. It is therefore encouraging to study the dynamics of EoS
parameter, which may arise purely due to geometric effects in the framework
of modified gravity theories such as teleparallel gravity.


\begin{figure}[H]
\centering
\includegraphics[width=8.5 cm]{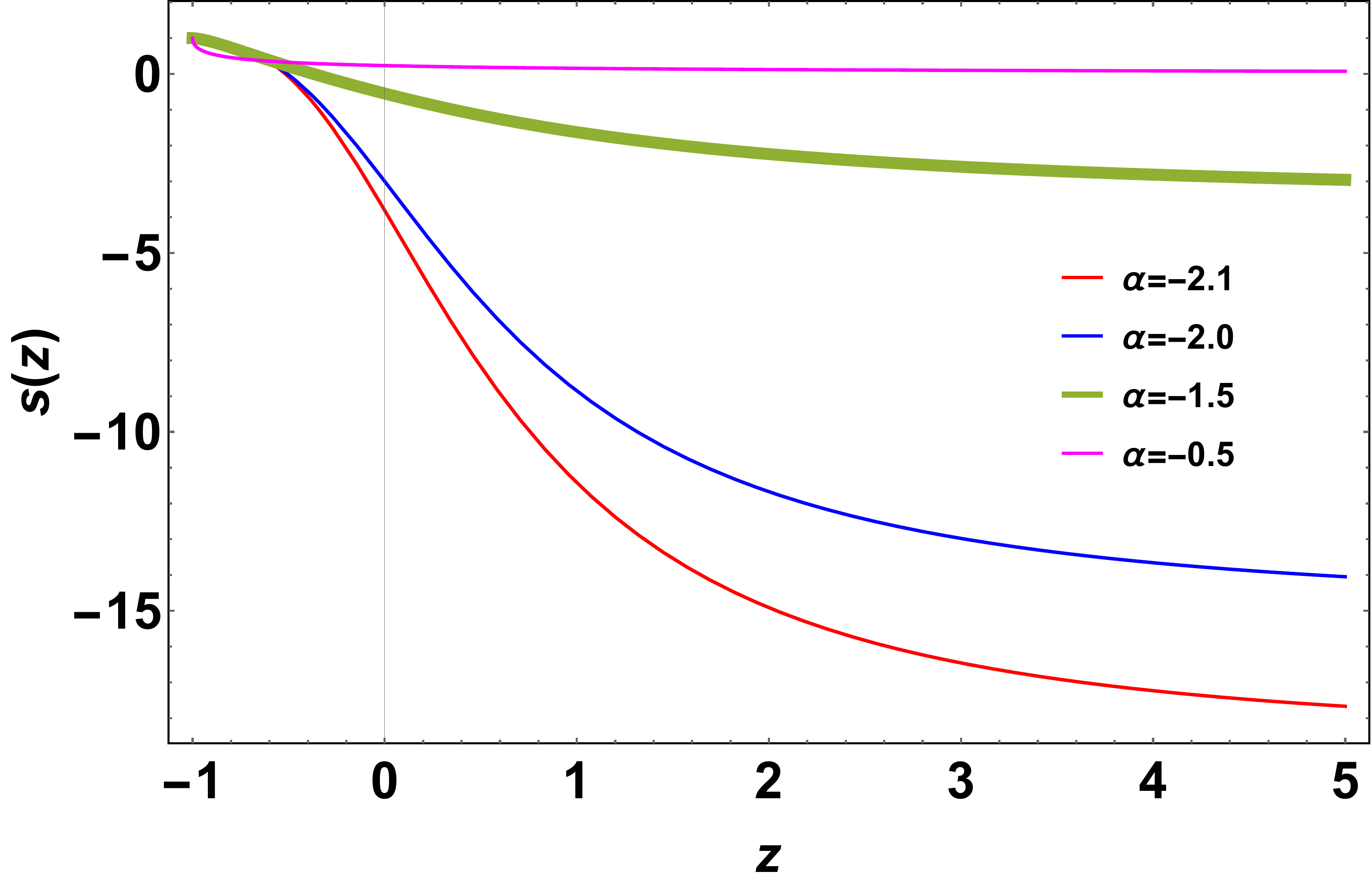}
\caption{Snap parameter as a function of redshift.}
\label{f3a}
\end{figure}

\section{Cosmology with Teleparallel Gravity}\label{IVa}

\subsection{ Hybrid Teleparallel Gravity}

For the first case, we presume the functional form of teleparallel gravity
to be 
\begin{equation}\label{a11}
f(\mathcal{T})=e^{m\mathcal{T}}\mathcal{T}^n,
\end{equation}
where $m\geq0$ and $n$ are constants. Interestingly, this model takes power-law
and exponential forms depending on the values of $n$ and $m$. Particularly:

\begin{itemize}
\item For $m=0$ Eq. \eqref{a11} reduces to $f(\mathcal{T})=\mathcal{T}^n$ (power law).

\item For $n=0$, Eq. \eqref{a11} reduces to $f(\mathcal{T})=e^{m\mathcal{T}}$ (exponential).
\end{itemize}

Using Eq. \eqref{a11} in Eq. \eqref{a2} and Eq. \eqref{a3}, the expressions of
energy density $\rho$, pressure $p$ and EoS parameter $\omega$ reads
respectively as,
\begin{equation} \label{a12}
\rho=3K+6^n(-K)^ne^{-6mK}\left(\frac{1}{2}-n+6nK\right),
\end{equation}
\begin{multline}\label{a13}
p=-2\left(\alpha K-\frac{e^{-t\alpha \beta}\alpha\beta^2}{-1+e^{-t\alpha\beta}}\right)\times
\left\lbrace -1+(-6K)^ne^{-6mK}\left[m+4mn-12Km^2-\frac{n}{6K}-\frac{n(n-1)}{3K}\right]\right\rbrace \\
-3K-6^n(-K)^ne^{-6mK}\left(\frac{1}{2}-n+6nK\right)
\end{multline}
\begin{multline}\label{a14}
\omega=-1-2\left(\alpha K-\frac{e^{-t\alpha \beta}\alpha\beta^2}{-1+e^{-t\alpha\beta}}\right)\times 
\left\lbrace -1+(-6K)^ne^{-6mK}\left[m+4mn-12Km^2-\frac{n}{6K}-\frac{n(n-1)}{3K}\right]\right\rbrace\\
\times\left\lbrace3K+6^n(-K)^ne^{-6mK}\left(\frac{1}{2}-n+6nK\right)\right\rbrace^{-1}
\end{multline}
where $K=\frac{e^{-2t\alpha\beta}\beta^2}{(-1+e^{-t\alpha\beta})^2}$. 
\begin{figure}[H]
\centering
\includegraphics[width=8.5 cm]{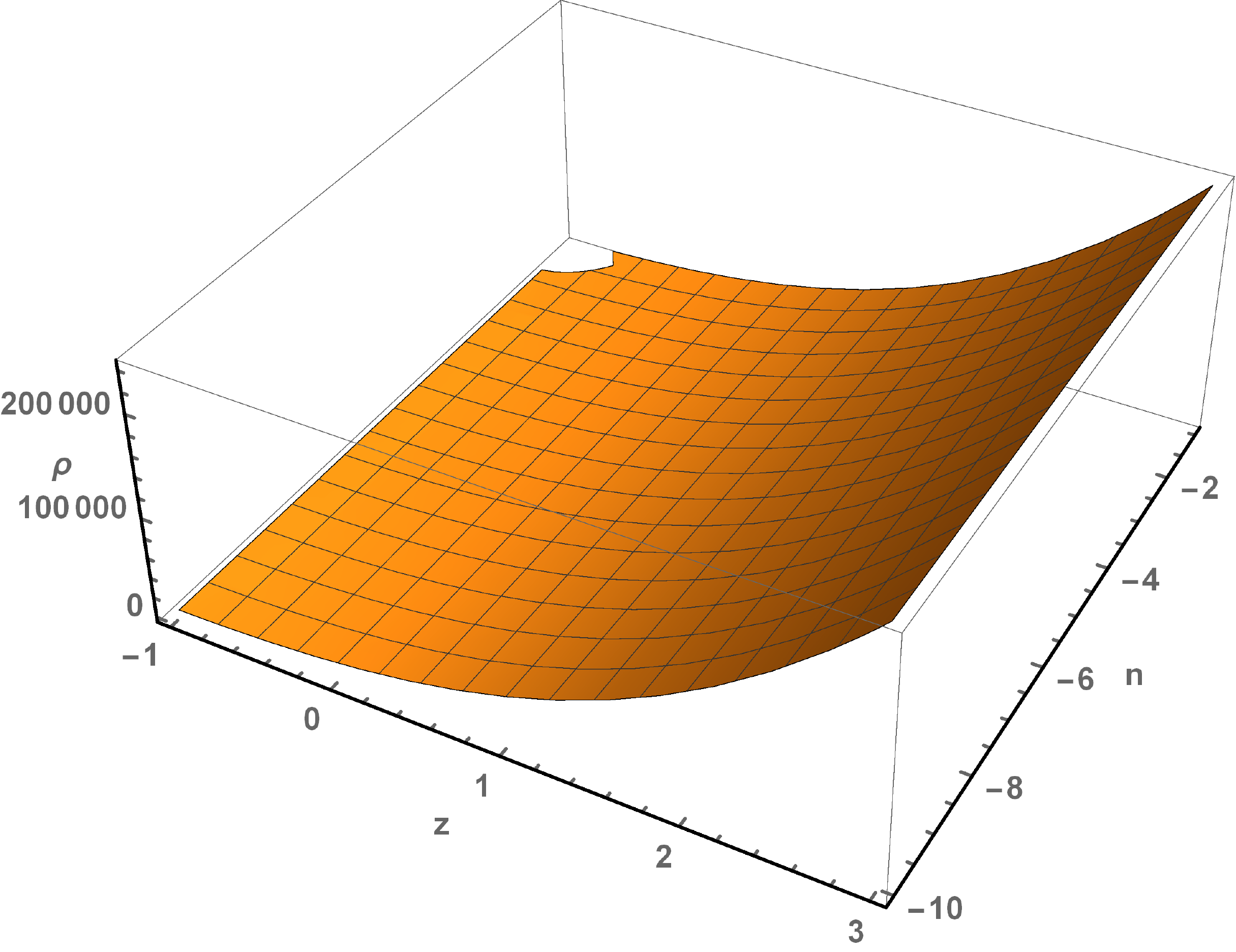}
\caption{Energy density as a function of redshift for $\protect\alpha=-1.5, 
\protect\beta=31.7455, m=0.00155$.}
\label{f5a}
\end{figure}
\begin{figure}[H]
\centering
\includegraphics[width=8.5 cm]{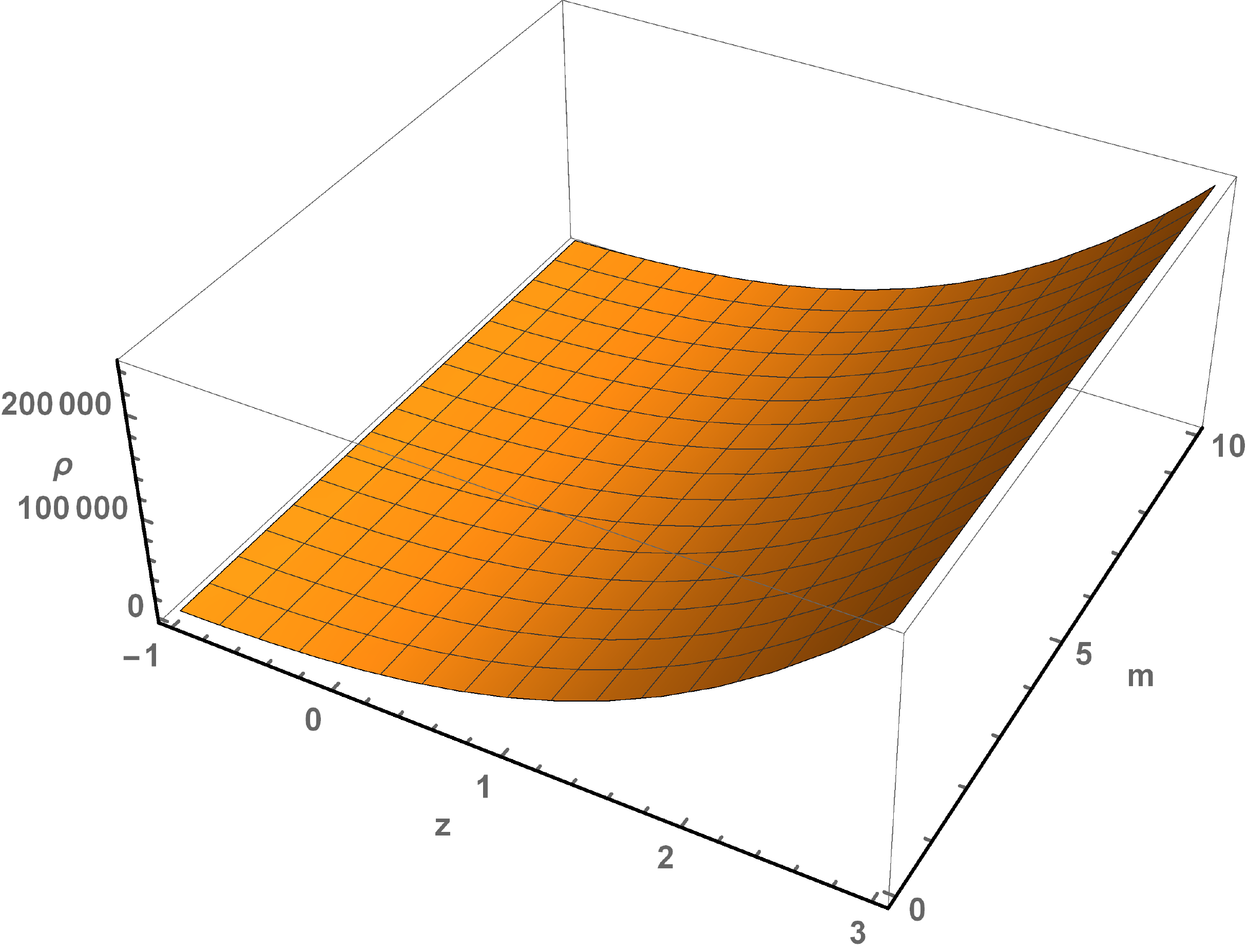}
\caption{Energy density as a function of redshift for $\protect\alpha=-1.5, 
\protect\beta=31.7455, \ \  \ \ n=5$.}
\label{f6a}
\end{figure}
\begin{figure}[H]
\centering
\includegraphics[width=8.5 cm]{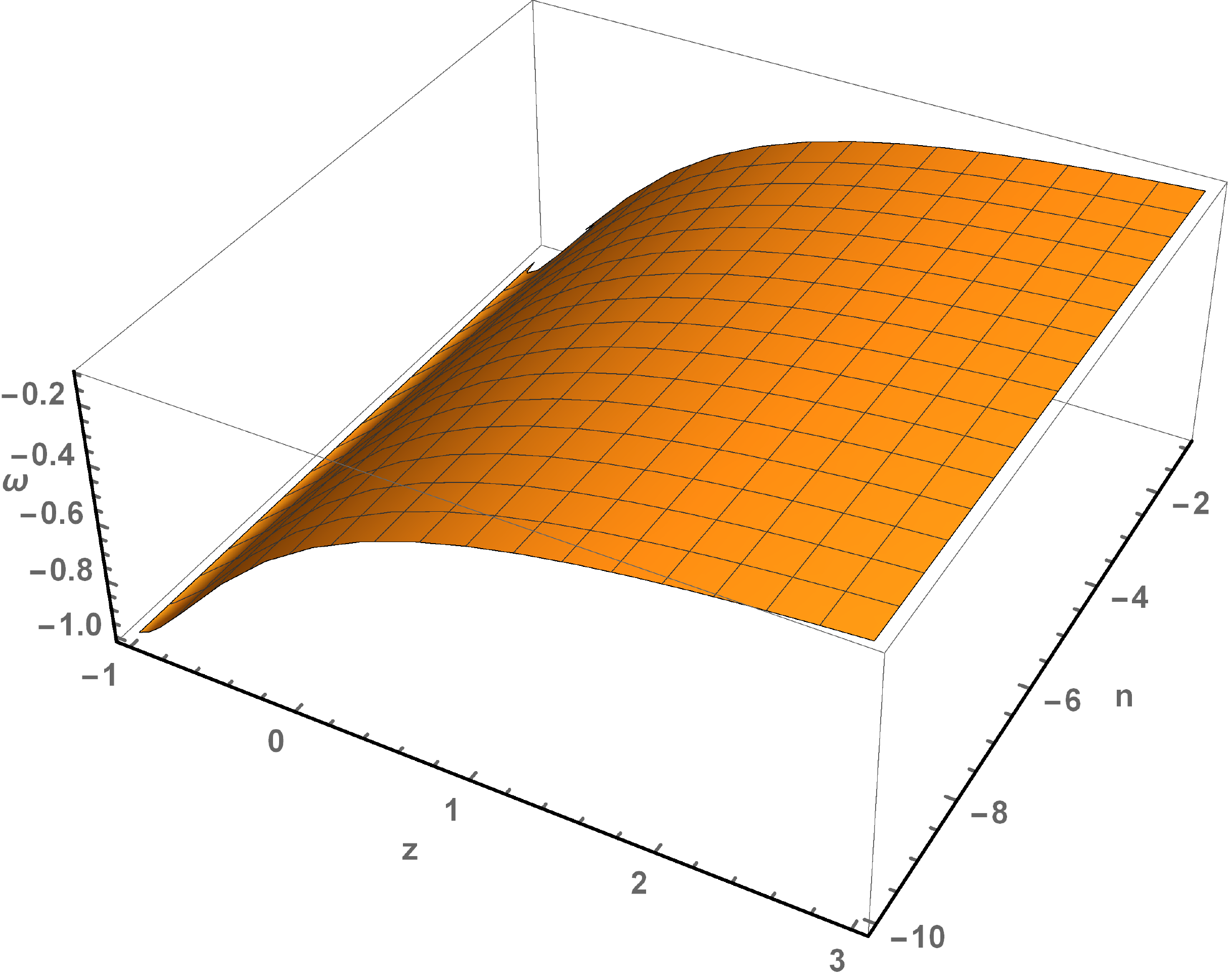}
\caption{EoS parameter as a function of redshift for $\protect\alpha=-1.5, 
\protect\beta=31.7455, m=0.00155$.}
\label{f7a}
\end{figure}

\begin{figure}[H]
\centering
\includegraphics[width=8.5 cm]{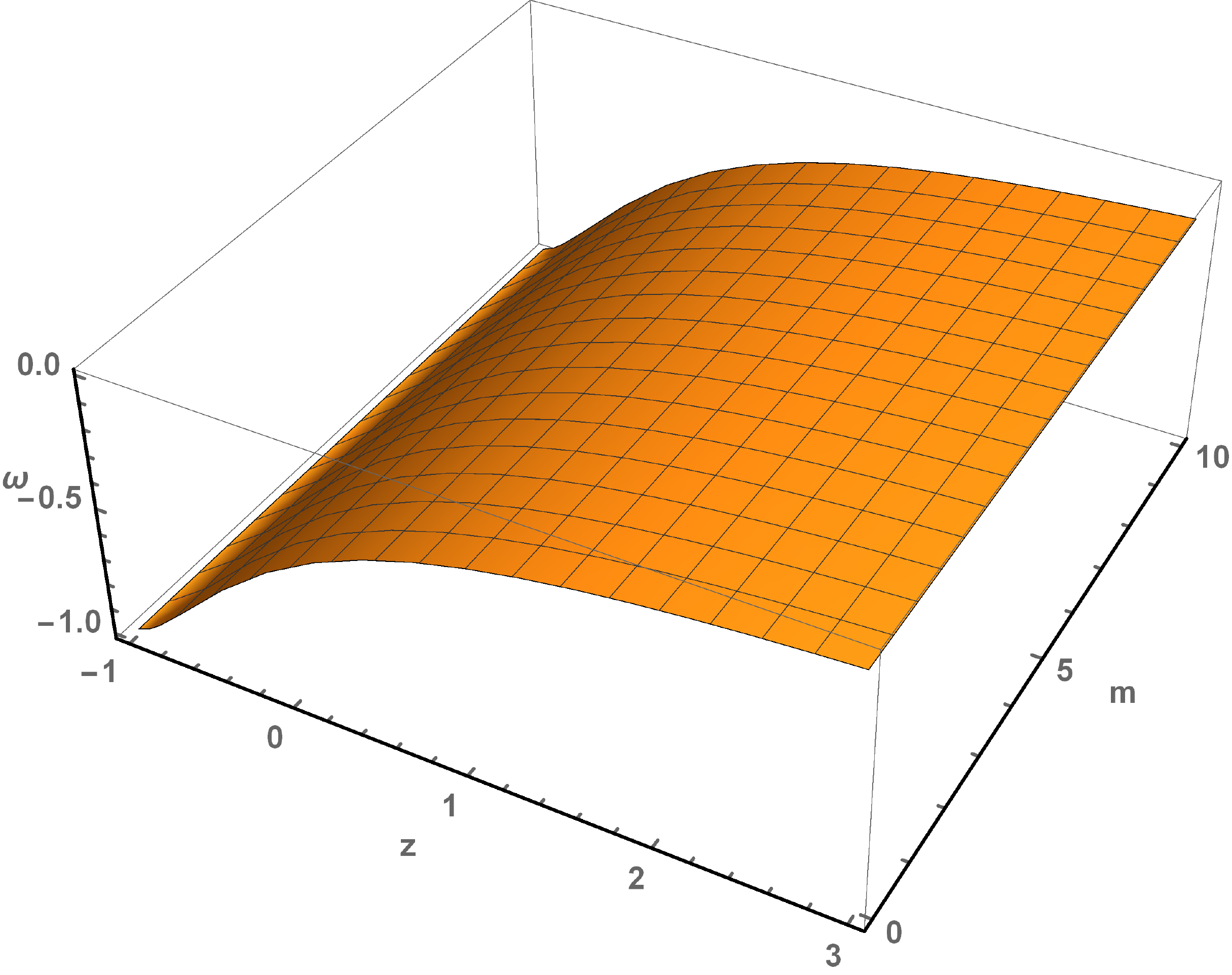}
\caption{EoS parameter as a function of redshift for $\protect\alpha=-1.5, 
\protect\beta=31.7455, n=5$.}
\label{f8a}
\end{figure}

\subsection{ Logarithmic Teleparallel Gravity}

For the second case, a logarithmic functional form of teleparallel gravity
to be presumed as,
\begin{equation}  \label{a15}
f(\mathcal{T})=D\log(b\mathcal{T}),
\end{equation}
where $D$ and $b<0$ are constants.\newline
Using Eq. \eqref{a15} in Eq. \eqref{a2} and Eq. \eqref{a3}, the expressions of
energy density $\rho$, pressure $p$ and EoS parameter $\omega$ reads
respectively as 
\begin{equation}  \label{a16}
\rho=-D+3K+\frac{D}{2}\log(6bK),
\end{equation}
\begin{equation}\label{a17}
p=D-3K-2\left(1+\frac{D}{6K}\right)\left(\alpha K-\frac{e^{-t\alpha\beta}\alpha\beta^2}{-1+e^{-t\alpha\beta}}\right)-\frac{D}{2}\log(6bK),
\end{equation}
\begin{equation}\label{a18}
\omega=-1-2\left(1+\frac{D}{6K}\right)\left(\alpha K-\frac{e^{-t\alpha\beta}\alpha\beta^2}{-1+e^{-t\alpha\beta}}\right)
\times \left\lbrace -D+3K+\frac{D}{2}\log(6bK)\right\rbrace^{-1}
\end{equation}

\begin{figure}[H]
\centering
\includegraphics[width=8.5 cm]{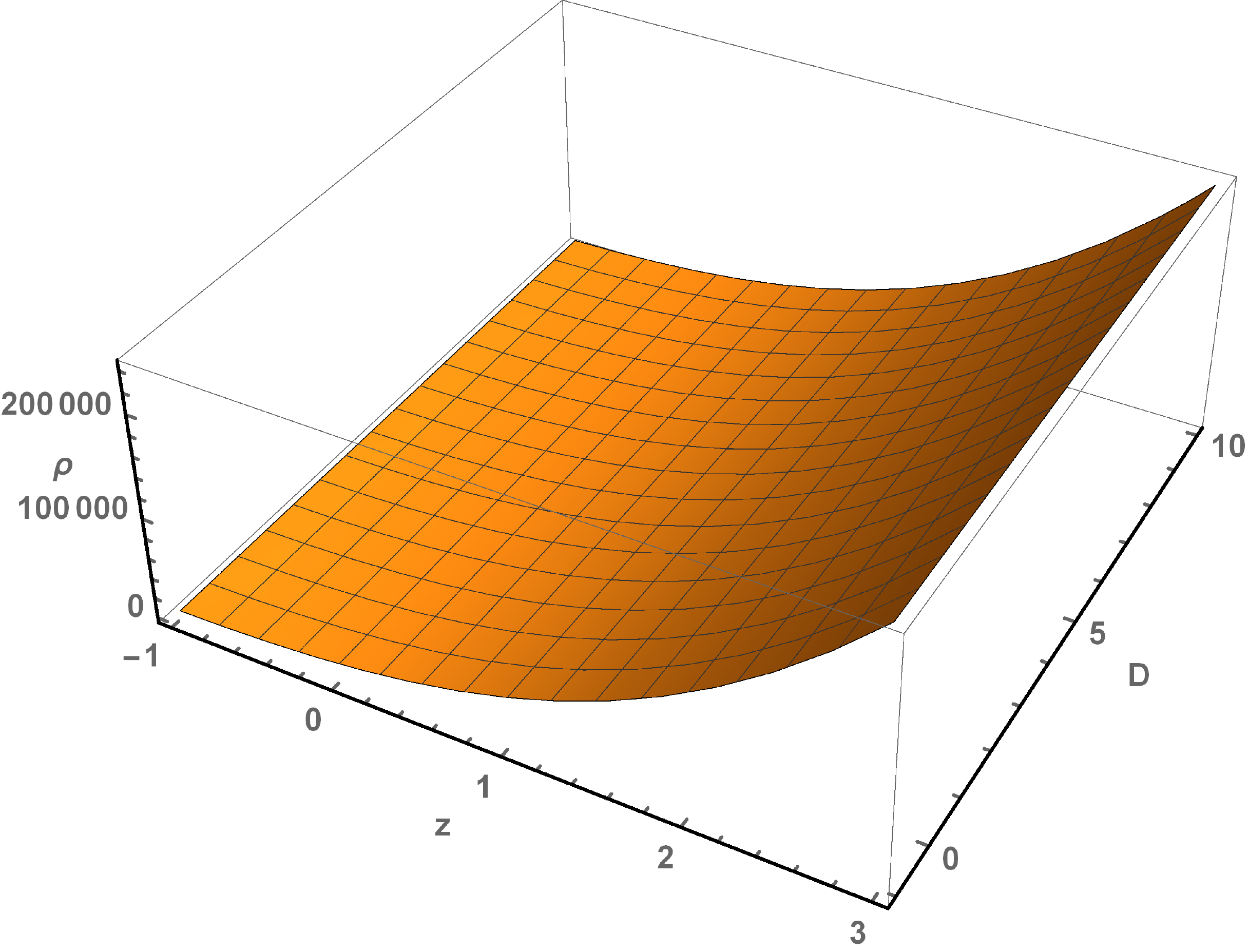}
\caption{Energy density as a function of redshift for $\protect\alpha=-1.5, 
\protect\beta=31.7455,  b=-2$.}
\label{f9a}
\end{figure}
\begin{figure}[H]
\centering
\includegraphics[width=8.5 cm]{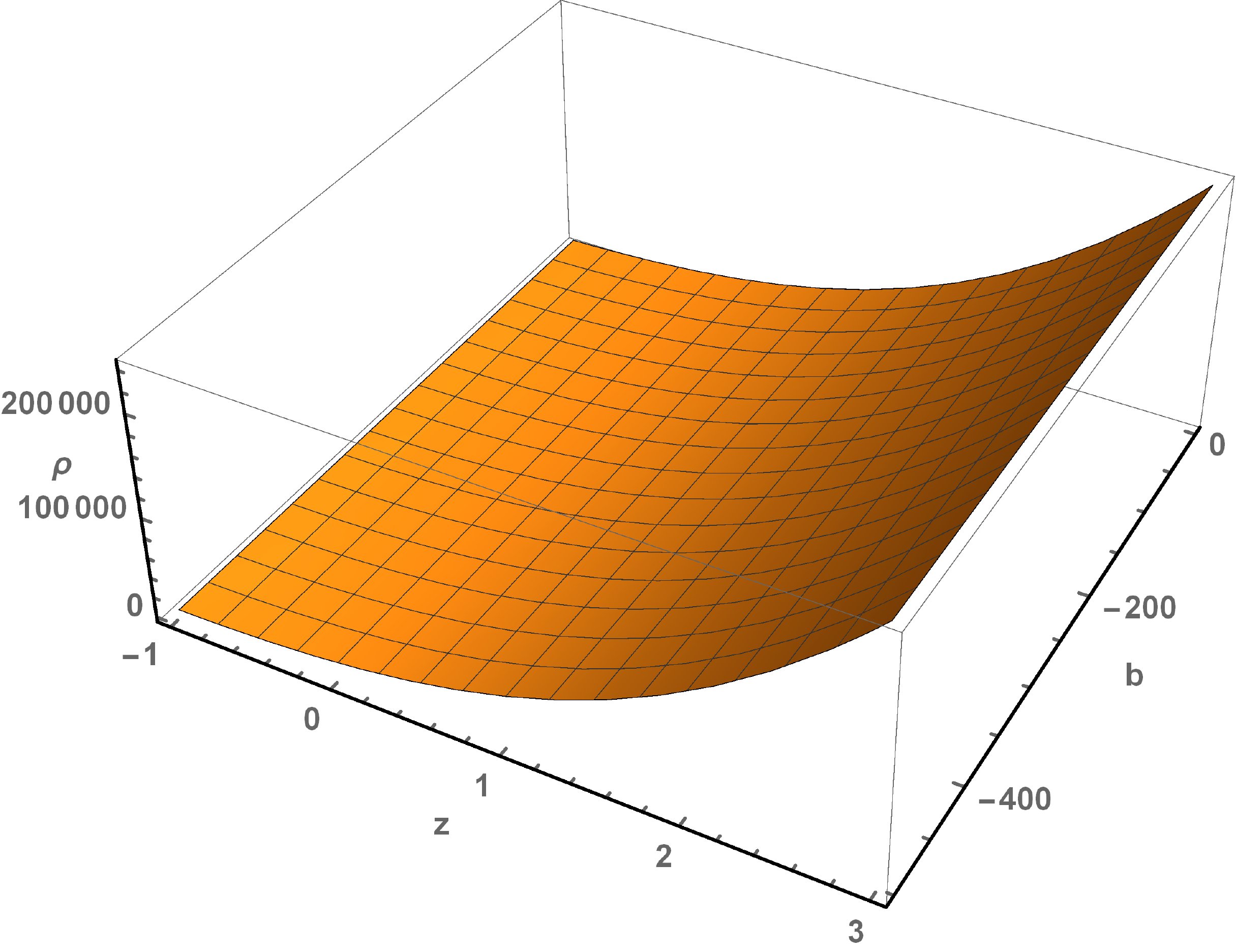}
\caption{Energy density as a function of redshift for $\protect\alpha=-1.5, 
\protect\beta=31.7455,  D=0.2$.}
\label{f10a}
\end{figure}
\begin{figure}[H]
\centering
\includegraphics[width=8.5 cm]{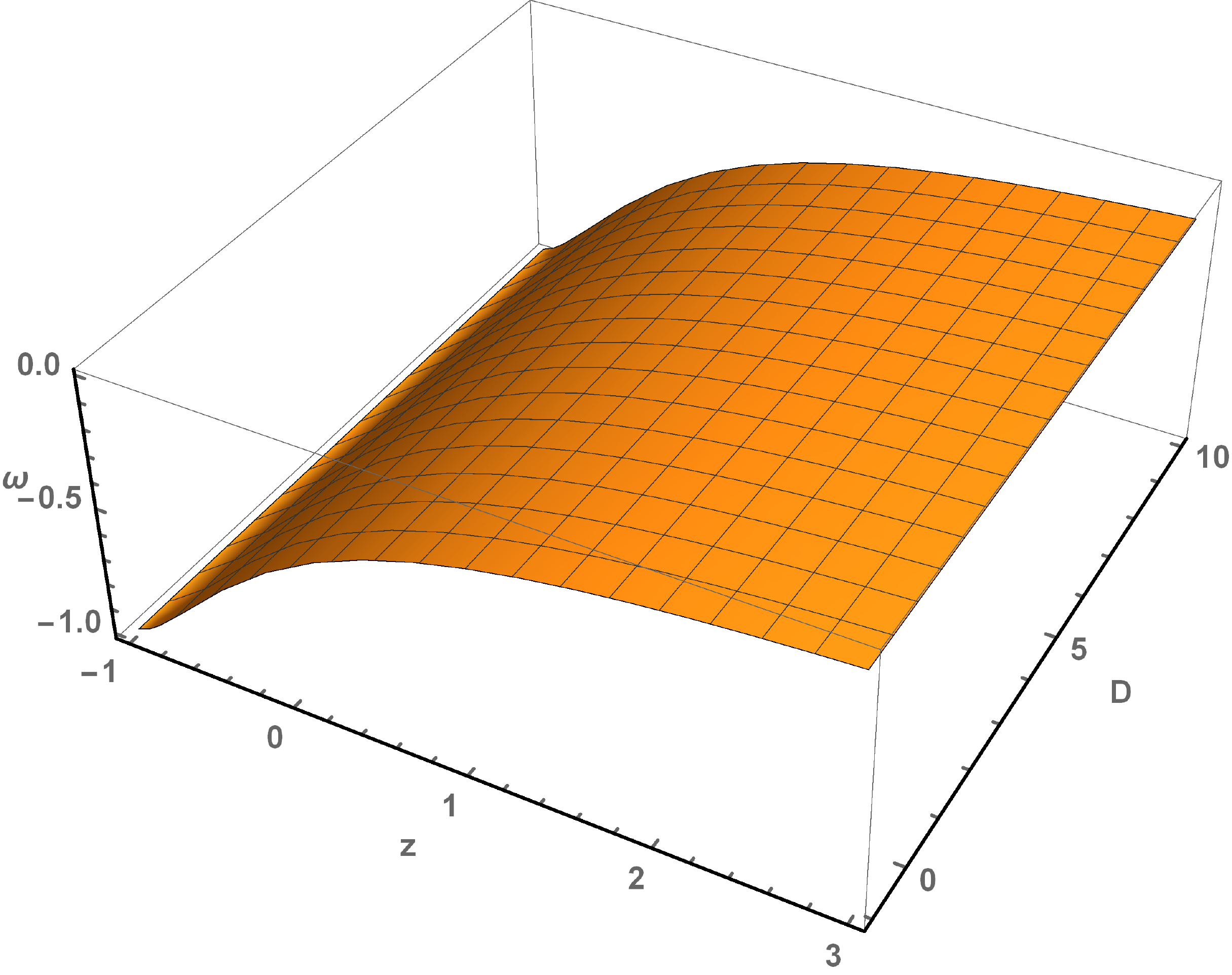}
\caption{EoS parameter as a function of redshift for $\protect\alpha=-1.5, 
\protect\beta=31.7455, b=-2$.}
\label{f11a}
\end{figure}
\begin{figure}[H]
\centering
\includegraphics[width=8.5 cm]{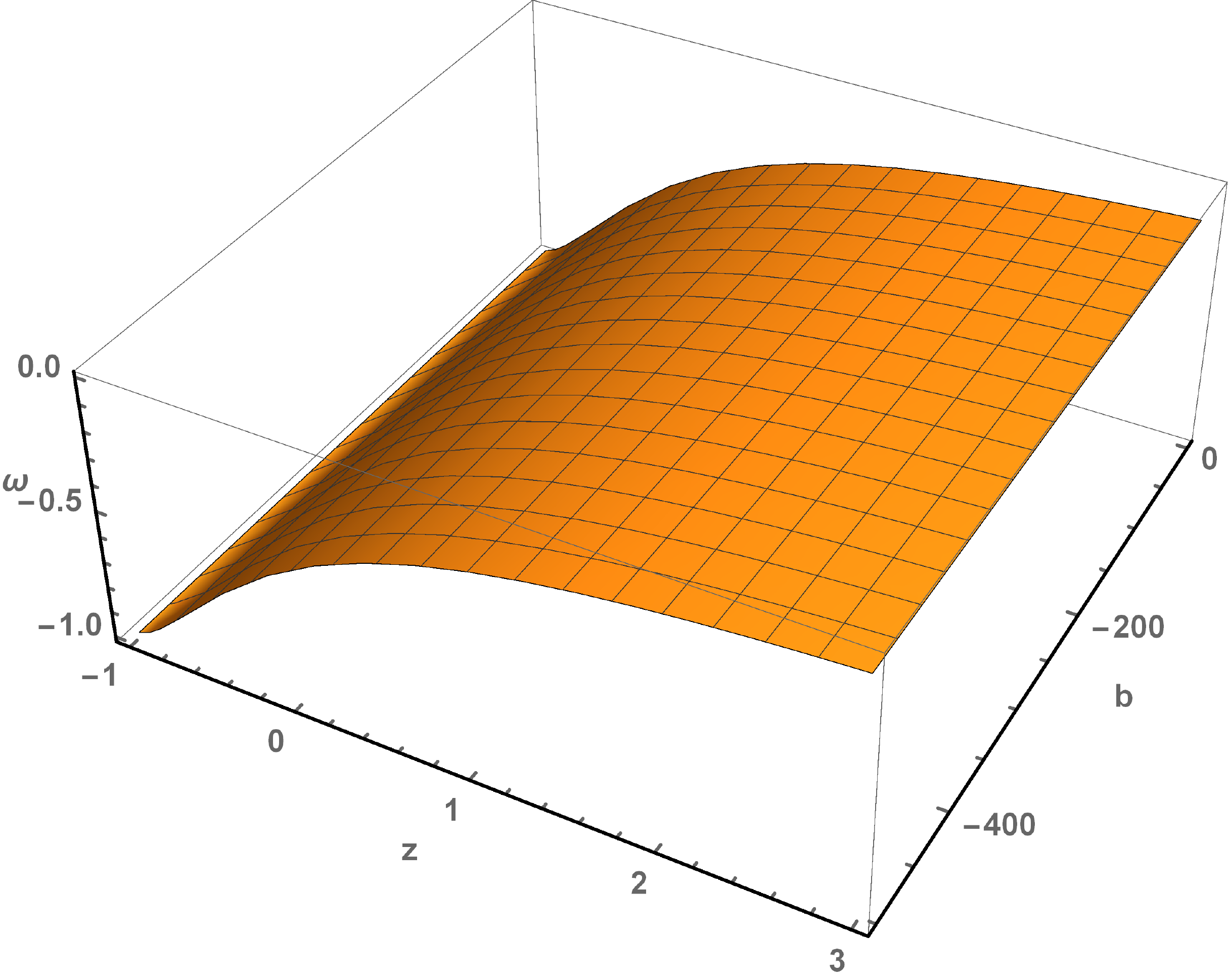}
\caption{EoS parameter as a function of redshift for $\protect\alpha=-1.5, 
\protect\beta=31.7455, D=0.2$.}
\label{f12a}
\end{figure}

\section{Geometrical Diagnostics}\label{Va}

\subsection{Statefinder Diagnostics}

Due to the fact that the number of dark energy models are quite large and
increasing on a daily basis, it becomes absolutely necessary to find a
method to distinguish a particular model from the well established DE models
like the $\Lambda$CDM, standard cold dark matter (SCDM), holographic dark energy (HDE), Chaplygin gas (CG) and Quintessence. With that reasoning, 
\cite{sahni} proposed the $\{r,s\}$ diagnostics where $r$ and $s$ are
defined as,

\begin{align*}
r=\frac{\dot{\ddot{a}}}{aH^3},
\end{align*}
\begin{align*}
s=\frac{r-1}{3\left(q-\frac{1}{2}\right)},\left(q\neq\frac{1}{2}\right).
\end{align*}
Different combinations of $r$ and $s$ represent different dark energy
models. Particularly:

\begin{itemize}
\item For $\Lambda$CDM$\rightarrow$ $(r=1,s=0)$.


\item For HDE$\rightarrow$ $(r=1,s=\frac{2}{3})$.

\item For CG$\rightarrow$ $(r>1,s<0)$.

\item For Quintessence $\rightarrow$ $(r<1,s>0)$.
\end{itemize}

The idea behind $\{r,s\}$ diagnostics tool is that different dark energy
models exhibit different trajectories in the $\{r,s\}$ plane. The deviation
from the point $\{r,s\} = \{0,1\}$ represent deviation from the well agreed $%
\Lambda$CDM model. Furthermore, the values of $r$ and $s$ could in principle
be inferred from observations \cite{sahni17to18} and therefore could be very
useful in discriminating dark energy models in the near future.\newline
The expression of $r$ and $s$ parameters for our model reads 
\begin{equation}\label{a19}
r=1+\frac{\alpha\left(\frac{1}{1+z}\right)^{\alpha}\left\lbrace
3+\alpha+\left(\frac{1}{1+z}\right)^{\alpha}(3+2\alpha)\right\rbrace}{%
\left\lbrace 1+\left(\frac{1}{1+z}\right)^{\alpha}\right\rbrace^2},
\end{equation}
\begin{equation}\label{a20}
s=\frac{\alpha}{3}\left\lbrace -2+\frac{1}{1+\left(\frac{1}{1+z}%
\right)^{\alpha}}+\frac{3}{3+\left(\frac{1}{1+z}\right)^{\alpha}(3+2\alpha)}%
\right\rbrace.
\end{equation}

\begin{figure}[H]
\centering
\includegraphics[width=8.5 cm]{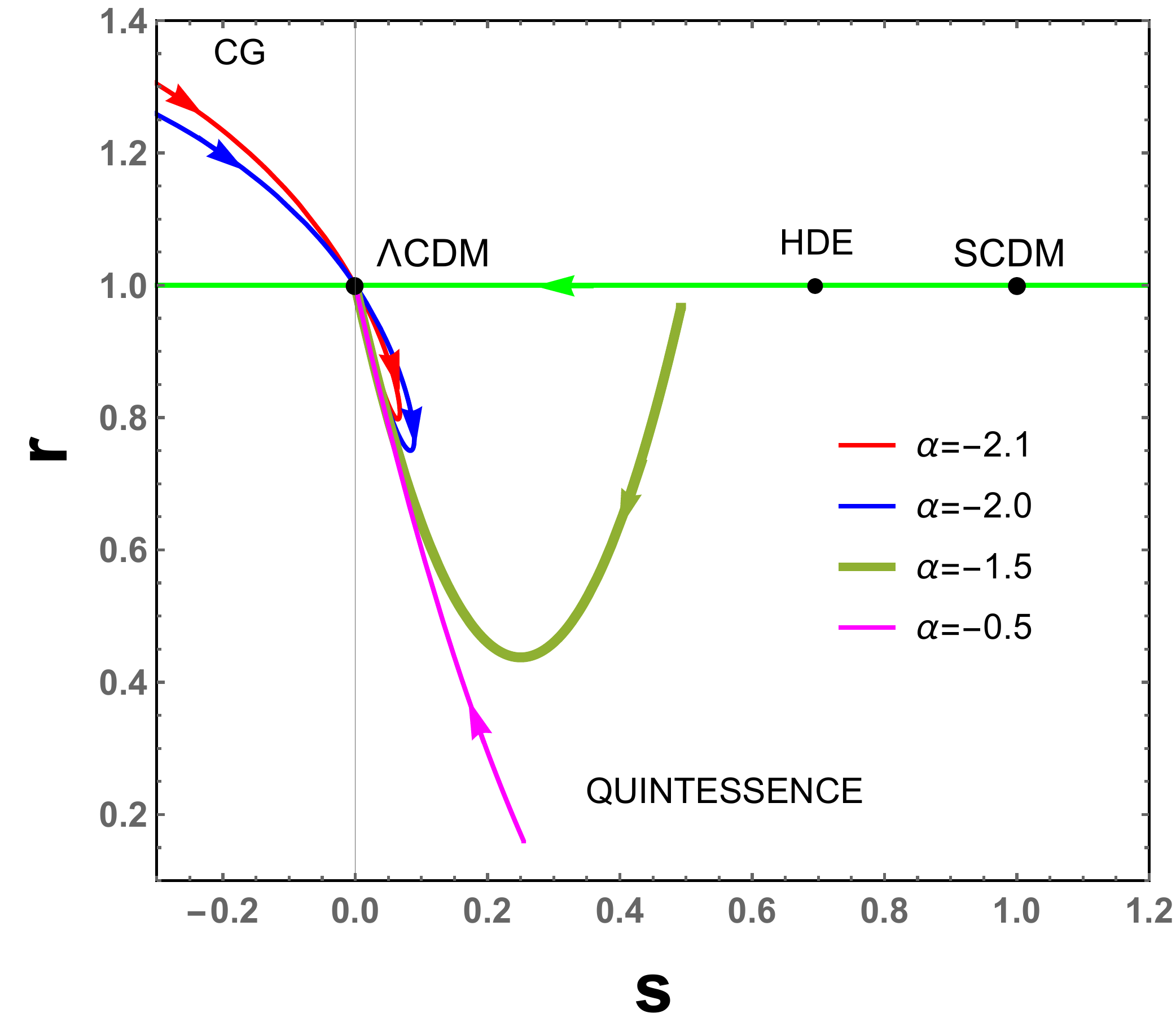}
\caption{$\{r,s\}$ plane for the redshift range $z \in [-1,5]$ for different
values of $\protect\alpha$.}
\label{f13a}
\end{figure}

In Fig. \ref{f13a}, the $\{r,s\}$ plane is shown for the parametrization %
\eqref{a8} where the arrows indicate the direction of temporal evolution.
The model is observed to deviate significantly from the point $(0,1)$
initially and is extremely sensitive to the value of $\alpha$. For $%
\alpha\leq-2$, the model initially starts its journey from the territory of
Chaplygin gas $(r>1,s<0)$ and approaches towards $\Lambda$CDM at late times.
For $\alpha>-1$, the model at high redshifts stays in the Quintessence
region but again approaches towards $\Lambda$CDM. Interestingly, for $%
\alpha=-1.5\sim-1.5$ which is the chi-square value, we observed at high
redshifts, the model to be very close to the point $(r=1,s=\frac{2}{3})$
which is the region of HDE. However, at late times the model is observed to
coincide with ($r=1,s=0$). Therefore, the parametrization used in this work
is interesting and warrants further attention. 
In addition to the $\{r,s\}$ plane, we construct the $\{r,q\}$ plane to get
additional understanding of the parametrization \eqref{a8}. In $\{r,q\}$
plane, the solid line in the middle represents the evolution of the $\Lambda$%
CDM universe and also divide the plane into two sections. The upper section
belong to Chaplygin gas model and the lower section to Quintessence model. 
\begin{figure}[H]
\centering
\includegraphics[width=8.5 cm]{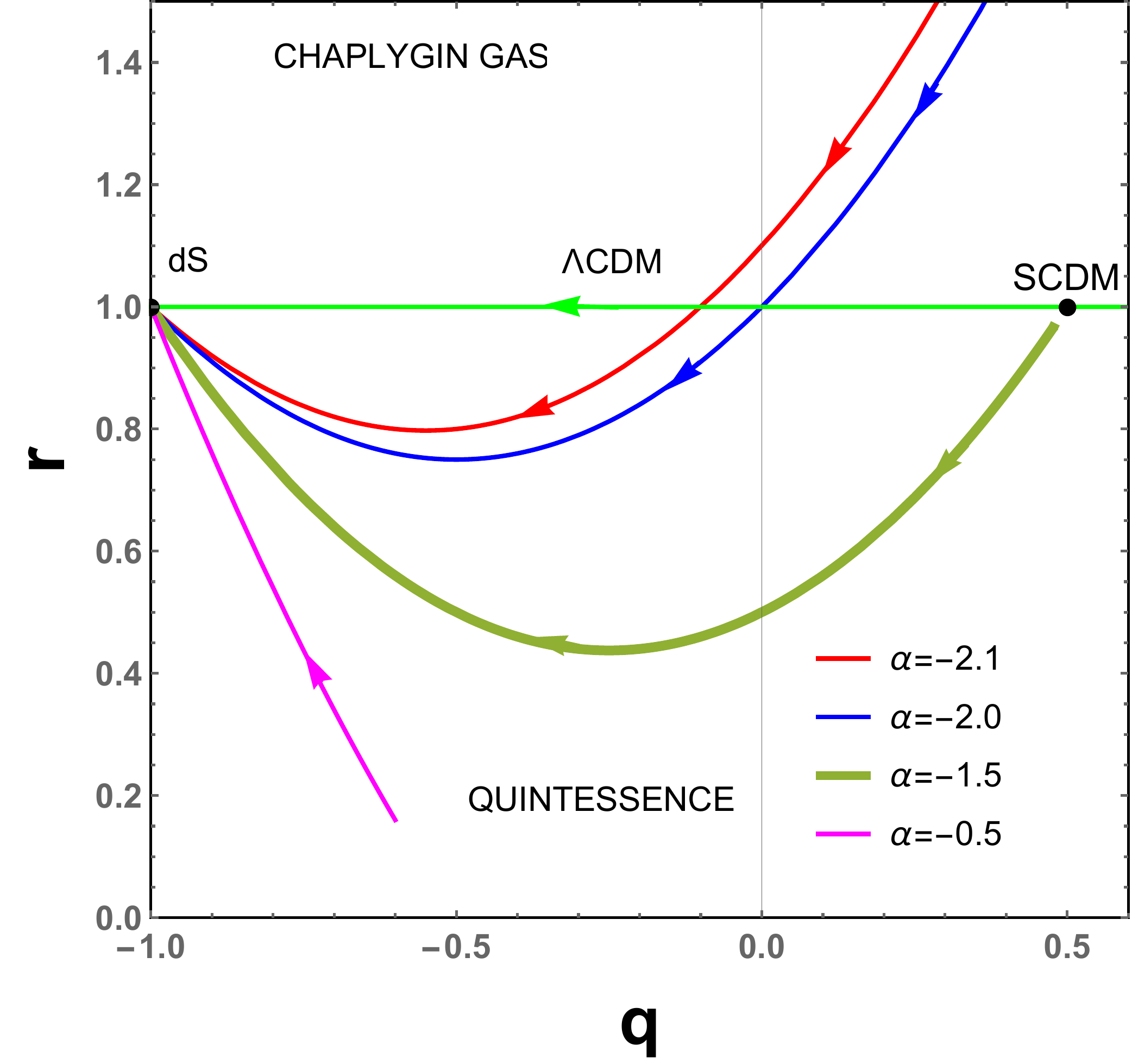}
\caption{$\{r,q\}$ plane for the redshift range $z \in [-1,5]$ for different
values of $\protect\alpha$.}
\label{f14a}
\end{figure}
From Fig. \ref{f14a}, it is observed that except for $\alpha=-0.5$, all the
profiles starts from $r>1,q>0$ which is very close to SCDM universe,
followed by the region $r<1,-1<q<0$ and finally approaches the de-Sitter
expansion with $r=1,q=-1$. However, for $\alpha<-1$, $q$ is always negative
and therefore the profile does not start from the SCDM universe.

\subsection{Om Diagnostic}

\begin{figure}[H]
\centering
\includegraphics[width=8.5 cm]{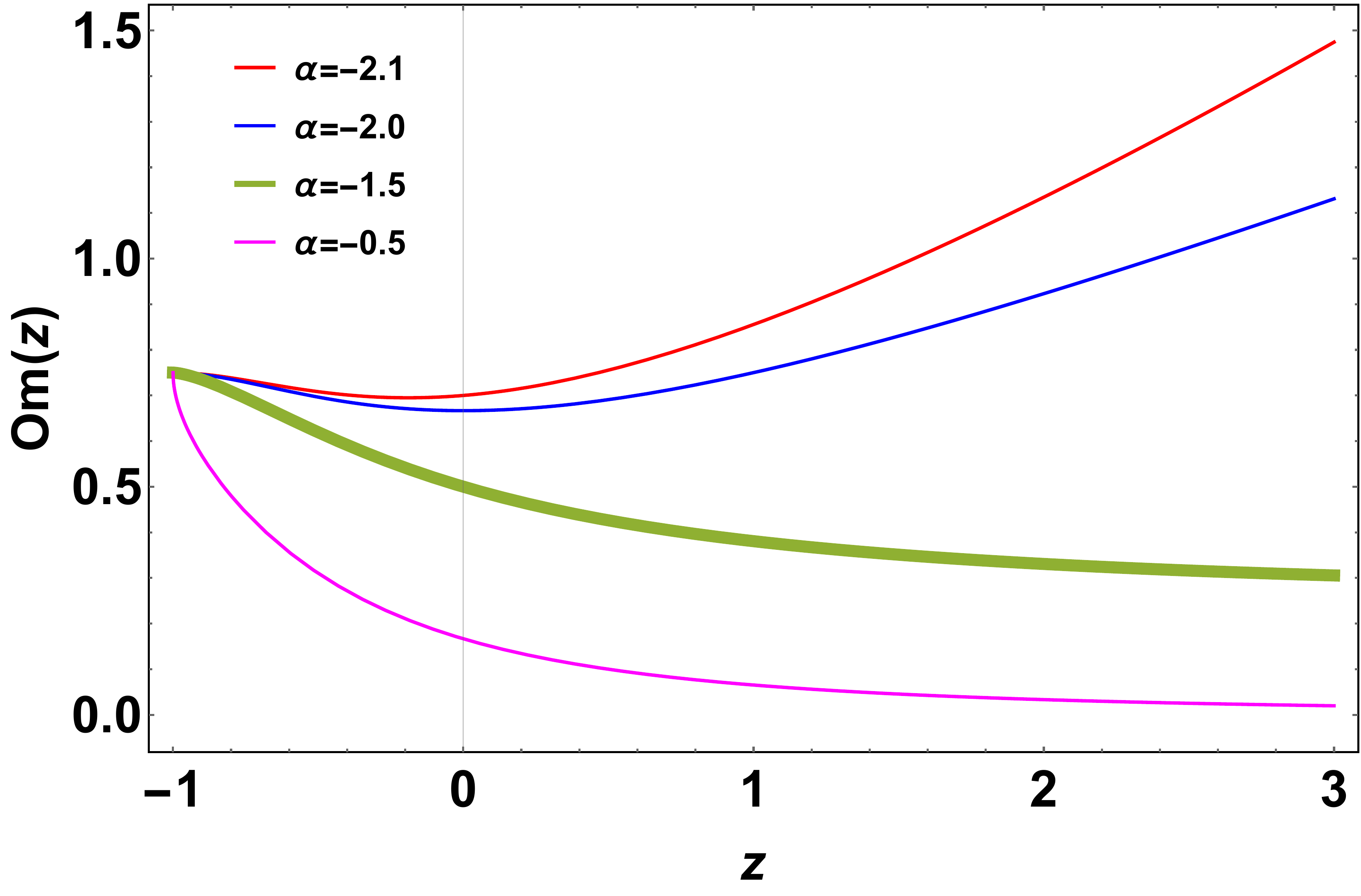}
\caption{ $Om(z)$ for different values of $\protect\alpha$.}
\label{f15a}
\end{figure}

Another very useful diagnostic tool constructed from the Hubble parameter is
the $Om$ diagnostic which essentially provides a null test of the $\Lambda$%
CDM model \cite{Omsahni}. This tool easily captures the dynamical nature of
dark energy models from the slope of $Om(z)$. If the slope of this
diagnostic tool were to be positive, it would imply a Quintessence nature ($%
\omega>-1$) whereas the opposite would prefer a Phantom nature ($\omega<-1$%
). Interestingly, only when the nature of the dark energy model coincides
with that of the cosmological constant ($\omega=-1$), the slope is constant
with respect to redshift. It is defined as,

\begin{equation}\label{a21}
Om(z)=\frac{\left(\frac{H(z)}{H_0}\right)^2-1}{z^3+3z^2+3z}
\end{equation}
From Fig. \ref{f15a}, one can observe a negative slope for $\alpha>-2$ and
therefore represents a dark energy model which is Phantom in nature.
Nonetheless, for $\alpha\leq-2$, $Om(z)$ increases with redshift and
therefore represents an Quintessence dark energy model. Hence, the value of $%
\alpha$ dictates the nature of the underlying dark energy model represented
by the parametrization \eqref{a8}.

\section{Energy Conditions}\label{VIa}

Based upon the Raychaudhuri equation, the energy conditions are essential to
describe the behavior of the compatibility of timelike, lightlike or
spacelike curves \cite{sahoo} and often used to understand the dreadful
singularities \cite{non39}. The energy conditions (ECs) are the essential tools to understand the geodesics of the universe. Such conditions can be derived from the well-known Raychaudhuri equations, whose forms are \cite{Poisson/2004}
\begin{equation}
\label{b10}
\frac{d\theta}{d\tau}=-\frac{1}{3}\theta^2-\sigma_{\mu\nu}\sigma^{\mu\nu}+\omega_{\mu\nu}\omega^{\mu\nu}-R_{\mu\nu}u^{\mu}u^{\nu}\,,
\end{equation}
\begin{equation}
\label{b11}
\frac{d\theta}{d\tau}=-\frac{1}{2}\theta^2-\sigma_{\mu\nu}\sigma^{\mu\nu}+\omega_{\mu\nu}\omega^{\mu\nu}-R_{\mu\nu}n^{\mu}n^{\nu}\,,
\end{equation}
where $\theta$ is the expansion factor, $n^{\mu}$ is the null vector, and $\sigma^{\mu\nu}$ and $\omega_{\mu\nu}$ are, respectively, the shear and the rotation associated with the vector field $u^{\mu}$. For attractive gravity, equations \eqref{b10}, and \eqref{b11} satisfy the following conditions
\begin{align}
\label{b12}
R_{\mu\nu}u^{\mu}u^{\nu}\geq0\,,\\
 R_{\mu\nu}n^{\mu}n^{\nu}\geq0\,.
\end{align}

Energy conditions in teleparallel gravity have
been studied in \cite{energy}. Energy conditions also provide the corners in
parameter spaces since they violate, for instance, in presence of
singularities. They are defined as:

\begin{itemize}
\item SEC: Gravity is always attractive and
therefore $\rho+3p\geq 0$;

\item WEC: Energy density should always be
positive, i.e., $\rho\geq 0, \rho+p\geq 0$;

\item NEC: Minimum requirement for the fulfilment of
SEC and WEC, i.e., $\rho+p\geq 0$;

\item DCE: Energy density is always positive
and independent of the observer's reference frame, i.e., $\rho\geq 0,
|p|\leq \rho$.
\end{itemize}

\begin{figure}[H]
\centering
\includegraphics[width=8.5 cm]{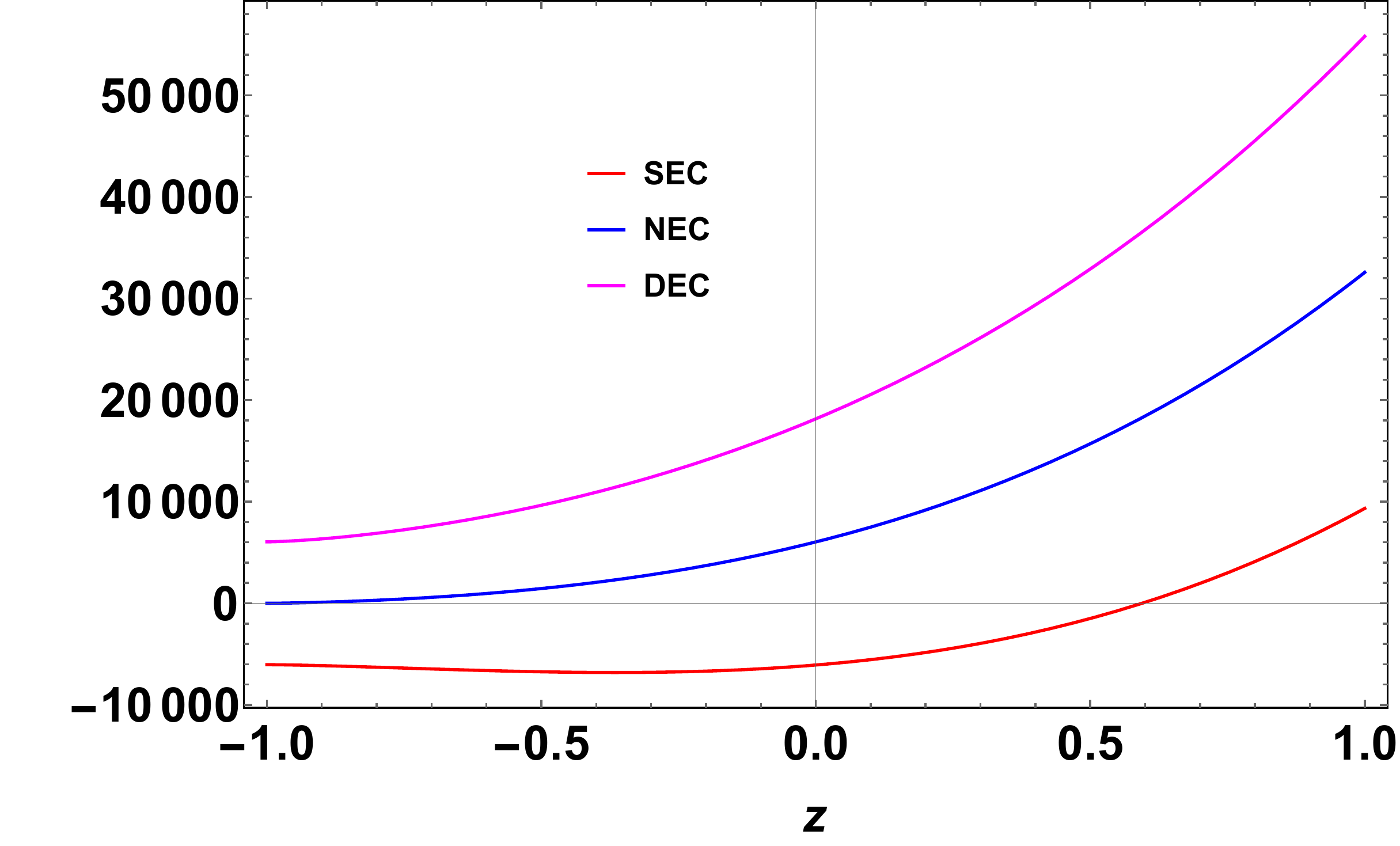}
\caption{ ECs as a function redshift $z$ for $\protect\beta=31.7455,\protect%
\alpha=-1.5$, $m=0.00155$ \& $n=-5$ for hybrid teleparallel gravity.}
\label{f16a}
\end{figure}

\begin{figure}[H]
\centering
\includegraphics[width=8.5 cm]{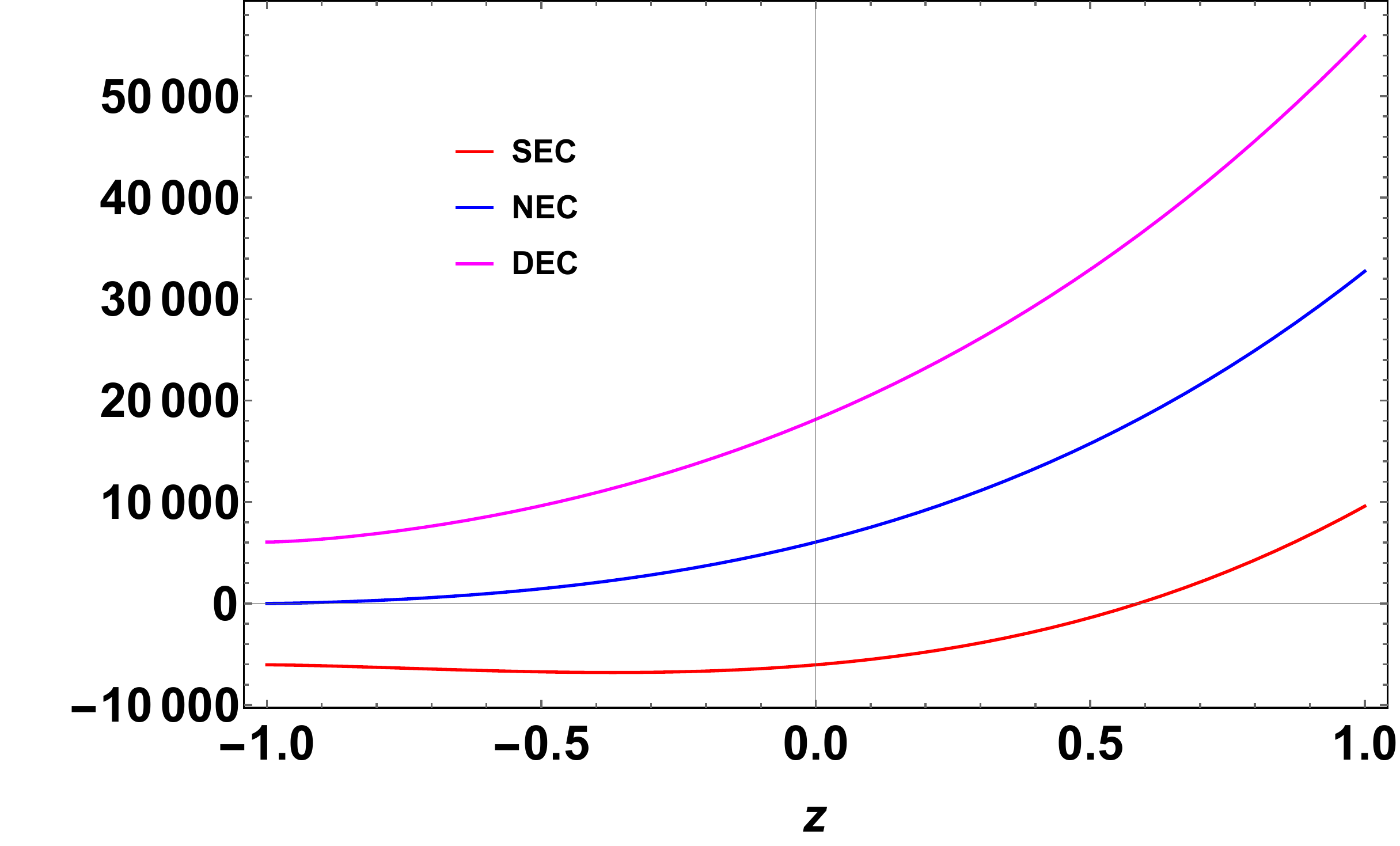}
\caption{ECs as a function redshift $z$ for $\protect\beta=31.7455,\protect%
\alpha=-1.5, b=-2$ \& $d=0.2$ for logarithmic teleparallel gravity.}
\label{f17a}
\end{figure}

Energy conditions for both the teleparallel gravity models are presented in
Fig. \ref{f16a}-\ref{f17a}.


\section{Observational Constraints}\label{VIIa}

In order to find the best fit value of the model parameters of our obtained
models, we need to constrain the parameters with some available datasets.
Here, in this chapter three datasets, namely, Hubble datasets with $57$ datapoints,
Supernovae datasets consisting of $580$ data points from Union$2.1$
compilation datasets and BAO datasets are used. For data analysis, the Bayesian statistics are used.

\subsection{Hubble Parameter}

Recently, Sharov and Vasiliev \cite{sharov} compiled a list of $57$ points
of measurements of the Hubble parameter at in the redshift range $%
0.07\leqslant z\leqslant 2.42$, measured by extraction of $H(z)$ from
line-of-sight BAO data including the analysis of correlation functions of
luminous red galaxies \cite{Hz1} and $H(z)$ estimations from differential
ages $\vartriangle t$ of galaxies (DA method) \cite{Hz2}. (See the Appendix
in \cite{sharov} for full list of tabulated datasets). Chi square test is
used to constrain the model parameters parameters given by,

\begin{equation}
\chi _{OHD}^{2}(p_{s})=\sum_{i=1}^{57}\frac{%
[H_{th}(p_{s},z_{i})-H_{obs}(z_{i})]^{2}}{\sigma _{H(z_{i})}^{2}}
\label{chi}
\end{equation}%
where $H_{th}(p_{s},z_{i})$ denotes the Hubble parameter at redshift $z_{i}$
predicted by the models with $p_{s}$ denoting the parameter space ($\alpha $
here in our model), $H_{obs}(z_{i})$ is the $i$-th measured one and $\sigma
_{H(z_{i})}$ is its uncertainty. One prior assumption is made as $H_{0}=67.8$ (Plank
result predicted value) for this analysis.

\subsection{Type Ia Supernova}

Further, the $580$ points of Union$2.1$ compilation supernovae
datasets \cite{SNeIa} is considered for the analysis for which the chi square formula is
given as,

\begin{equation}
\chi _{SN}^{2}(\mu _{0},p_{s})=\sum\limits_{i=1}^{580}\frac{[\mu _{th}(\mu
_{0},p_{s},z_{i})-\mu _{obs}(z_{i})]^{2}}{\sigma _{\mu (z_{i})}^{2}},
\label{chisn}
\end{equation}%
where $\mu _{th}$ and $\mu _{obs}$ are correspondingly the theoretical and
observed distance modulus for the model and the standard error is $\sigma
_{\mu (z_{i})}$. The distance modulus $\mu (z)$ is defined to be $\mu
(z)=m-M=5LogD_{l}(z)+\mu _{0},$ where $m$ and $M$ are the apparent and
absolute magnitudes of any standard candle (supernovae of type \textit{Ia}
here) respectively. Luminosity distance $D_{l}(z)$ and the nuisance
parameter $\mu _{0}$ are given by $D_{l}(z)=(1+z)H_{0}\int_{0}^{z}\frac{1}{%
H(z^{\ast })}dz^{\ast }$ and $\mu _{0}=5Log\Big(\frac{H_{0}^{-1}}{Mpc}\Big)%
+25$ respectively. In order to calculate luminosity distance, this analysis
restricted the series of $H(z)$ up to tenth term only and then integrated
the approximate series to obtain the luminosity distance.

\subsection{Baryon Acoustic Oscillations}

Finally, a sample of BAO distances measurements from surveys of
SDSS(R) \cite{padn}, 6dF Galaxy survey \cite{6df}, BOSS CMASS \cite{boss}
and three parallel measurements from WiggleZ \cite{wig} is considered. In the context of
BAO measurements, the distance redshift ratio $d_{z}$ is given as, 
\begin{equation}
d_{z}=\frac{r_{s}(z_{\ast })}{D_{v}(z)},  \label{drr}
\end{equation}%
where $r_{s}(z_{\ast })$ is the co-moving sound horizon at the time photons
decouple and $z_{\ast }$ indicates the photons decoupling redshift i.e. $%
z_{\ast }=1090$ \cite{planck/2015}. Moreover, $r_{s}(z_{\ast })$ is assumed same as
considered in the reference \cite{waga} together with the dilation scale is
given by $D_{v}(z)=\big(\frac{d_{A}^{2}(z)z}{H(z)}\big)^{\frac{1}{3}}$,
where $d_{A}(z)$ is the angular diameter distance. The $\chi _{BAO}^{2}$
values corresponding to BAO measurements are discussed in details in \cite%
{gio} and the chi square formula is given by,%
\begin{equation}
\chi _{BAO}^{2}=A^{T}C^{-1}A,  \label{OBAO}
\end{equation}%
where the matrix $A$ is given by,
\[
A=\left[ {%
\begin{array}{cc}
\frac{d_{A}(z_{\ast })}{D_{v}(0.106)}-30.84 &  \\ 
\frac{d_{A}(z_{\ast })}{D_{v}(0.35)}-10.33 &  \\ 
\frac{d_{A}(z_{\ast })}{D_{v}(0.57)}-6.72 &  \\ 
\frac{d_{A}(z_{\ast })}{D_{v}(0.44)}-8.41 &  \\ 
\frac{d_{A}(z_{\ast })}{D_{v}(0.6)}-6.66 &  \\ 
\frac{d_{A}(z_{\ast })}{D_{v}(0.73)}-5.43 &  \\ 
& 
\end{array}%
}\right] ,
\]%
and $C^{-1}$ representing the inverse of covariance matrix $C$ given as in
the reference \cite{gio} adopting the correlation coefficients presented in 
\cite{hing} as 
\[
C^{-1}=\left[ {%
\begin{array}{cccccc}
0.52552 & -0.03548 & -0.07733 & -0.00167 & -0.00532 & -0.00590 \\ 
-0.03548 & 24.97066 & -1.25461 & -0.02704 & -0.08633 & -0.09579 \\ 
-0.07733 & -1.25461 & 82.92948 & -0.05895 & -0.18819 & -0.20881 \\ 
-0.00167 & -0.02704 & -0.05895 & 2.91150 & -2.98873 & 1.43206 \\ 
-0.00532 & -0.08633 & -0.18819 & -2.98873 & 15.96834 & -7.70636 \\ 
-0.00590 & -0.09579 & -0.20881 & 1.43206 & -7.70636 & 15.28135 
\end{array}%
}\right] .
\]

Below, a comparision of the obtained model with the $\Lambda $%
CDM model together with error bars due to the $57$ points of $H(z)$ datasets
and the $580$ points of Union$2.1$ compilation datasets is shown.

\begin{figure}[H]
\label{Error-sn-teleparallel}
\par
\begin{center}
$%
\begin{array}{c@{\hspace{.1in}}c}
\includegraphics[width=3.0 in, height=2.5 in]{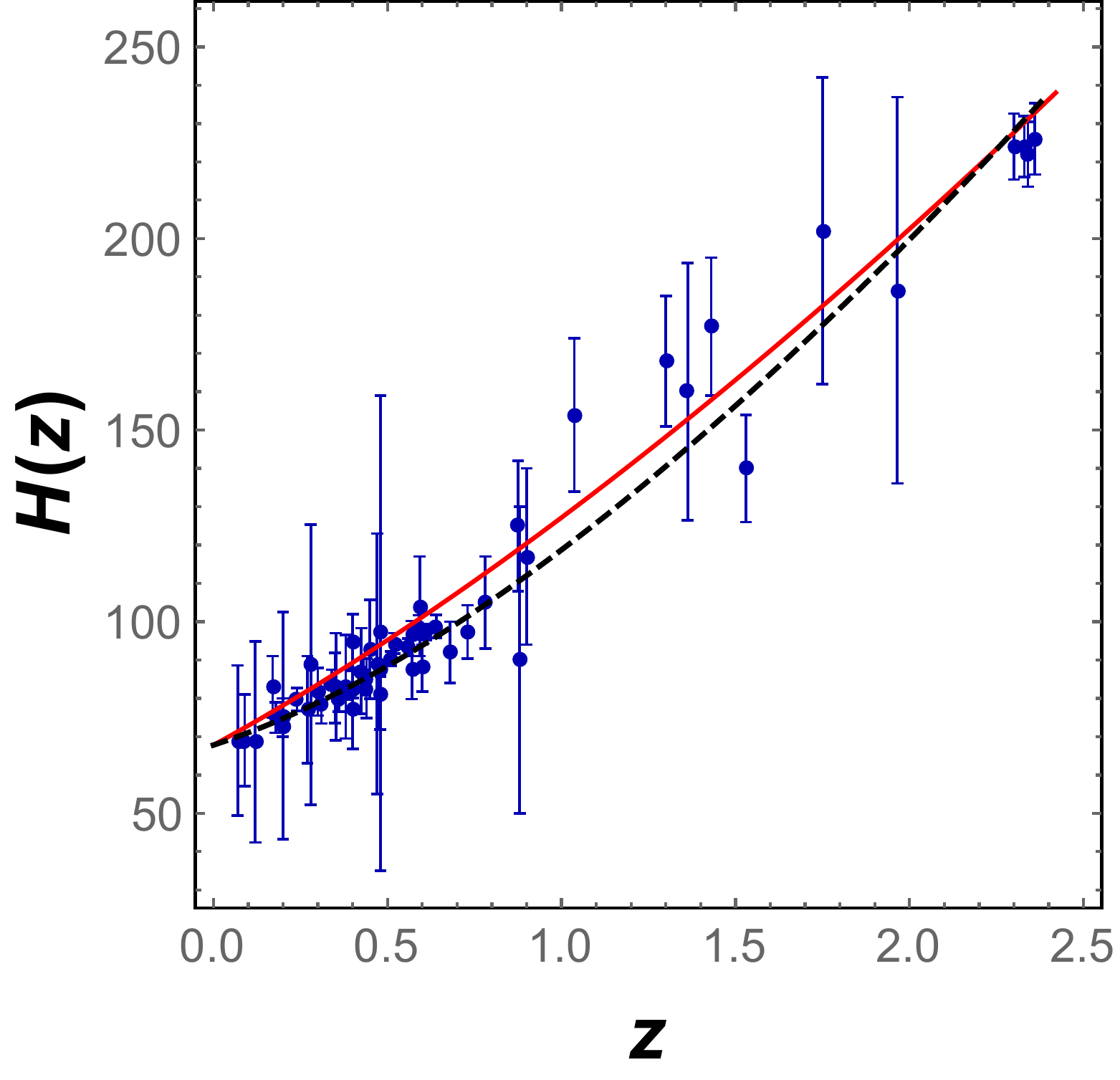} & %
\includegraphics[width=3.0 in, height=2.5 in]{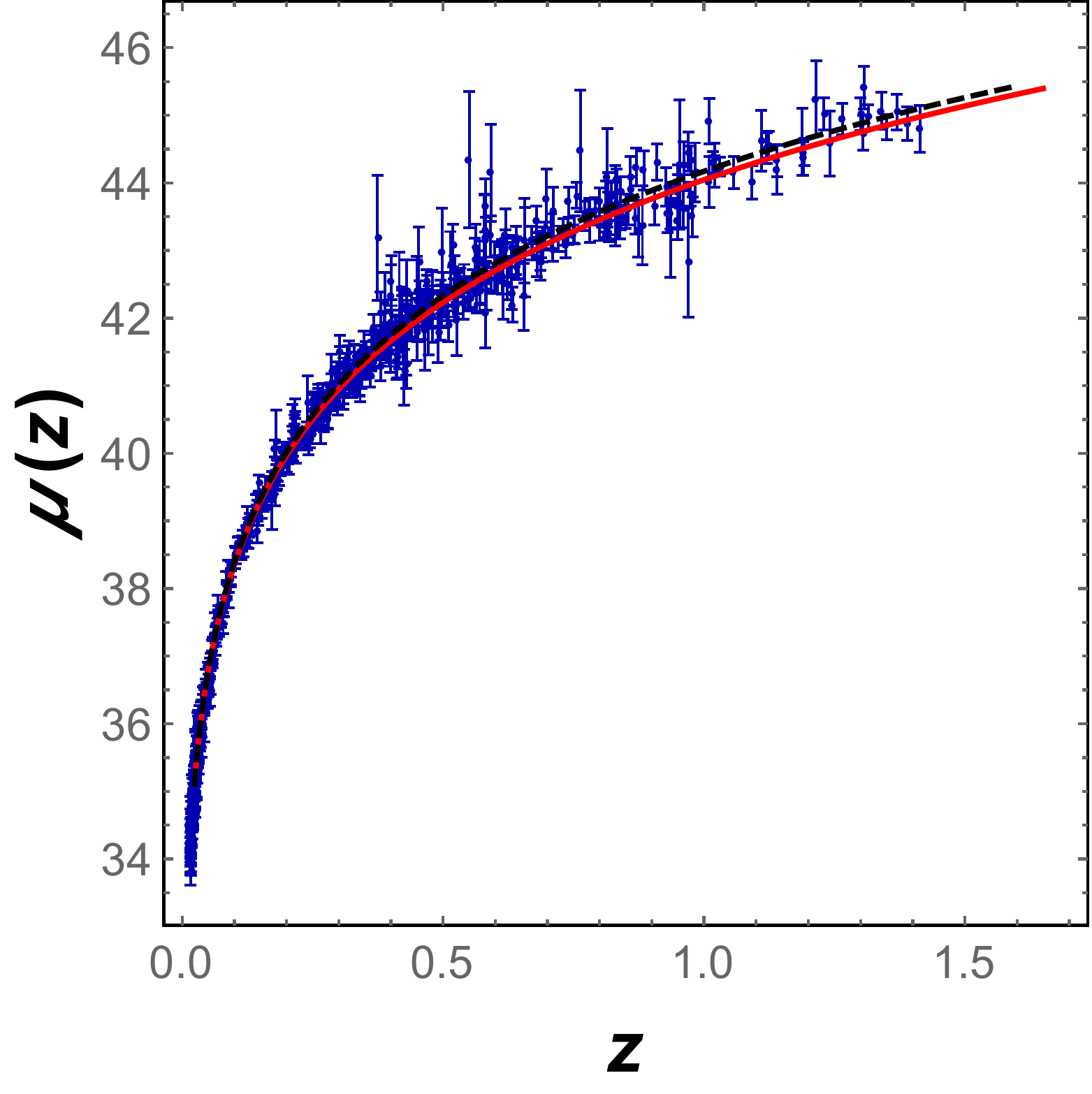} \\ 
\mbox (a) & \mbox (b)%
\end{array}%
$%
\end{center}
\caption{ Figures (a) and (b) are the error bar plots of $57$
points of $H(z)$ datasets and $580$ points of Union $2.1$ compilation
supernovae datasets together with our obtained model (solid red lines) and $%
\Lambda$CDM model (black dashed lies), respectively.}\label{f18a}
\end{figure}

Next, the likelihood contours for the model parameter $\alpha $
and Hubble constant $H_{0}$ with errors at $1$-$\sigma $, $2$-$\sigma $ and $%
3$-$\sigma $ levels in the $\alpha $-$H_{0}$ plane are shown. The best fit constrained
values of $\alpha $ and $H_{0}$ are found to be $\alpha =-1.497294$ \& $%
H_{0}=63.490604$ due to $H(z)$ datasets only with $\chi _{\min
}^{2}=31.333785$ and $\alpha =-1.503260$ \& $H_{0}=63.361612$ due to joint
datasets $H(z)$ + SNeIa + BAO with $\chi _{\min }^{2}=650.312968$
respectively.

\begin{figure}[H]
\begin{center}
$%
\begin{array}{c@{\hspace{.1in}}c}
\includegraphics[width=3.0 in, height=2.5 in]{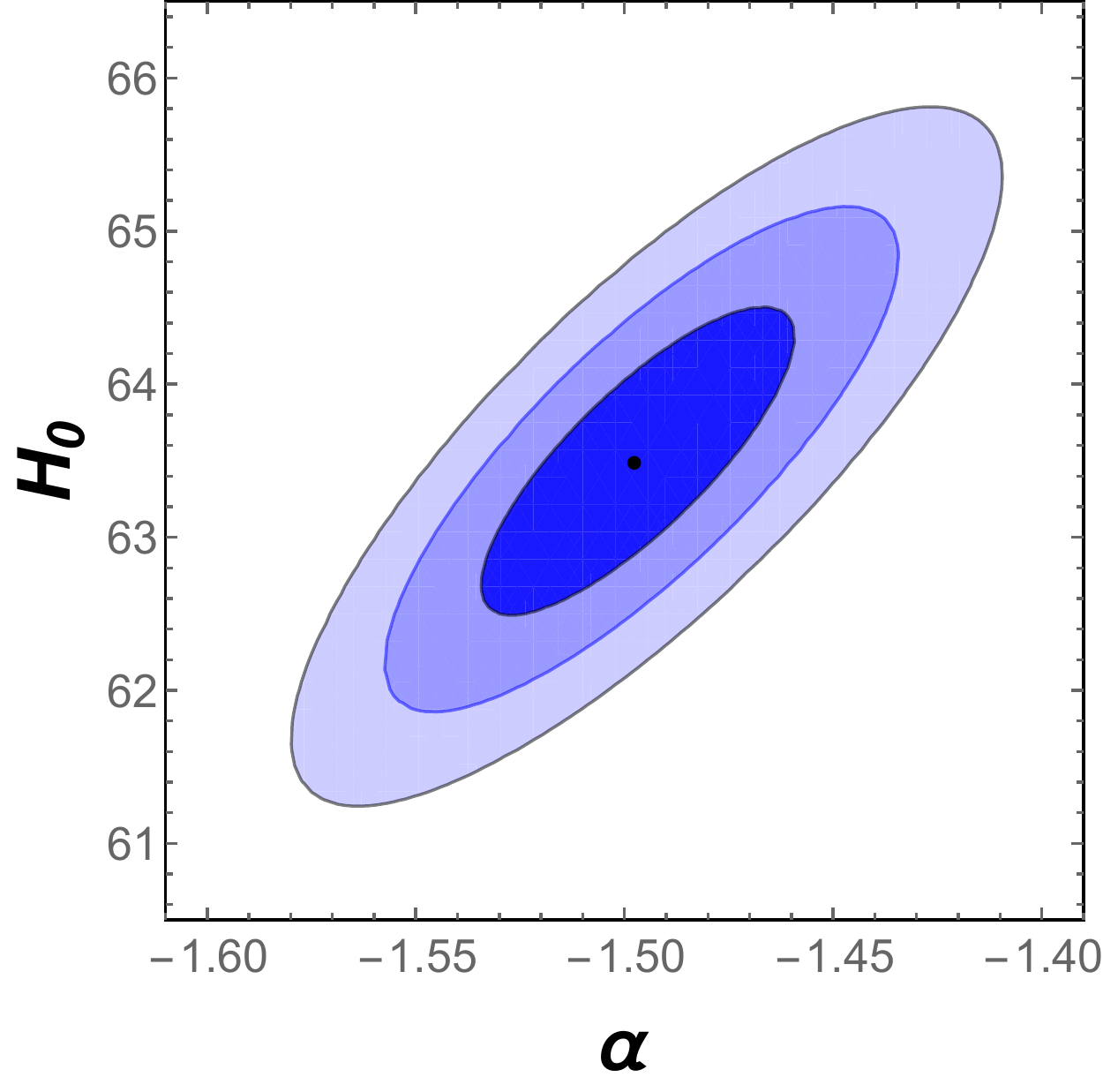} & %
\includegraphics[width=3.0 in, height=2.5
in]{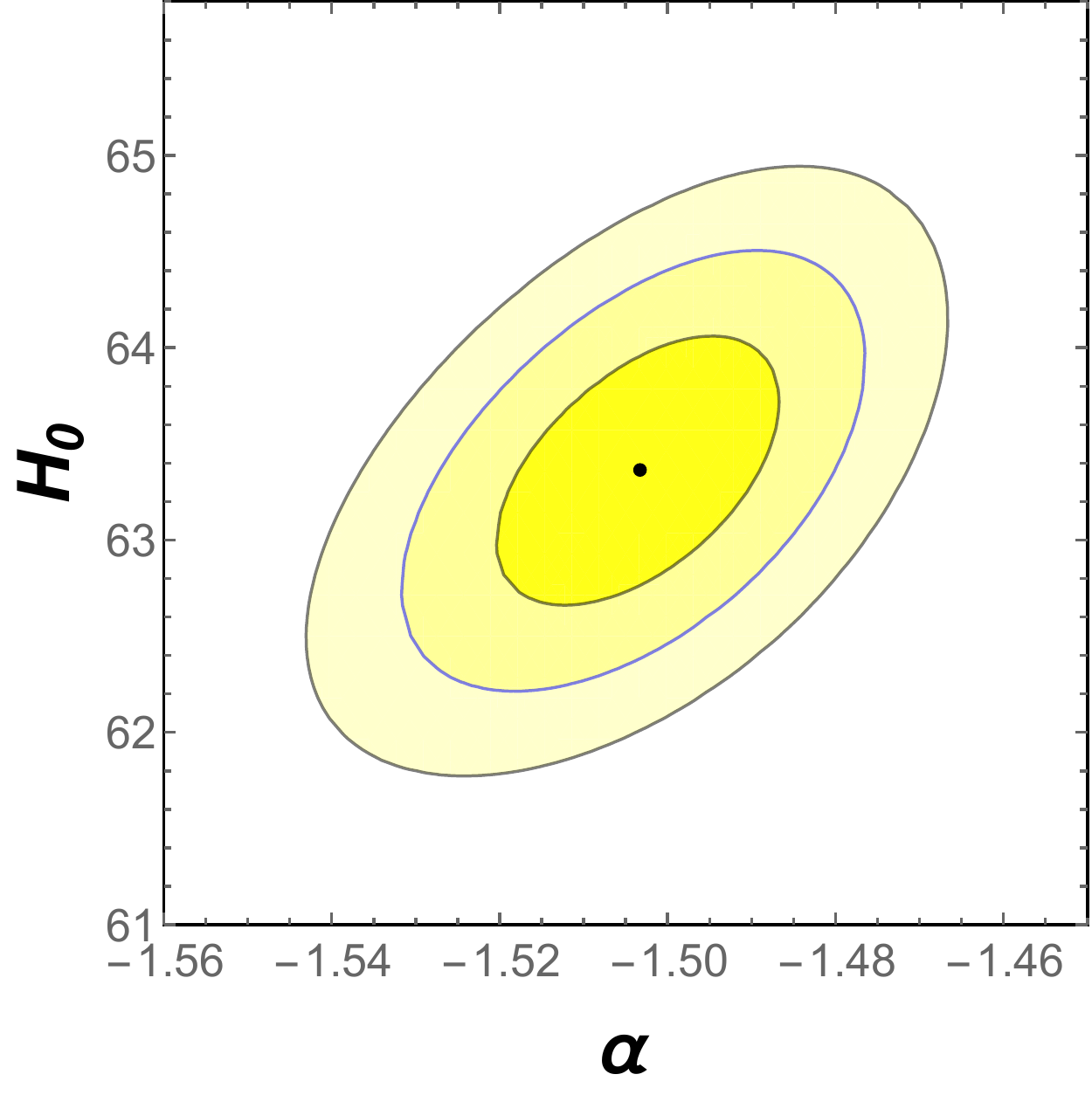} \\ 
\mbox (a) & \mbox (b)%
\end{array}%
$%
\end{center}
\caption{Figures (a) shows the maximum likelihood contours in the $\protect%
\alpha $-$H_{0}$ plane for $H(z)$ datasets only while figure (b) shows the
maximum likelihood for $H(z)$ + SNeIa + BAO datasets jointly. The three
contour regions shaded with dark, light shaded and ultra lightshaded in both
the plots are with errors at $1$-$\protect\sigma $, $2$-$\protect\sigma $
and $3$-$\protect\sigma $ levels. The black dots represent the best fit
values of model parameter $\protect\alpha $ and $H_{0}$ in both the plots.}\label{f19a}
\end{figure}

\section{Results and Discussions}\label{VIIIa}

The chapter communicates the phenomena of late time acceleration in the
framework of hybrid and logarithmic teleparallel gravity. To obtain the
exact solutions of the field equations, this chapter employs a parametrization of
the deceleration parameter first proposed in \cite{banerjee}. This chapter discusses the energy conditions and the cosmological viability of the underlying teleparallel gravity models.

In Section \ref{VIa}, this chapter shows the temporal evolution of SEC, NEC, and WEC for
both the teleparallel gravity models. Note that in order to suffice the late
time acceleration, the SEC has to violate \cite{non41to42}. This is due to
the fact that for an accelerating universe compatible with observations \cite%
{riess/1998,perlmutter/1999}, the EoS parameter $\omega \simeq -1$ \cite{Planck/2018}, and
therefore $\rho (1+3\omega )<0$ always. From Fig.\ref{f16a}-\ref{f17a}, one
can clearly observe that SEC violates for both the models, whereas NEC and WEC
do not violate. Interestingly, SEC is also violated for curvature coupled
, and minimally coupled scalar field theories \cite{non41to42}.

To understand the cosmological viability of both the teleparallel gravity
models, this chapter show in Fig. \ref{f1a}, the deceleration parameter ($q$) as a
function of redshift. The plot of the deceleration parameter $q(z)$ clearly
shows that our model successfully generates late time cosmic acceleration
along with a decelerated expansion in the past for $-1>\alpha >-2$. The
deceleration parameter undergoes a signature flipping at the redshift $%
z_{tr}\simeq0.6$ for the chi-square value of $\alpha =-1.5$ which is
compatible with latest Planck measurements \cite{Planck/2018}. The obtained values of $%
q_{0}=-0.251355$ and $z_{tr}$ are consistent with values reported by other
authors \cite{Jaime/2019}. From Figs. \ref{f8a} \& \ref{f12a} the values of EoS 
$\omega $ at $z=0$ for both our models are obtained as $-0.500903$ and $%
-0.500935$ respectively. These values of $\omega $ behave in concordance
with standard cosmological model predictions (with Plank data, $\omega
_{eff}^{\Lambda \text{CDM}}\sim -0.68$ at $z=0$ as in Ref. \cite{Planck/2018}).

In Fig. \ref{f5a},\ref{f6a},\ref{f9a} and \ref{f10a}, we plot the energy density
for both the models as a function of redshift. In this chapter, we choose the model parameters
so as to satisfy the WEC. Fulfillment of WEC ensures the cosmological
pressure has to be negative to account for the negative EoS parameter and, therefore, the cosmic acceleration. It is interesting to note that no known
entity has the remarkable property of negative pressure and can only be
achieved by exotic matter or by modifications to general relativity.

The EoS parameter is an important cosmological parameter that has sparked
a great deal of interest among cosmologists. Owing to the mysterious nature of
the cosmological entity responsible for this acceleration, various dark
energy models have been devised to suffice the observations. To investigate
the nature of the dark energy model represented by the equation \eqref{a8},
we study in Section \ref{VIa}, the $\{r,s\}$ and $\{r,q\}$ plane and $Om(z)$.
We observe that the value of $\alpha $ dictates the evolution of the $r$-$s$
and $r$-$q$ trajectories. We find the model to deviate significantly from
the $\Lambda $CDM at early times. However, at late times the model is
observed to coincide with ($r=1,s=0$) and therefore consistent with $\Lambda 
$CDM cosmology. This result is further re-assured from the $r$-$q$ plane in
Fig. \ref{f15a}. However, discrepancy arises from Fig. \ref{f16a}, where for
none of the values of $\alpha $, we obtain a constant $Om$ which clearly does
not reflect dark energy that is time-independent. Furthermore, the nature
of dark energy represented by the equation \eqref{a8} changes from being an
Quintessence to Phantom as $\alpha $ changes from $\alpha \leq -2$ to $%
\alpha >-2$ respectively. Finally, we have discussed our obtained models in
the light of some observational datasets. The obtained model has a nice fit
to the $57$ points of Hubble datasets and the $580$ points of Union$2.1$
compilation supernovae datasets. We have used Bayesian statistics to
find the constraints on the model parameters. The maximum likelihood
contours for the model parameter $\alpha $ and Hubble constant $H_{0}$ with
errors at $1$-$\sigma $, $2$-$\sigma $ and $3$-$\sigma $ levels in the $%
\alpha $-$H_{0}$ plane is shown separately for $H(z)$ datasets only and
joint datasets $H(z)$ + SNeIa + BAO. The best fit constrained values of $%
\alpha $ and $H_{0}$ are found to be $\alpha =-1.497294$ \& $H_{0}=63.490604$
due for $H(z)$ datasets with $\chi _{\min }^{2}=31.333785$ and $\alpha
=-1.503260$ \& $H_{0}=63.361612$ due to $H(z)$ + SNeIa + BAO datasets
with $\chi _{\min }^{2}=650.312968$, respectively.

In further study, it would be interesting to find the range of the model parameters for which corresponding cosmological models will show the accelerated expansion for the present time.



\chapter{Energy Conditions in $f(Q)$ Gravity} 

\label{Chapter4} 

\lhead{Chapter 3. \emph{Energy Conditions in $f(Q)$ Gravity}} 

\vspace{8 cm}
* The work, in this chapter, is covered by the following publication: \\
 
\textit{Energy Conditions in $f(Q)$ Gravity}, Physical Review D, \textbf{102}, 024057 (2020).

\clearpage
In this chapter, we aim to discuss one promising approach that lies in a new class of teleparallel theory of gravity named $f(Q)$, where the non-metricity $Q$ is responsible for the gravitational interaction. The important role any of these alternative theories should obey are the energy condition constraints. Such constraints establish the compatibility of a given theory with the causal and geodesic structure of space-time. This chapter presents a complete test of energy conditions for $f(Q)$ gravity models. The energy conditions allowed us to fix our free parameters, restricting the families of $f(Q)$ models compatible with the accelerated expansion our Universe passes through. Our results show the viability of the $f(Q)$ theory, leading us close to the dawn of a complete theory for gravitation.

\section{Introduction}\label{I}


In literature, there are several motivations to explore theories beyond the standard formulation of gravity. Among this effort, we highlight models based on the so-called symmetric teleparallel gravity or $f(Q)$ gravity, introduced by Jimenez et al. \cite{Jimenez/2018}, where the non-metricity $Q$ is responsible for the gravitational interaction. Investigations on $f(Q)$ gravity have been rapidly developed as well as observational constraints to confront it against standard GR formulation. 

An interesting set of constraints on $f(Q)$ gravity was done by Lazkoz et al. \cite{Lazkoz/2019}, where the $f(Q)$ Lagrangian is written as polynomial functions of the redshift $z$. The constraints for these models were successfully derived using data from the expansion rate, Type Ia Supernovae, Quasars, Gamma-Ray Bursts, Baryon Acoustic Oscillations data, and Cosmic Microwave Background distance. Another relevant analysis about $f(Q)$ gravity consists in understanding its behavior under different energy conditions. 

As it is known, the energy conditions represent paths to implement the positiveness of the stress-energy tensor in the presence of matter. Moreover, they can be used to describe the attractive nature of gravity, besides assigning the fundamental causal and the geodesic structure of space-time \cite{Capozziello/2018}. This chapter studies the strong, the weak, the null, and the dominant energy conditions for $f(Q)$ gravity, working with a perfect fluid matter distribution. The actual accelerating phase of our Universe passes through constraints that the strong energy condition should be violated. This constraint, together with the actual values of Hubble and deceleration parameters, allowed us to test the viability of different forms of $f(Q)$ gravity. 

The ideas presented in this chapter are organized in the following way: section \ref{IIb} presents the generalities on $f(Q)$ gravity, the field equations, as well as the energy conservation equation for a perfect fluid. Section \ref{IIIb} shows the explicit forms of the energy conditions derived from the Raychaudhuri equations. The two scenarios for $f(Q)$ gravity and their constraints are carefully analyzed through section \ref{IVb}. This chapter also verified the deviations between our scenarios and $\Lambda$CDM cosmological model in section \ref{Vb}. The final remarks and perspectives are discussed in section \ref{VIb}.

\section{Motion Equations in $f(Q)$ Gravity}\label{IIb}

The FLRW line element enable us to write the trace of the non-metricity tensor as
\begin{align*}
Q=6H^2\,.
\end{align*} 
Now, let us take the energy-momentum tensor for a perfect fluid, or
\begin{equation}
\label{b1}
T_{\mu\nu}=(p+\rho)u_{\mu}u_{\nu}+pg_{\mu\nu}\,,
\end{equation}
Therefore, by substituting \eqref{flrw}, and \eqref{b1} in \eqref{64} one can find 
\begin{equation}
\label{b2}
3H^2=\frac{1}{2f_Q}\left(-\rho+\frac{f}{2}\right)\,,
\end{equation}
\begin{equation}
\label{b3}
\dot{H}+3H^2+\frac{\dot{f_Q}}{f_Q}H=\frac{1}{2f_Q}\left(p+\frac{f}{2}\right)\,,
\end{equation}
as the modified Friedmann equations for $f(Q)$ gravity.
The modified Friedmann equations enable us to write the density and the pressure for the universe as,
\begin{equation}
\label{b4}
\rho=\frac{f}{2}-6H^2f_Q\,,
\end{equation}
\begin{equation}
\label{b5}
p=\left(\dot{H}+3H^2+\frac{\dot{f_Q}}{f_Q}H\right)2f_Q-\frac{f}{2}\,.
\end{equation}

In analogy with GR, we can rewrite Eq.\eqref{b2}, \eqref{b3} as,
\begin{equation}\label{b6}
3H^2=-\frac{1}{2}\tilde{\rho}\,,
\end{equation}
\begin{equation}\label{b7}
\dot{H}+3H^2=\frac{\tilde{p}}{2}\,.
\end{equation}
where
\begin{equation}
\label{b8}
\tilde{\rho}=\frac{1}{f_Q}\left(\rho-\frac{f}{2}\right)\,,
\end{equation}
\begin{equation}
\label{b9}
\tilde{p}=-2\,\frac{\dot{f_Q}}{f_Q}\,H+\frac{1}{f_Q}\,\left(p+\frac{f}{2}\right)\,.
\end{equation}
The previous equations are going to be components of a modified energy-momentum tensor $\tilde{T}_{\,\mu\nu}$, embedding the dependence on the trace of the non-metricity tensor.
\section{Energy Conditions}\label{IIIb}

Since, this chapter works with a perfect fluid matter distribution, the energy conditions recovered from standard GR are
\begin{itemize}
\item SEC if  $\tilde{\rho}+3\tilde{p}\geq 0\,$;

\item WEC if  $\tilde{\rho}\geq 0, \tilde{\rho}+\tilde{p}\geq 0\,$;

\item NEC if  $\tilde{\rho}+\tilde{p}\geq 0\,$;

\item DEC if $\tilde{\rho}\geq 0, |\tilde{p}|\leq \rho\,$.
\end{itemize}

Taking Eq.  \eqref{b8} and \eqref{b9} into WEC, NEC and DEC constraints, one can able to prove that

\begin{itemize}
\item WEC if  $\rho\geq 0, \rho+p\geq 0\,$;

\item NEC if  $\rho+p\geq 0\,$;

\item DEC if $\rho\geq 0, |p|\leq \rho\,$,
\end{itemize}
corroborating with the work from Capozziello et al.\cite{Capozziello/2018}. In the case of SEC condition, we yield to the constraint
\begin{equation}\label{b13}
\rho+3\,p-6\,\dot{f_Q}\,H+f \geq 0\,.
\end{equation}

Moreover, let us consider the Hubble, deceleration, jerk, and snap parameters, whose forms are
\begin{eqnarray}
\label{b14}
&&
H=\frac{\dot{a}}{a}\,,\qquad q=-\frac{1}{H^2}\frac{\ddot{a}}{a}\,,\\ \nonumber
&&
j=\frac{1}{H^3}\frac{\dot{\ddot{a}}}{a}\,, \qquad s=\frac{1}{H^4}\frac{\ddot{\ddot{a}}}{a}\,.
\end{eqnarray}
Such parameters enable us to represent the time derivatives of $H$ as,
\begin{equation}
\label{b15}
\dot{H}=-H^2(1+q)\,,
\end{equation}
\begin{equation}
\label{b16}
\ddot{H}=H^3(j+3q+2)\,,
\end{equation}
\begin{equation}
\label{b17}
\dot{\ddot{H}}=H^4(s-2j-5q-3)\,.
\end{equation}
So, by using Eq. \eqref{b14}-\eqref{b15}, we can rewrite \eqref{b4}, and \eqref{b5} as
\begin{equation}
\label{b18}
\rho=\frac{f}{2}-6H^2f_Q\,,
\end{equation}
\begin{equation}
\label{b19}
p=\left(-H^2(1+q)+3H^2+\frac{\dot{f_Q}}{f_Q}H\right)2f_Q-\frac{f}{2}\,,
\end{equation}
which are the density and the pressure for the $f(Q)$ gravity. Therefore, the previous equations establish the following constraints for the energy conditions:
\begin{equation}
\label{b20}
\text{\textbf{SEC :}} \ \rho+3p-6\dot{f_Q}H+f=\frac{f}{2}-6H^2f_Q+3\left(-H^2(1+q)+3H^2+\frac{\dot{f_Q}}{f_Q}H\right)2f_Q-3\frac{f}{2}-6\dot{f_Q}H+f\geq0 \,,
\end{equation}
\begin{equation}
\label{b21}
\text{\textbf{NEC :}} \ \rho+p=-6H^2f_Q+\left(-H^2(1+q)+3H^2+\frac{\dot{f_Q}}{f_Q}H\right)2f_Q\geq0\, ,
\end{equation}
\begin{equation}\label{b22}
\text{\textbf{WEC :}} \ \rho=\frac{f}{2}-6H^2f_Q\geq0,
\rho+p=-6H^2f_Q+\left(-H^2(1+q)+3H^2+\frac{\dot{f_Q}}{f_Q}H\right)2f_Q\geq0\, ,
\end{equation}
\begin{equation}\label{b23}
\text{\textbf{DEC :}} \ \rho=\frac{f}{2}-6H^2f_Q\geq0,
\rho\pm p=\frac{f}{2}-6H^2f_Q\pm3\left(-H^2(1+q)+3H^2+\frac{\dot{f_Q}}{f_Q}H\right)2f_Q-3\frac{f}{2}\geq0\, .
\end{equation}

\section{Constraining $f(Q)$ Theories}\label{IVb}

This section discusses the viability of the functional form of $f(Q)$ in FLRW spacetime. In order to do so,  the present values for the Hubble and the decelerating parameters are considered as $H_0=67.9\,,$ and $q_0=-0.503$, respectively \cite{Planck/2018,Capozziello/2019}. Moreover, several observations confirm that the universe is going through an accelerated phase \cite{Riess/1998}, carried by a negative pressure regime. Such a scenario imposes that SEC needs to be violated \cite{Planck/2018}.

There are several approaches in the literature deriving energy conditions beyond Einstein's GR, we can see for instance, EC constraints in $f(R)$ theory \cite{Santos/2007,Bertolami/2009}, $f(G)$ theory \cite{Gracia/2011,Bamba/2017}, $f(\mathcal{T})$ theory \cite{Liu/2012}, $f(\mathcal{G},T)$ theory \cite{Sharif/2016}, $f(R,T,R_{\mu\nu}T^{\mu\nu})$ theory \cite{Sharif/2013}, $f(R,\mathcal{G})$ theory \cite{Atazadeh/2014}, $f(R,\square R,T)$ gravity \cite{Yousaf/2018}, $f(R,T)$ theory \cite{Moraes/2019} etc. However, the previous studies are mainly focused on the WEC energy condition, whereas our intent in this paper is to study the constraint of all the ECs in $f(Q)$ theory. To investigate the ECs with the present values of the geometrical parameters in $f(Q)$ theory, we need to fix the functional form of $f(Q)$. Once this form is fixed, it will be easy to investigate the cosmological scenarios. In their beautiful work, T. Harko, et al. \cite{Harko/2018} discussed the coupling matter in modified $Q$ gravity by assuming a power-law and an exponential form of $f(Q)$.  This investigation, motivated us to work with a polynomial form for $f(Q)$ gravity. Moreover, this chapter also introduce a logarithmic dependence of $f(Q)$, which we are going to analyze thoroughly.

\subsection{$f(Q)= Q+mQ^n$}

\begin{figure}[H]
\begin{center}
$%
\begin{array}{c@{\hspace{.1in}}c}
\includegraphics[width=3.0 in, height=2.5 in]{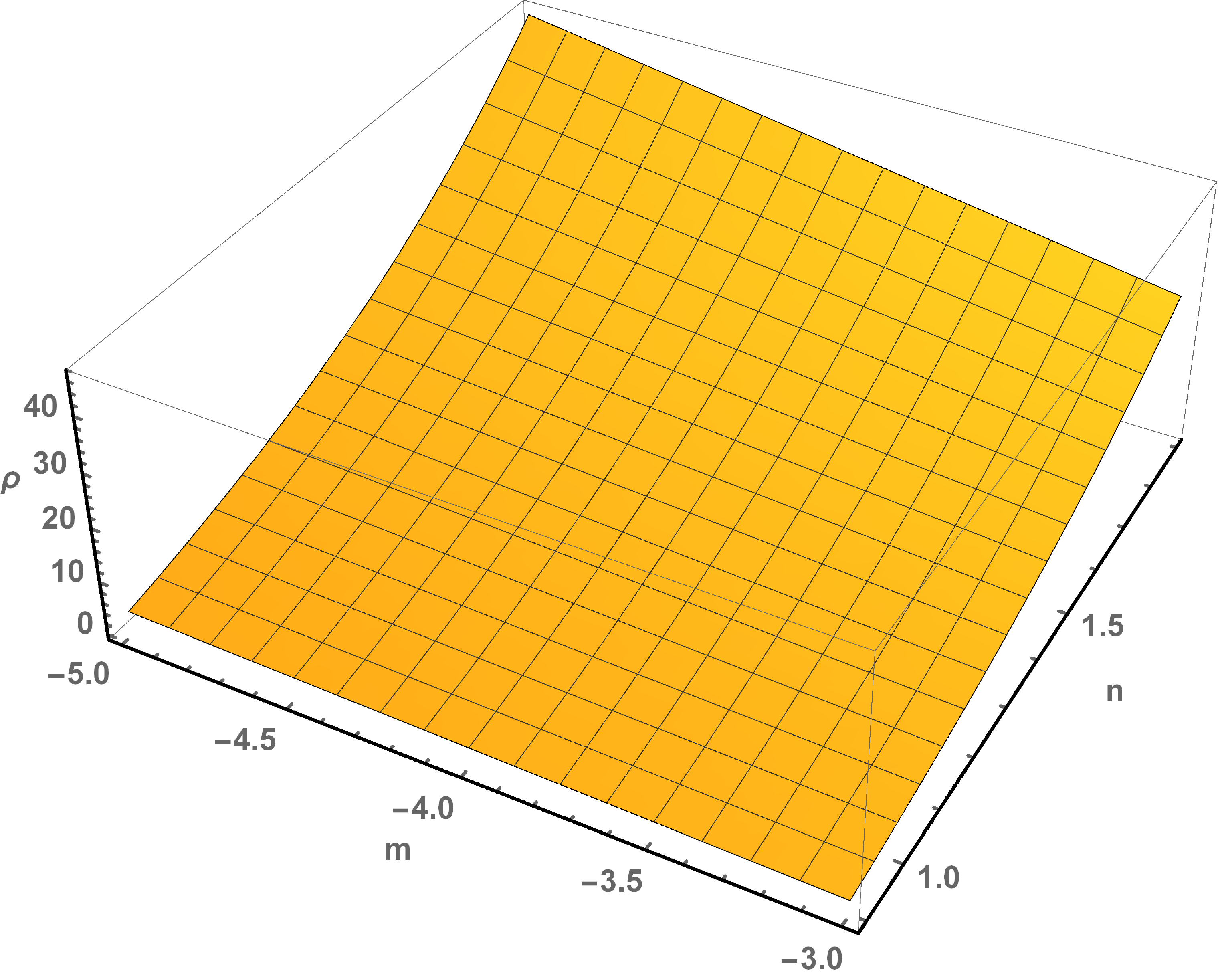} & %
\includegraphics[width=3.0 in, height=2.5 in]{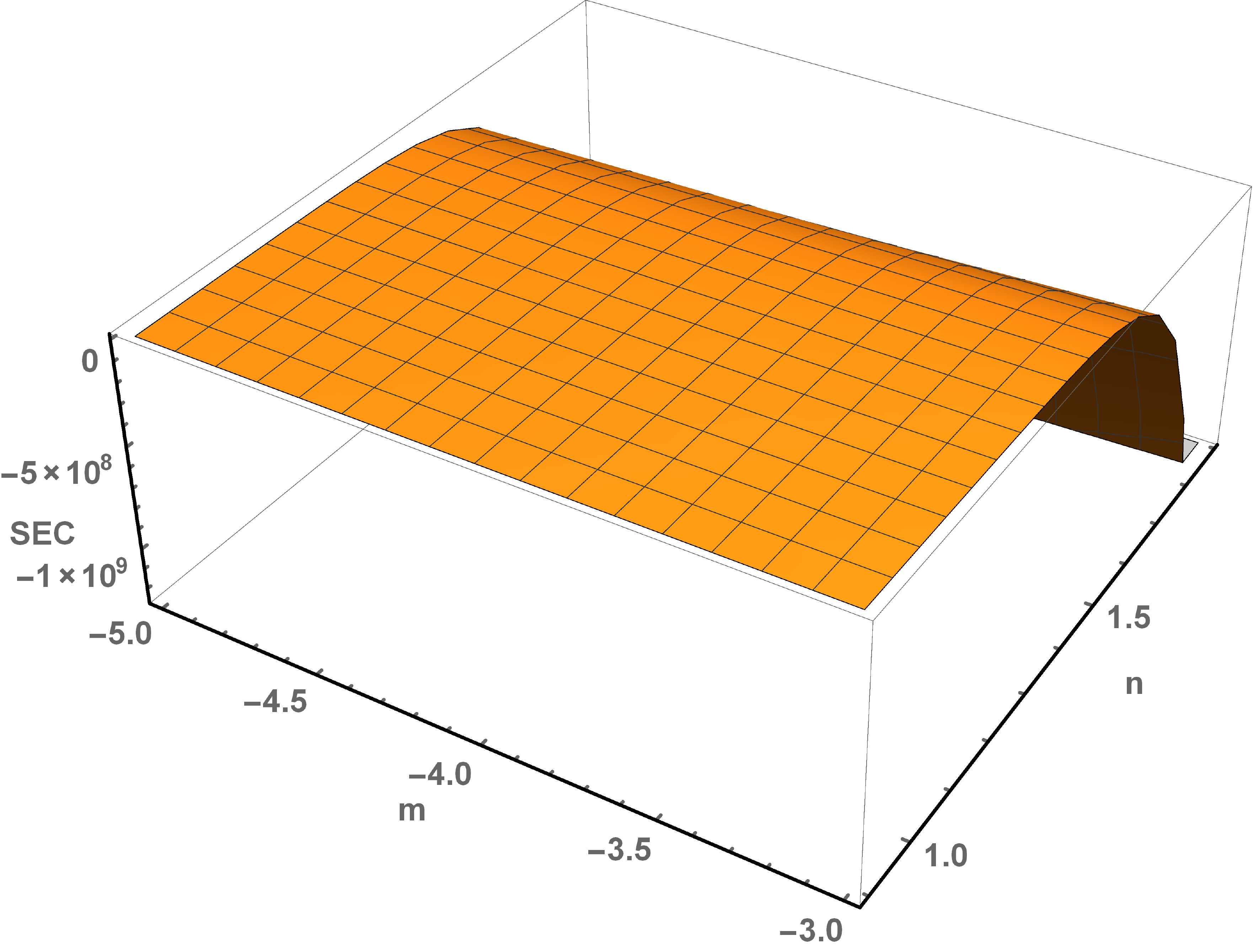}\\
\includegraphics[width=3.0 in, height=2.5 in]{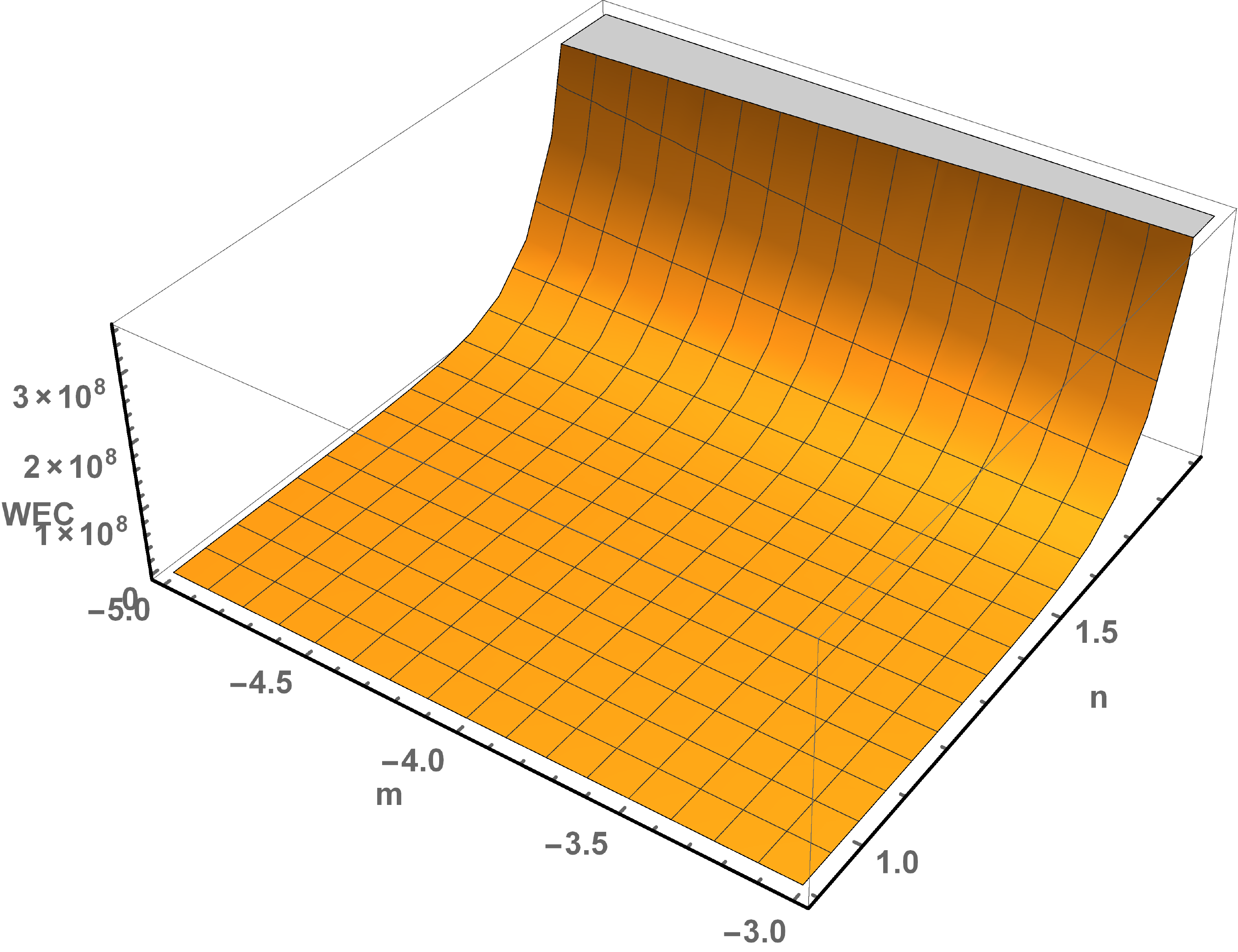} & %
\includegraphics[width=3.0 in, height=2.5 in]{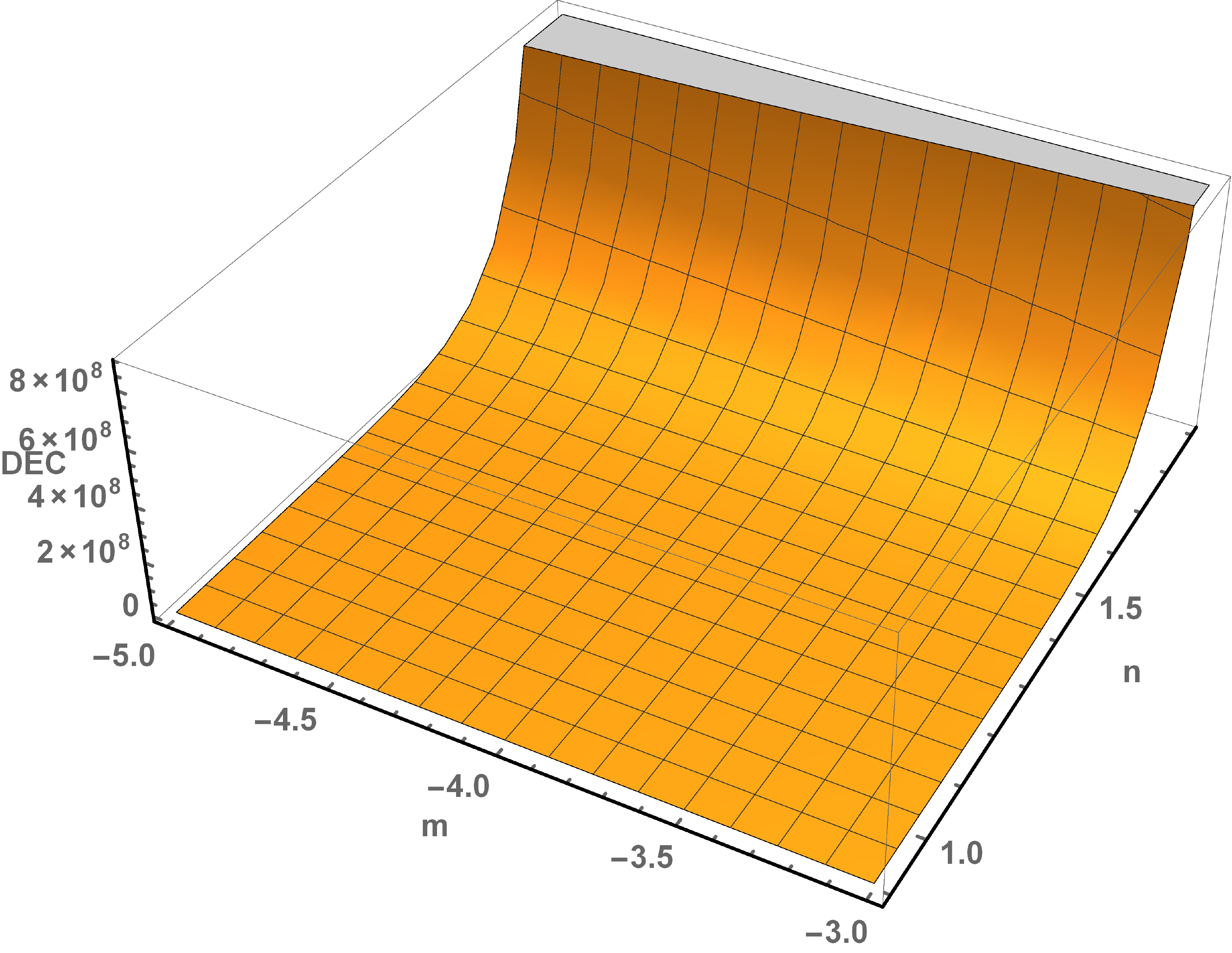}\\
\end{array}%
$%

%
%
%

\caption{Energy conditions for $f(Q)= Q+mQ^n$ derived with the present values of $H_0$ and $q_0$ parameters.}
\label{f1b}

\end{center}
\end{figure}

This subsection presume the $f(Q)$ as a algebraic polynomial function of $Q$ with free parameters $m$ and $n$. The previous function establishes that the ECs need to satisfy the following conditions:
\begin{multline}\label{b24}
\text{\textbf{SEC : }}3 H_0^2-m 6^n \left\lbrace 2^{-1} (2 n-1)-1\right\rbrace H_0^{2n}+\frac{1}{2} \left[H_0^2 (6-12 q_0)-m 6^n (2 n-1) H_0^{2n} \left\lbrace n (q_0+1)-3\right\rbrace\right]+ \\
 2^{1 + n} 3^n H_0^{2n} m (-1 + n) n (1 + q_0)\geq0\,,
\end{multline}
\begin{equation}\label{b25}
\text{\textbf{NEC : }}-3 H_0^2-m2^{-1}6^n (2 n-1) H_0^{2n}+\frac{1}{6} \left[H_0^2 (6-12 q_0)-m 6^n (2 n-1) H_0^{2n} \left\lbrace n (q_0+1)-3\right\rbrace\right]\geq0\,,
\end{equation}
\begin{multline}\label{b26}
\text{\textbf{WEC : }}-3 H_0^2-m2^{-1}6^n (2 n-1) H_0^{2n}\geq0,\\ \text{and }
-3 H_0^2-m2^{-1}6^n (2 n-1) H_0^{2n}+\frac{1}{6} \left[H_0^2 (6-12 q_0)-m 6^n (2 n-1) H_0^{2n} \left\lbrace n (q_0+1)-3\right\rbrace\right]\geq0\,,
\end{multline}
\begin{multline}\label{b27}
\text{\textbf{DEC : }}-3 H_0^2-m2^{-1}6^n (2 n-1) H_0^{2n}\geq0,\\ \text{and }
-3 H_0^2-m2^{-1}6^n (2 n-1) H_0^{2n}\pm\frac{1}{6} \left[H_0^2 (6-12 q_0)-m 6^n (2 n-1) H_0^{2n} \left\lbrace n (q_0+1)-3\right\rbrace\right]\geq0\,.
\end{multline}
From \eqref{b24}-\eqref{b27}, one can easily observe that the ECs directly depend on the free parameters $m$ and $n$. Nevertheless, one cannot take the values of $m$ and $n$ arbitrarily, which may violate the ECs as well as the current scenario of the universe dominated by the dark energy. Therefore, using \eqref{b24}-\eqref{b27}, some restrictions on the model parameters $m$ and $n$ are found. Using WEC, we found that $m$ should be less than or equal to $-1$ ($m\leq -1$), and $n$ should be greater than or equal to $0.9$ ($n\geq 0.9$). Also, equations \eqref{b24}, \eqref{b25}, \eqref{b27} reduces the range of the model parameter to $0.9\leq n\leq 2$, in order to proper describe SEC, NEC and DEC. Finally, this model conclude that for $m\leq -1$ and $0.9\leq n\leq 2$, represents the current stage of the universe. In addition to this, the profile of all energy conditions for some range of model parameters $m$, and $n$ are shown. From Fig. \ref{f1b}, one can observe that WEC, NEC, DEC are satisfied, while SEC is violated, corroborating with an accelerated expansion \cite{Visser/2000,Moraes/2017}.

%
%
%

\subsection{$f(Q)=\alpha+\beta\log Q$}

Here, this subsection introduce $f(Q)$ as a logarithmic function of the non-metricity with free parameters $\alpha$, and $\beta$. Therefore, the ECs are impelled to obey the constraints
\begin{equation}\label{b30}
\text{\textbf{SEC : }}-\beta-2 \beta  (q_0+1) +\frac{3}{2} \left[\alpha +\beta  \log \left(6 H_0^2\right)\right]-\frac{3 H_0^2 \left[\alpha -2 \beta +\beta  \log \left(6 H_0^2\right)\right]-2 \beta  H_0^2 (q_0+1)}{2 H_0^2}\geq 0\,,
\end{equation}
\begin{equation}\label{b31}
\text{\textbf{NEC : }}-\beta +\frac{1}{2} \left[\alpha +\beta  \log \left(6 H_0^2\right)\right]-\frac{3 H_0^2 \left[\alpha -2 \beta +\beta  \log \left(6 H_0^2\right)\right]-2 \beta  H_0^2 (q_0+1)}{6 H_0^2}\geq0\,,
\end{equation}
\begin{multline}\label{b32}
\text{\textbf{WEC : }}-\beta+\frac{1}{2} \left[\alpha +\beta  \log \left(6 H_0^2\right)\right]\geq0,\\ \text{and }
-\beta +\frac{1}{2} \left[\alpha +\beta  \log \left(6 H_0^2\right)\right]-\frac{3 H_0^2 \left[\alpha -2 \beta +\beta  \log \left(6 H_0^2\right)\right]-2 \beta  H_0^2 (q_0+1)}{6 H_0^2}\geq0\,,
\end{multline}
\begin{multline}\label{b33}
\text{\textbf{DEC : }}-\beta+\frac{1}{2} \left[\alpha +\beta  \log \left(6 H_0^2\right)\right]\geq0,\\  \text{and }
-\beta +\frac{1}{2} \left[\alpha +\beta  \log \left(6 H_0^2\right)\right]\pm\frac{3 H_0^2 \left[\alpha -2 \beta +\beta  \log \left(6 H_0^2\right)\right]-2 \beta  H_0^2 (q_0+1)}{6 H_0^2}\geq0\,.
\end{multline}

\begin{figure}[ ]
\begin{center}
$%
\begin{array}{c@{\hspace{.1in}}c}
\includegraphics[width=3.0 in, height=2.5 in]{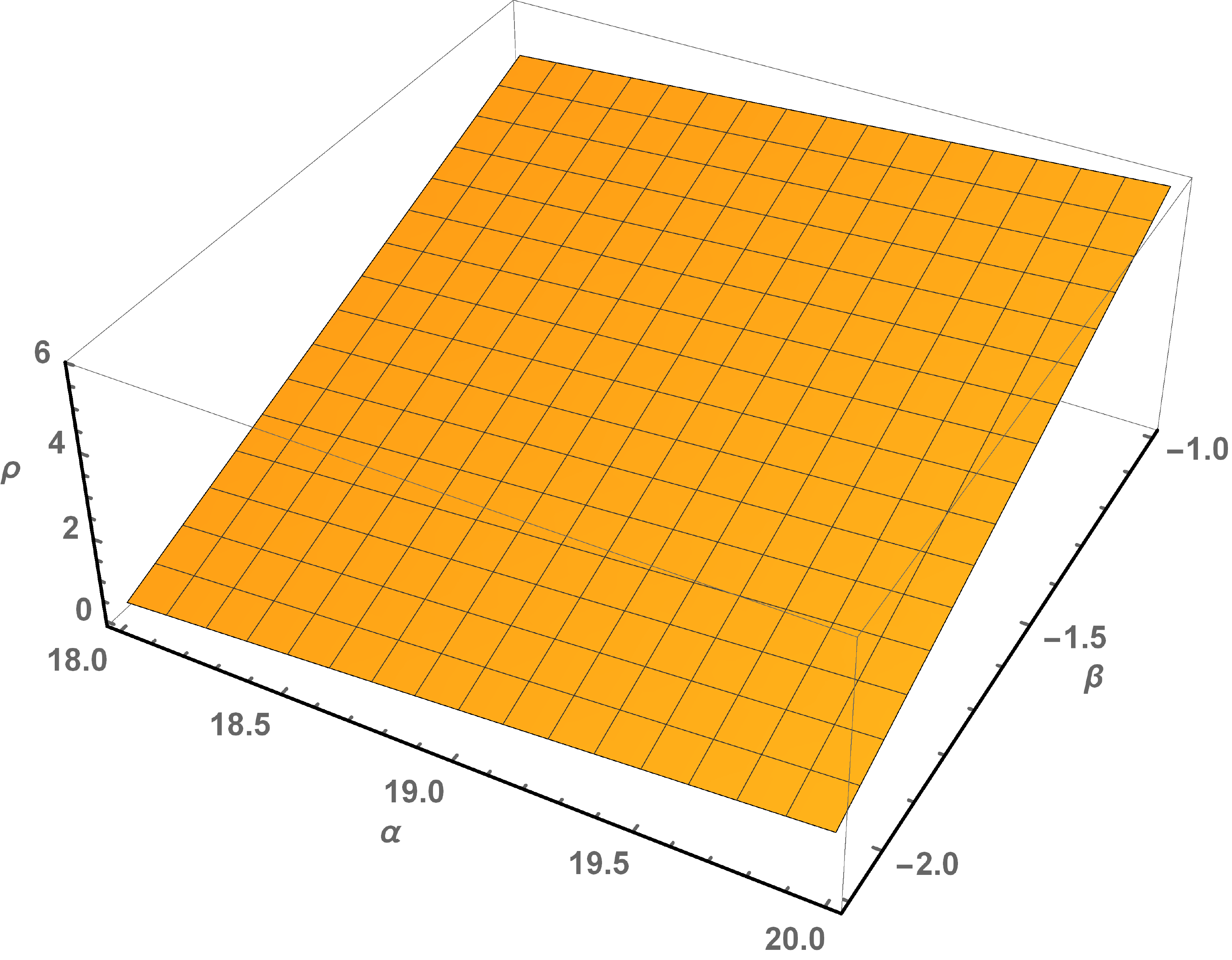} & %
\includegraphics[width=3.0 in, height=2.5 in]{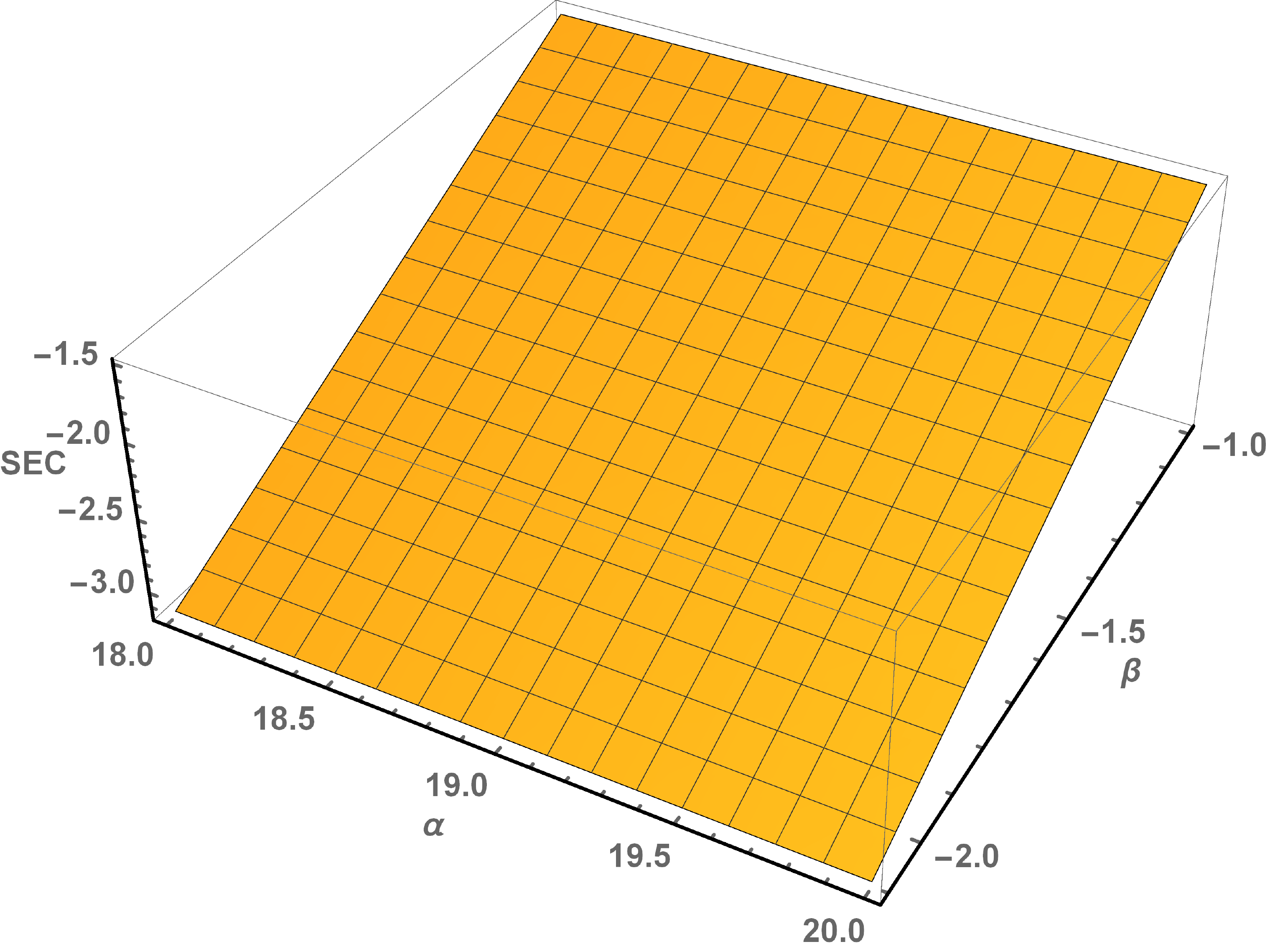}\\
\includegraphics[width=3.0 in, height=2.5 in]{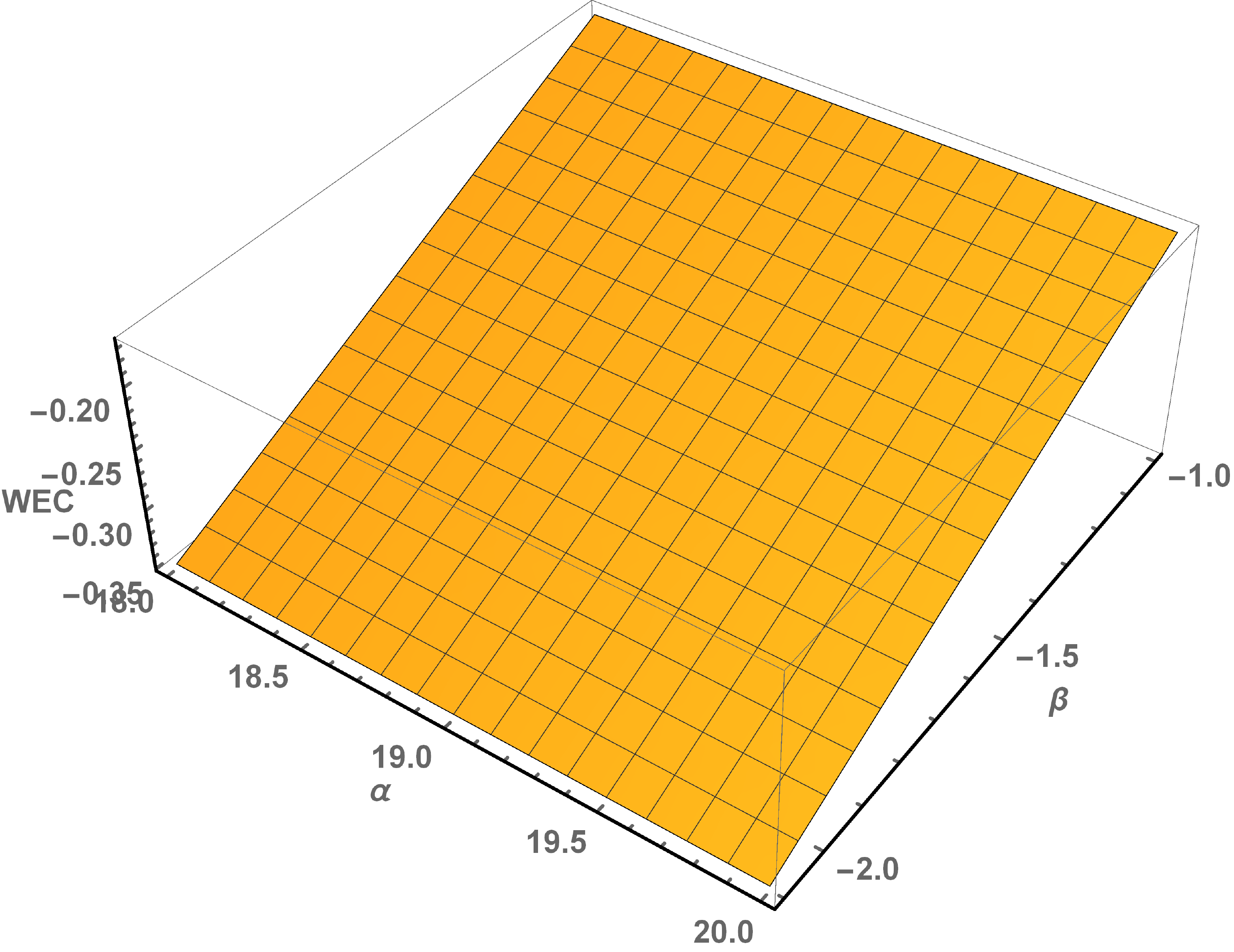} & %
\includegraphics[width=3.0 in, height=2.5 in]{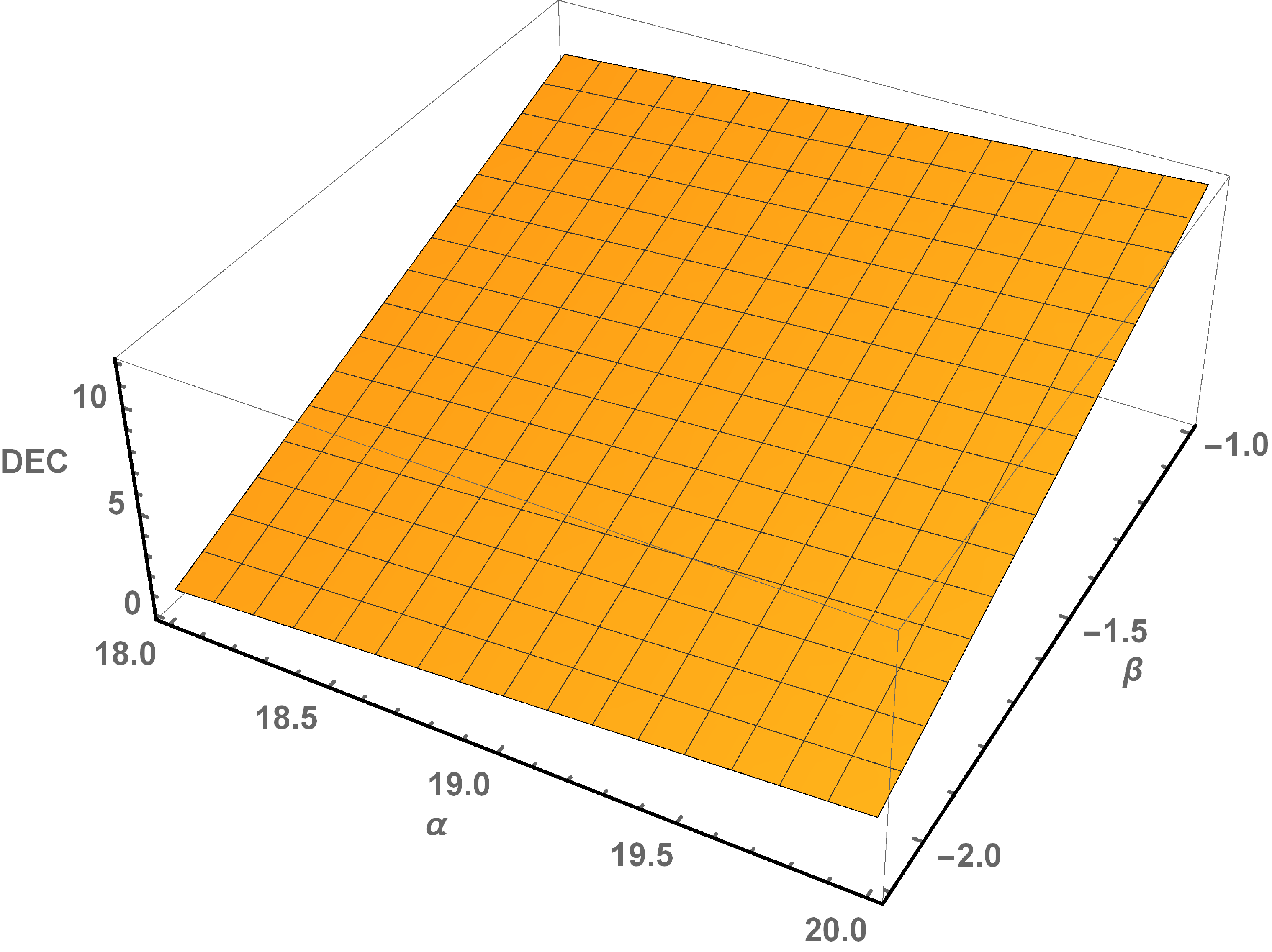}\\
\end{array}%
$%
%
%
%

\caption{Energy conditions for $f(Q)=\alpha+\beta\log Q$ derived with the present values for $H_0$, and $q_0$  parameters.}
\label{f2b}

\end{center}
\end{figure}

The ECs showed in Eq. \eqref{b30}-\eqref{b33} unveil their direct dependence on free parameters $\alpha$ and $\beta$.  The previous equations also established that we cannot choose arbitrary values for these free parameters, as observed in our previous model. Through Eq. \eqref{b30}, \eqref{b31}, \eqref{b32}, and \eqref{b33}, the numerical analysis found that condition SEC is violated and condition WEC is partially satisfied ($\rho>0$) if $\alpha\geq -9\beta, (\beta\leq -1)$, besides NEC, and DEC are still obeyed. This violation of WEC with positive density notably makes this $f(Q)$  theory behaves like scalar-tensor gravity models \cite{Whinnett/2004}, and such a violation can be interpreted as natural contributions from quantum effects to classical gravity \cite{Calcagni_book}. The features of these conditions can be appreciated in Fig. \ref{f2b}, where the graphics were depicted considering a specific range of values for parameters $\alpha$, and $\beta$.

\section{Deviation from the Standard $\Lambda$CDM Model}\label{Vb}

So far, the $\Lambda$CDM is the most well-succeeded model used to describe the actual observations of the universe. Such a model has been broadly tested by several different surveys along the past few years, such as WMAP, Planck, and the Dark Energy Survey (DES). As pointed by Lazkoz et al. \cite{Lazkoz/2019}, the $f(Q)$ model can mimics the $\Lambda$CDM one by taking $f(Q)=-Q$. Therefore, considering this specific mapping for $f(Q)$, one can find the following energy conditions 
\begin{itemize}
\item \textbf{SEC : }$6H^2 (q-1) \geq 0$,
\item \textbf{NEC : }$2H^2(1+q)\geq 0$,
\item \textbf{WEC : }$3H^2\geq 0$ and  $2H^2(1+q)\geq 0$,
\item \textbf{DEC : }$3H^2\geq 0$ and  $2H^2(1+q)\geq 0$ or, $-2H^2(-2+q)\geq 0$.
\end{itemize}

By taking the actual values of $H_0$ and $q_0$ in the above conditions, one can prove that WEC, NEC, DEC are satisfied, however, SEC condition is violated. This is the expected behavior for a standard accelerated phase for the universe. Moreover, one can realize an analogous description in respect to energy conditions between the first constructed model for $f(Q)$ gravity, and the $\Lambda$CDM.

\begin{figure}[H]
\begin{center}
\includegraphics[width=7.5 cm]{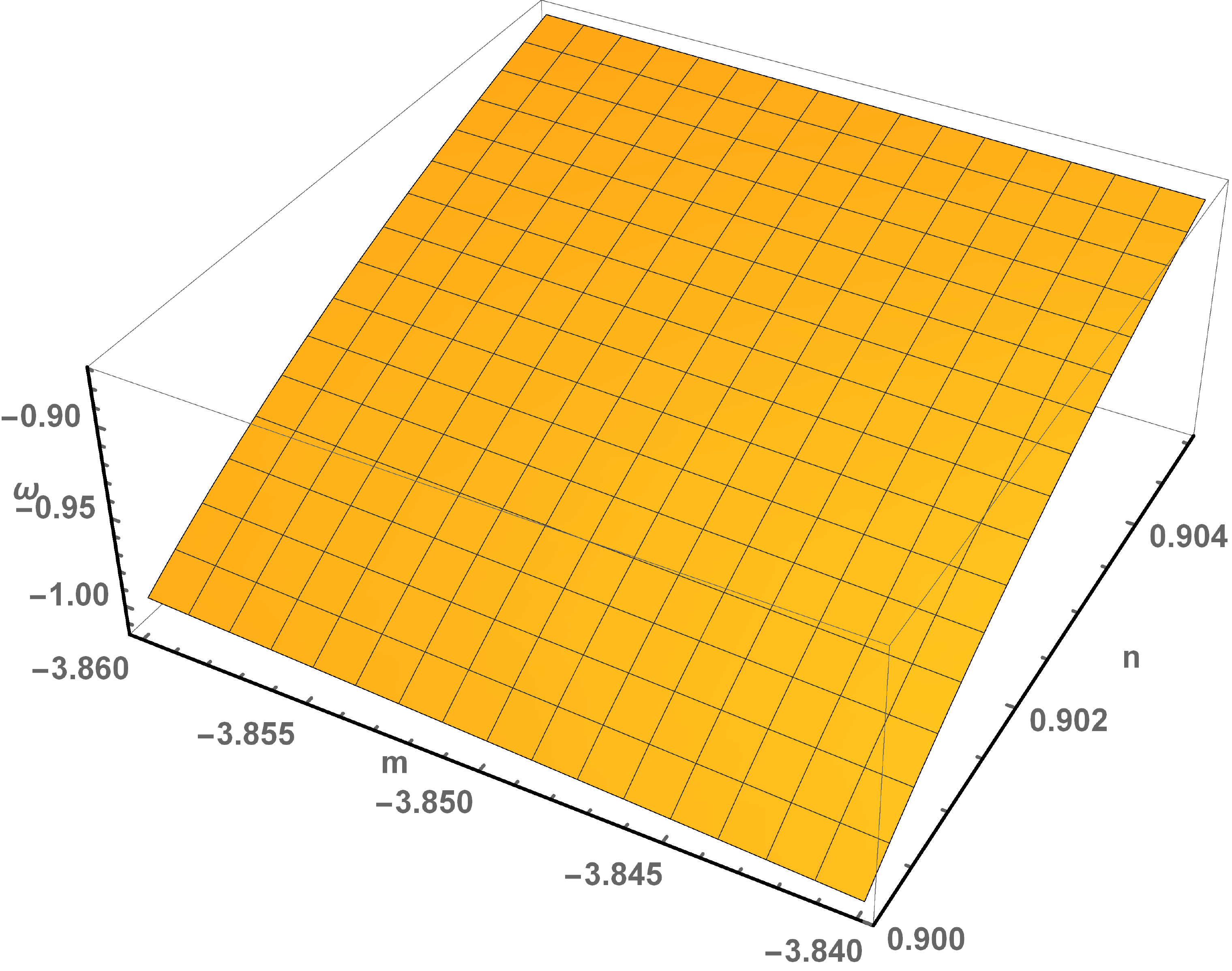}
\caption{EoS parameter for $f(Q)=Q+mQ^n$ derived with the present values for $H_0$, and $q_0$  parameters.}
\label{f3b}
\end{center}
\end{figure}

\begin{figure}[H]
\begin{center}
\includegraphics[width=7.5 cm]{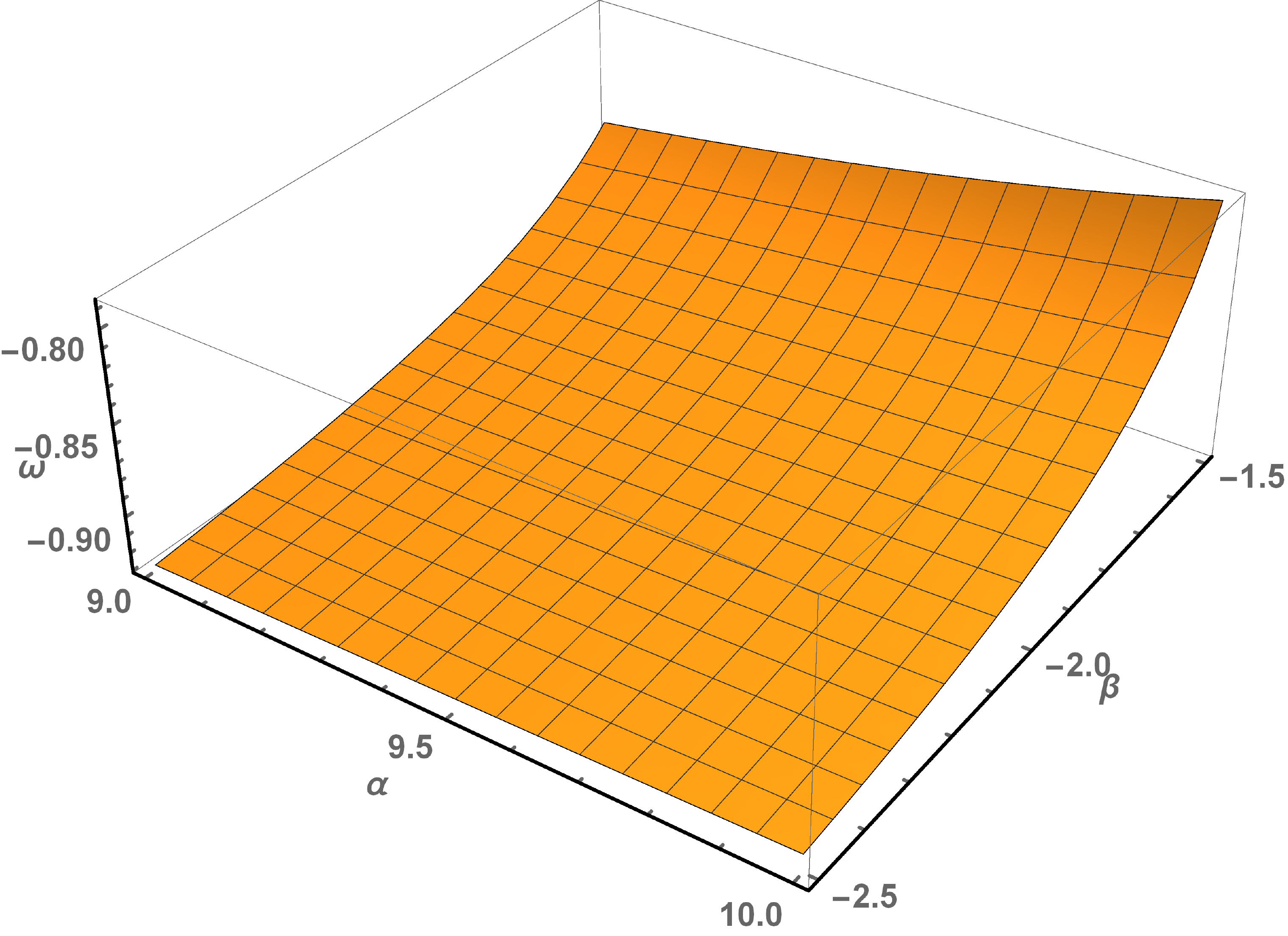}
\caption{EoS parameter for $f(Q)=\alpha+\beta\log Q$ derived with the present values for $H_0$, and $q_0$  parameters.}
\label{f4b}
\end{center}
\end{figure}

Besides, the recent observations from Planck collaboration, as well as the $\Lambda$CDM model, confirm that the equation of state parameter is $\omega\simeq -1$ \cite{Planck/2018}. This behavior corresponds to a negative pressure regime for the universe, which configures its current accelerated phase. Therefore, $\omega$ parameter presents as a suitable candidate to compare our models with $\Lambda$CDM. Our previous models yields to the following forms of $\omega$:
\begin{equation}
\label{b35}
 \omega=-1+\frac{2 (q_0+1) \left\lbrace m 6^n n (2 n-1) H_0^{2n}+6 H_0^2\right\rbrace}{3 m 6^n (2 n-1) H_0^{2n}+18 H_0^2}\,,
\end{equation}
for $f(Q)=Q+mQ^n$ gravity, and
\begin{equation}
\label{b36}
 \omega=-1+\frac{2 \beta  (q_0+1)}{3\alpha -6 \beta +3\beta  \log \left(6 H_0^2\right)}
\end{equation}
for $f(Q)=\alpha+\beta\log Q$. In Figs. \ref{f3b} and \ref{f4b}, the profiles of the equation of state parameter for both $f(Q)$ models here introduced have shown. The graphics were depicted considering the energy conditions constraints for free parameters $m$, $n$, $\alpha$, and $\beta$. From these figures, one can observe that the values of $\omega$ are very close to $-1$, which is in agreement with the recent observational results. Consequently, the constructed models fit the equation of state parameter as good as $\Lambda$CDM, corroborating with the violation of SEC, and confirming their viability to describe an accelerated universe.


\section{Conclusion}\label{VIb}

There are several theories of gravity beyond Einstein's GR, however, one critical role to define their self-consistencies is the energy condition. The physical motivation for a new theory of gravity is related to its compatibility with the causal and geodesic structure of space-time, which can be addressed through different sets of energy conditions. In the present study, this chapter derived the strong, the weak, the null, and the dominant energy conditions for two different $f(Q)$ gravity models. Inspired by the work of Harko et al. \cite{Harko/2018}, the first model was a polynomial function of the non-metricity $Q$ and has two free parameters $m$, and $n$. The energy conditions established $m\leq -1$, and $0.9\leq n\leq 2$ as constraints to describe an accelerated expansion of the universe. 

As a second approach, this chapter introduced a gravity model with a logarithmic dependence on the non-metricity. Such a model means that the $f(Q)$ smoothly tends to the Einstein-Hilbert model ($f(Q) \varpropto Q$), and had two free parameters named $\alpha$ and $\beta$. The graphics presented in Fig. \ref{f2b} unveil a desired accelerating universe for $18\leq \alpha \leq 20$, and $-2\leq \beta \leq -1$. Moreover, such a theory violates both SEC and WEC with positive density, exhibiting a behavior analogous to scalar-tensor field gravity models \cite{Whinnett/2004}. 

As a matter of completeness, this chapter compared the energy constraints with those from the $\Lambda$CDM model. In the $\Lambda$CDM gravity, all energy conditions are satisfied except SEC. This behavior is compatible with our first proposed model where $f(Q)=Q+mQ^n$, strengthening its potential as a promising new description for gravity.    

Moreover, the equation of state parameters, derived from the two $f(Q)$ approaches, are compatible with a current phase of negative pressure, presenting values close to $-1$. This behavior also corroborates with $\Lambda$CDM description for dark energy, as well as with current experimental observations \cite{Planck/2018}. 

These previous results allowed us to verify the viability of different families of $f(Q)$ gravity models, lighting new routes for a complete description of gravity compatible with the dark energy era. In further study, it would be interesting to constraint the Lagrangian function $f(Q)$.


\chapter{Cosmography in $f(Q)$ Gravity} 

\label{Chapter5} 

\lhead{Chapter 4. \emph{Cosmography in $f(Q)$ Gravity}} 
\vspace{10 cm}
* The work presented in this chapter is covered by the following two publications: \\
 
\textit{Cosmography in $f(Q)$ Gravity}, Physical Review D, \textbf{102} (12):124029 (2020).

\clearpage
In this chapter, we discuss the cosmography idea and its application to modern cosmology. Cosmography is an ideal tool to investigate the cosmic expansion history of the universe in a model-independent way. The equations of motion in modified theories of gravity are usually very complicated; cosmography may select practical models without imposing arbitrary choices a priori. This study uses the model-independent way to derive $f(z)$ and its derivatives up to the fourth-order in terms of measurable cosmographic parameters. Then we write the luminosity distance in terms of cosmographic functions. And perform the MCMC simulation by considering three different sets of cosmographic functions. For this purpose, the largest Pantheon data points for supernovae Ia are used, and constraints on the Hubble parameter $H_0$ and the cosmographic functions are estimated. The best fits for the functions of cosmographic sets for three statistical models are estimated.

\section{Introduction}\label{sec1}

In this chapter, we will work on symmetric teleparallel gravity in which the gravitational interaction is completely described by the nonmetricity $Q$ with torsion and curvature-free geometry. As this is a novel approach to exploring some universe insights, so far, a few works have been done in this approach. Exploring this formulation will hopefully provide some insight into the current scenario of the universe. Lazkoz et al. have analyzed the different forms of $f(Q)$ by transferring it to redshift form $f(z)$ with observational data. They proposed various polynomial forms of $f(z)$ including additional terms, which causes the deviation from $\Lambda$CDM model and checks their validity \cite{Lazkoz/2019}. In the previous chapter, we studied the energy conditions in order to check the stability of their assumed cosmological models and constraints of the model parameters with the present values of cosmological parameters in $f(Q)$ gravity \cite{Mandal/2020}. Lu et al. studied symmetric teleparallel gravity comparing with the $f(\mathcal{T})$ and $f(R)$ gravity and found some interesting results. Besides, they investigated one $f(Q)$ model and showed five critical points in the STG model \cite{jianbo/2019}. R\"unkla and Vilson \cite{mikhel/2018}, Gakis et al. \cite{viktor/2020}, studied the extension of symmetric teleparallel gravity in which they have reformulated the scalar non-metricity theories, derived the field equations, and discussed their properties. Harko et al., in their interesting work, have proposed the extension of symmetric teleparallel gravity by considering the Lagrangian of the form of non-minimal coupling between the non-metricity $Q$ and the matter Lagrangian. Besides this, they have studied several cosmological aspects by presuming power law and exponential forms of $f_1(Q)$ and $f_2(Q)$. They also found that their model shows the accelerated expansion \cite{Harko/2018}. The motivation for working in symmetric teleparallel gravity is that in this approach, the field equations are in second-order, which is easy to solve. Furthermore, the advantage is that it overcomes the problem which is generated by the higher derivative property of the scalar $R$ such as for a density of a canonical scalar field $\phi$, the non-minimal coupling between geometry and matter Lagrangian produces an additional kinetic term which is not an agreement with the stable Horndeski class \cite{olmo/2015}. This study focuses on constraint of the functions of the cosmographic set using the cosmographic idea, which provides the maximum amount of information from the luminosity distances of SNe Ia. To constraint those functions, a Bayesian statistical analysis is adopted using MCMC simulation with the latest large Pantheon dataset. 

The outline of the chapter is detailed as: Sec. \ref{sec2c} discuss the Einstein Lagrangian for the symmetric teleparallel geometry. Sec. \ref{sec4c} has discussed the cosmographic parameters with their origin and use. After this, the $f(Q)$ and its derivatives in terms of cosmographic parameters have been expressed in Sec. \ref{sec5c}. Then, we have constraints three models of function of cosmographic set in Sec. \ref{sec6c}. There, MCMC simulation is used to constraint the parameters with the latest Pantheon dataset. Finally, the conclusions are provided in Sec. \ref{sec7c}.

\section{Covariant Einstein Lagrangian}\label{sec2c}

Albert Einstein presented a simple Lagrangian for his motion equations using the Levi-Civita connection defined in Eq. \eqref{4}, in 1916 \cite{einstein/1916}, which is given by
\begin{equation}\label{c1}
L_E=g^{\mu\nu}\left(\left\lbrace{^{\alpha}}_{\beta\mu}\right\rbrace \left\lbrace{^{\beta}}_{\nu\alpha}\right\rbrace - \left\lbrace{^{\alpha}}_{\beta\alpha}\right\rbrace \left\lbrace{^{\beta}}_{\mu\nu}\right\rbrace\right)
\end{equation}
Nevertheless, the standard Lagrangian formulation was proposed by Hilbert in 1915. The Lagrangian is described by the Ricci scalar $\mathcal{R}$, which contains the metric tensor's second-order derivatives. Moreover, the Ricci scalar for this formulation can be written as
\begin{equation}\label{c2}
\mathcal{R}=L_E+L_B,
\end{equation}
where $L_B$ is the boundary term, and it is given by
\begin{equation}\label{c3}
L_B=g^{\alpha\mu}\mathcal{D}_{\alpha}\left\lbrace{^{\nu}}_{\mu\nu}\right\rbrace-g^{\mu\nu}\mathcal{D}_{\alpha}\left\lbrace{^{\alpha}}_{\mu\nu}\right\rbrace
\end{equation}
The symbol $\mathcal{D}_\alpha$ represents the covariant derivative with the Levi-Civita connection \eqref{5}. The Lagrangian defined in Eq. \eqref{c1} is not a covariant one; therefore, the higher-order derivative mechanism results in the standard one. Also, one can upgrade the Christoffel symbol to a covariant one using partial derivatives. Hence, one can write Eq. \eqref{7} with the covariant derivative $\nabla_\alpha$ as
\begin{equation}\label{c4}
L^{\alpha}_{\beta\gamma}=-\frac{1}{2}g^{\alpha\lambda}\left(\nabla_{\gamma}g_{\beta\gamma}+\nabla_{\beta}g_{\lambda\gamma}-\nabla_{\lambda}g_{\beta\gamma} \right).
\end{equation}
Now, the non-metricity, $Q$, can be written as
\begin{equation}\label{c5}
Q=-g^{\mu\nu}\left({L^{\alpha}}_{\beta\mu}{L^{\beta}}_{\nu\alpha}-{L^{\alpha}}_{\beta\alpha}{L^{\beta}}_{\mu\nu}\right).
\end{equation}
Whenever the covariant derivative reduces to the partial derivative at that time, the non-metricity $Q$ will be equivalent to the negative of the Einstein Lagrangian \eqref{c1} i.e.
\begin{equation}\label{c6}
\nabla_{\alpha}\circeq {\partial} _{\alpha}, \hspace{0.5cm} Q\circeq -L_{E},
\end{equation}
where $`\circ$' in the above expressions was called the \textit{gauge coincident}, and it is consistent in the symmetric teleparallel geometry \cite{Jimenez/2018}.\newline
In symmetric teleparallel geometry, the connection ${\Gamma^{\alpha}}_{\mu\nu}$ does not depend on the curvature and torsion. However, the connection in Eq. \eqref{5} and its curvature still show their physical roles. Remember that the Dirac Lagrangian, connected with the connection ${\Gamma^{\alpha}}_{\mu\nu}$ in the symmetric teleparallel geometry, filters out everything but the Christoffel symbols \ref{5} from ${\Gamma^{\alpha}}_{\mu\nu}=\left\lbrace{^{\alpha}}_{\mu\nu}\right\rbrace+{L^{\alpha}}_{\mu\nu}$. As a consequence, the symmetric teleparallel mechanism is a good and stable modification of GR. Since (minimally coupled) fermions are still metrically connected  \cite{tomi/2018}, and although the pure gravity field is now trivially interconnected, nothing actually changes, but only the higher-derivative boundary term $L_B$ disappears from this operation.

\section{Cosmographic Parameters}\label{sec4c}
Modern cosmology is growing by a prominent number of observations. Therefore, the reconstruction of the Hubble diagram (i.e., the redshift- distance relation) is possible for a higher redshift. The parametrization technique is a good method for studying cosmological models. But, this type of procedure is entirely dependent on the models, and we check the viability by contrasting it against the observational data and putting limits on its model parameters. So, there are some unclear doubts about its characterizing parameters for the present-day values of the age of the universe and the cosmological quantities. To overcome all these issues, one may adopt cosmography. Cosmography is the study of scale factor by expanding it through the Taylor series with respect to cosmic time. This type of expansion gives us a distance-redshift relation and is also independent of the solution of the equations of motion of the cosmological models. 


After some algebraic computation on the cosmographic parameter (Hubble parameter, deceleration parameter, jerk parameter, snap parameter, and lerk parameter), one can derive the following relations:
\begin{equation}
\label{c12}
\dot{H}=-H^2(1+q),
\end{equation}
\begin{equation}
\label{c13}
\ddot{H}=H^3(j+3q+2),
\end{equation}
\begin{equation}
\label{c14}
\dddot{H}=H^4[s-4j-3q(q+4)-6],
\end{equation}
\begin{equation}
\label{c15}
H^{(iv)}=H^5[l-5s+10(q+2)j+30(q+2)q+24],
\end{equation}
and $H^{(iv)}=\frac{d^4H}{dt^4}$.\\
The degeneracy problem is one of the most common issues of the cosmological models. This problem is cosmological models with parameters that are little bit different from those given by $\Lambda$CDM model but fits equally well the angular power spectrum of the CMB data. Cosmography is one of the best methods to deal with it. Furthermore, another advantage of cosmography is the luminosity distance can relate to the cosmographic parameters. In this concern, the direct measurement of luminosity distance can overcome the statistical error propagations, which has discussed in \cite{Cattoen/2008}. Therefore, the theoretical predictions are directly comparable to the observed data, without assuming a priori form of $H$ and $f(Q)$ \cite{will/2006}.\\
The series expansion of the scale factor $a(t)$ up to its 5th order in terms of cosmographic set is
\begin{multline}
\label{c16}
a(t)=a(t_0)\left[1+ H_0(t-t_0)-\frac{q_0}{2}H_0^2(t-t_0)^2 \right. \left.
+\frac{j_0}{3!}H_0^3(t-t_0)^3+\frac{s_0}{4!}H_0^4(t-t_0)^4 \right. \\ \left.
+\frac{l_0}{5!}H_0^5(t-t_0)^5+\mathcal{O}[(t-t_0)^6]\right].
\end{multline}
By definition, 
the luminosity distance reads
\begin{equation}
\label{c18}
d_L=\sqrt{\frac{\mathcal{L}}{4\pi\mathcal{F}}}=\frac{r_0}{a(t)},
\end{equation}
where $\mathcal{L}$ and $\mathcal{F}$ are the luminosity and flux, respectively. And,
\begin{equation}
\label{c19}
r_0=\int_t^{t_0}\frac{d\eta}{a(\eta)},
\end{equation}
its physical meaning is the distance travelled by a photon from a source at $r=r_0$ to the observer at $r=0$. Now, one can express the luminosity distance as series expansion of redshift $z$ with the cosmographic set and, also in terms of $f(z)$ and its derivatives, those are written in equation \eqref{ca2}. Moreover, one can express $f(Q)=f(Q(z))=f(z)$ in terms of cosmographic set i.e., $f(z)=f(H(z),q(z),j(z),s(z),l(z))$. To do this,
we rewrite $Q$ in terms of redshift $z$ as,
\begin{equation}
\label{c20}
Q(z)=6H(z)^2.
\end{equation}
Using definition of redshift in terms of cosmic time
\begin{equation}
\label{c21}
\frac{d\log (1+z)}{dt}=-H(z).
\end{equation}
Therefore, one is able to calculate $Q$ and its derivatives with respect to $z$  and presented them in $z=0$. We ended up with the following results
\begin{equation}
\label{c22}
Q_{0}=6 H_0,
\end{equation}
\begin{equation}
\label{c23}
Q_{z0}=12 H_0 H_{z0},
\end{equation}
\begin{equation}
\label{c24}
Q_{2z0}=12[H_{z0}^2+H_0H_{2z0}],
\end{equation}
\begin{equation}
\label{c25}
Q_{3z0}=12[3H_{z0}H_{2z0}+H_0H_{3z0}],
\end{equation}
\begin{equation}
\label{c26}
Q_{4z0}=12[3H_{2z0}^2+4H_0H_{3z0}+H_0H_{4z0}]
\end{equation}
Here, $Q_0=Q(z)|_{z=0}$, $Q_{z0}=\frac{dQ}{dz}|_{z=0}$, $Q_{2z0}=\frac{d^2Q}{dz^2}|_{z=0}$, etc. Similarly, $H_0=H(z)|_{z=0}$, $H_{z0}=\frac{dH}{dz}|_{z=0}$, $H_{2z0}=\frac{d^2H}{dz^2}|_{z=0}$, etc.

In order to express $Q$ and its derivatives in terms of cosmographic parameters, we have to evaluate the derivatives of $H(z)$ in terms of cosmographic parameters. To do so, one can use \eqref{c21} in \eqref{c12}-\eqref{c15} and get the following results
\begin{equation}
\label{c27}
\frac{H_{z0}}{H_0}=1+q_0,
\end{equation}
\begin{equation}
\label{c28}
\frac{H_{2z0}}{H_0}=j_0-q_0^2,
\end{equation}
\begin{equation}
\label{c29}
\frac{H_{3z0}}{H_0}=-3j_0-4j_0q_0+q_0^2+3q_0^3-s_0,
\end{equation}
\begin{equation}
\label{c30}
\frac{H_{4z0}}{H_0}=12j_0-4j_0^2+l_0+32j_0q_0-12q_0^2+25j_0q_0^2
-24q_0^3-15q_0^4+8s_0+7q_0s_0,
\end{equation}
Then using above equations, we are able to express $Q$ and its derivatives in terms of cosmographic set.
\section{$f(z)$ Derivatives vs Cosmography}\label{sec5c}
As discussed above, the study of cosmological models by presuming an arbitrary form of $f(Q)$ and then solving the modified Friedmann equations creates doubt on its model parameters. So, in this section we try to express the derivatives of $f(z)$ in terms of the present values of the cosmographic parameters $(q_0,j_0,s_0,l_0)$. Doing this, gives us a hint about the functional form of $f(Q)$ which could be able to compit the observation.\\

\begin{figure*}[htbp]
	\centering
	\includegraphics[scale=0.6]{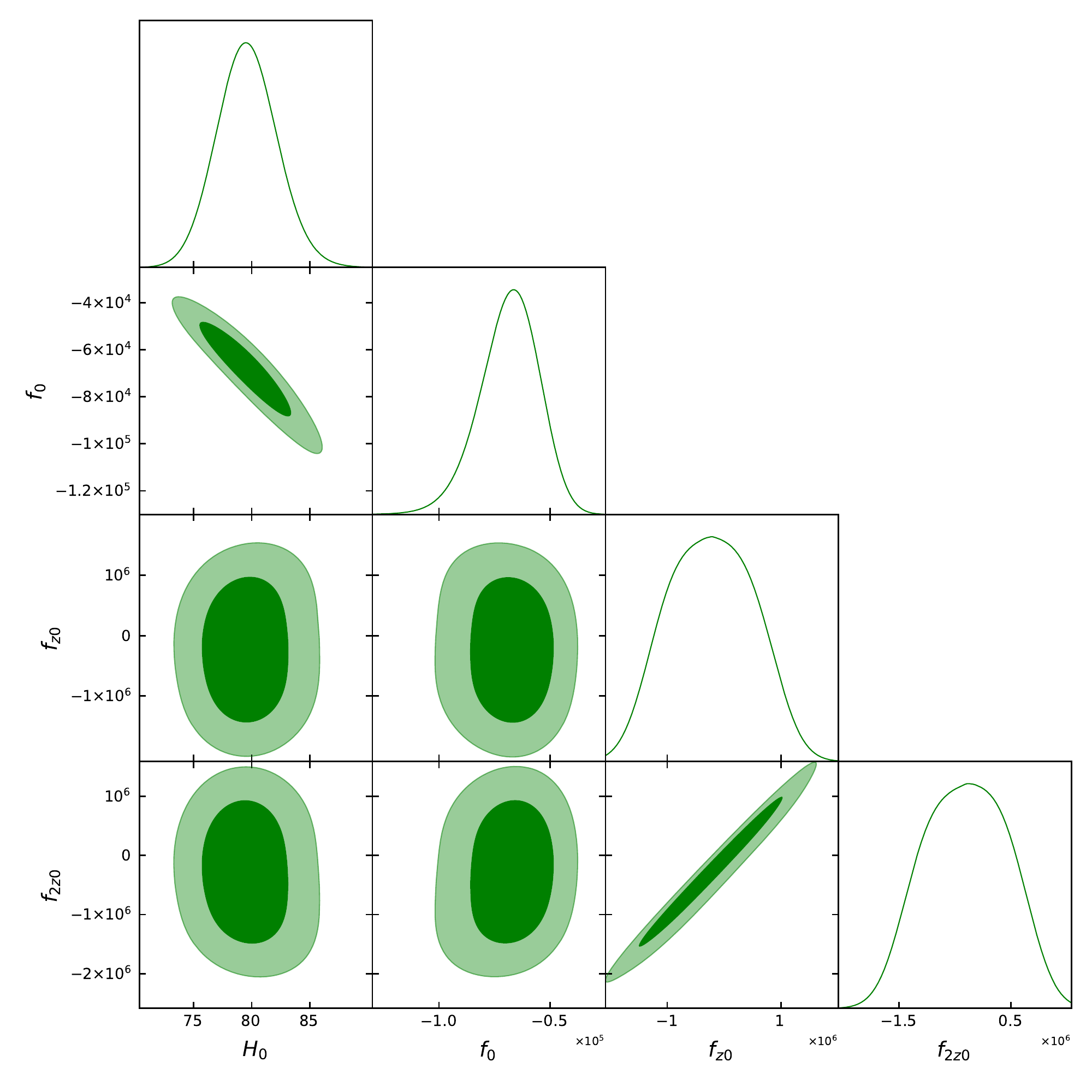}
	\caption{The marginalized constraints on the cosmographic parameters of M1 are shown by using the Pantheon SNe Ia sample. }
	\label{f1c}
\end{figure*}

\begin{figure*}[htbp]
	\centering
	\includegraphics[scale=0.6]{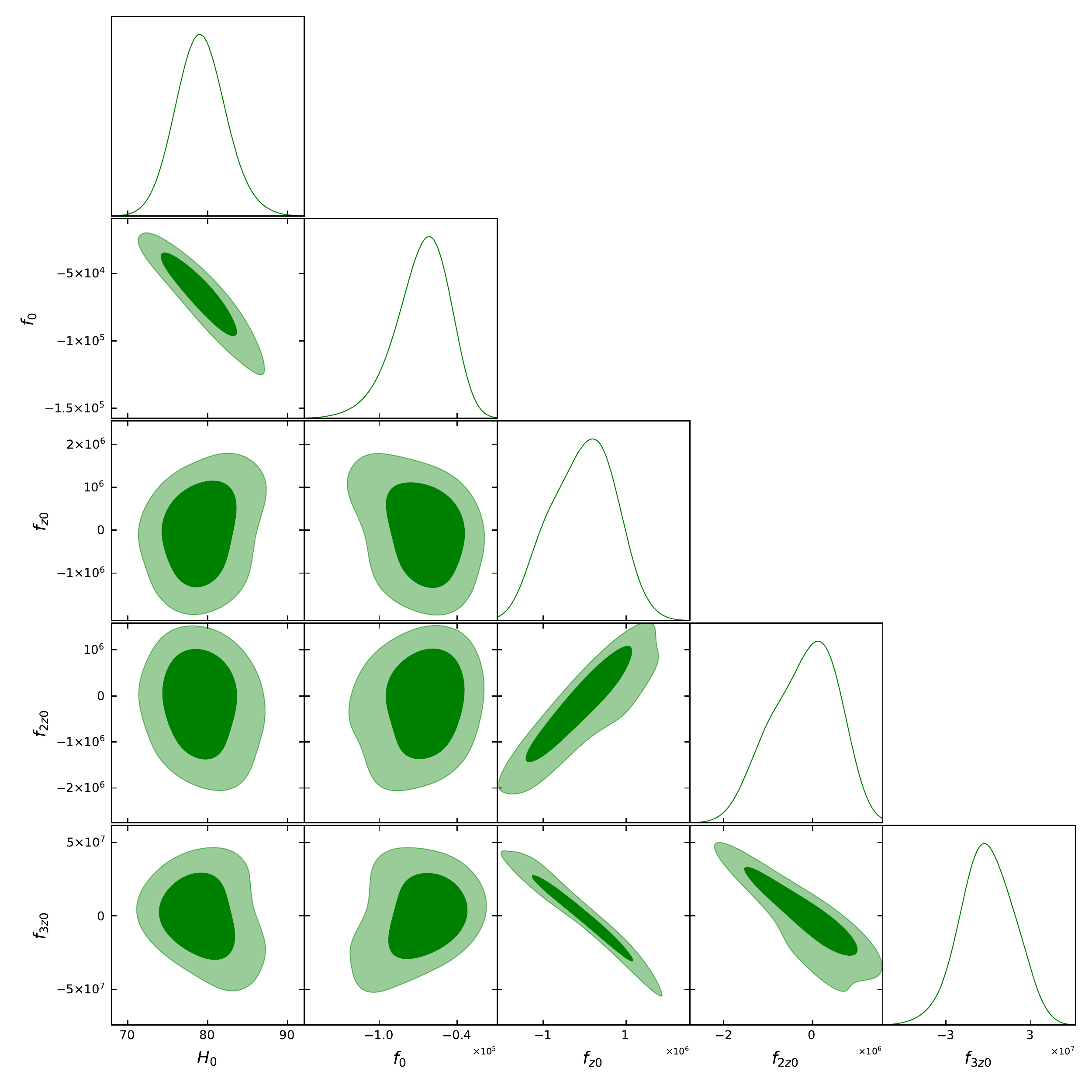}
	\caption{The marginalized constraints on the cosmographic parameters of M2 are shown by using the Pantheon SNe Ia sample. }
	\label{f2c}
\end{figure*}

\begin{figure*}[htbp]
	\centering
	\includegraphics[scale=0.46]{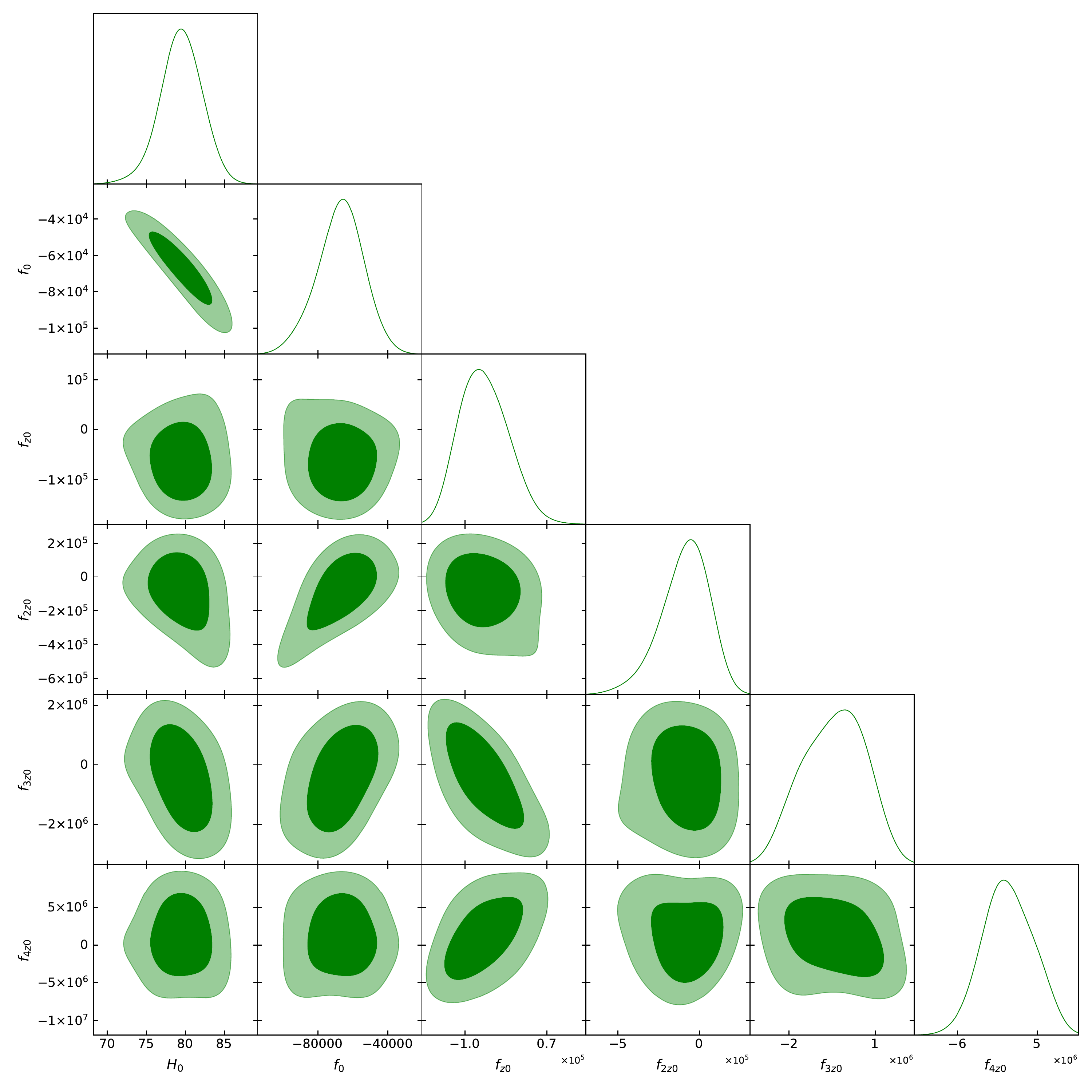}
	\caption{The marginalized constraints on the cosmographic parameters of M3 are shown by using the Pantheon SNe Ia sample. }
	\label{f3c}
\end{figure*}

The modified motion Eq. \eqref{b2} and \eqref{b3} can be rewritten as
\begin{equation}
\label{c31}
H^2=\frac{1}{12 f'(Q)}[-Q\Omega_m+f(Q)],
\end{equation}
\begin{equation}
\label{c32}
\dot{H}=\frac{1}{4f'(Q)}[Q\Omega_m-4H\dot{Q}f''(Q)],
\end{equation}
where $\Omega_m$ represents the dimensionless matter density parameter.\\
The $f(z)$ derivatives can be written as the functional dependence

\begin{align*}
f_z=f'(Q)Q_z,
\end{align*}
\begin{align*}
f_{2z}=f''(Q)Q_z^2+f'(Q)Q_{2z},
\end{align*}
\begin{align}
\label{c33}
f_{3z}=f'''(Q)Q_z^3+3f''(Q)Q_zQ_{2z}+f'(Q)Q_{3z}.
\end{align}
and so on. Furthermore, following \cite{Lazkoz/2019,Mandal/2020}, we know that $f(Q)=-Q$ mimic $\Lambda$CDM. Now, one can compare the obtained results with the $\Lambda$CDM by fixing the bounds on $\Lambda$CDM
\begin{equation}
\label{c34}
\Omega_{m0}=\frac{2}{3}(1+q_0),\hspace{0.5cm} f'(Q_0)=-1
\end{equation}
Using \eqref{c34} in \eqref{c31} we get
\begin{equation}
\label{c35}
\frac{f_0}{6H_0^2}=\Omega_{m0}-2,
\end{equation}
\begin{equation}
\label{c36}
\frac{f_{z0}}{6H_0^2}=-\frac{Q_{z0}}{6H_0^2},
\end{equation}
\begin{equation}
\label{c37}
\frac{f_{2z0}}{6H_0^2}=-\frac{Q_{2z0}}{6H_0^2},
\end{equation}
and so on. Now, one can write the $f(z)$ and its derivatives as
\begin{equation}
\label{c38}
\frac{f_0}{4H_0^2}=-2+q_0,
\end{equation}
\begin{equation}
\label{c39}
\frac{f_{z0}}{12H_0^2}=-1-q_0,
\end{equation}
\begin{equation}
\label{c40}
\frac{f_{2z0}}{12H_0^2}=-1-2q_0-j_0,
\end{equation}
\begin{equation}
\label{c41}
\frac{f_{3z0}}{12H_0^2}=3q_0+q_0j_0-q_0^2+s_0,
\end{equation}
\begin{multline}
\label{c42}
\frac{f_{4z0}}{12H_0^2}=-j_0^2-12q_0^4+19j_0q_0^2+16j_0q_0
-8q_0^2-12q_0^3+4s_0+l_0+7q_0s_0,
\end{multline}
Now, the aim is to put constraints on the values of $f_0$, $f_{z0}$, $f_{2z0}$, $f_{3z0}$, and $f_{4z0}$. In order to do this, we have expressed the luminosity distance in terms of cosmographic parameters as well as $f(z)$ and its derivatives for the present time. Now, the expression of $d_L$ reads
\begin{multline}
\label{ca1}
d_L(z)=\frac{1}{H_0}\biggl[z+\frac{1}{2}(1-q_0)z^2-\frac{1}{6}(1-q_0+j_0-3q_0^2)z^3+\frac{1}{24}(2+5j_0-2q_0\\
+10j_0q_0-15q_0^2-15q_0^3+s_0)z^4+\biggl(-\frac{1}{20}-\frac{9j_0}{40}+\frac{j_0^2}{12}-\frac{l_0}{120}+\frac{q_0}{20}-\frac{11 j_0q_0}{12}\\
+\frac{27q_0^2}{40}-\frac{7j_0q_0^2}{8}+\frac{11q_0^3}{8}+\frac{7q_0^4}{8}-\frac{11s_0}{120}-\frac{q_0s_0}{8}\biggr)z^5+\mathcal{O}(z^6)\biggr]
\end{multline}
The above equation can write in terms of $f(z)$ and its derivatives as
\begin{multline}
\label{ca2}
d_L(z)=\frac{1}{H_0}\biggl[z-\frac{4H_0^2+f_0}{8H_0^2}z^2+\frac{1}{288 H_0^4}(9 f_0^2+168 f_0 H_0^2-4 f_{z0} H_0^2+4 f_{2z0} H_0^2+720 H_0^4)z^3\\ 
+\frac{1}{4608H_0^6}(-45 f_0^3+16H_0^4 (-846 f_0+23 f_{z0}-23 f_{2z0}+f_{3z0})-12H_0^2 f_0 (113 f_0-3 f_{z0}+3 f_{2z0})\\
-44160H_0^6)z^4 
+\frac{1}{92160 H_0^8}[279 f_0^4+16 H_0^4 (11268 f_0^2-417 f_0 f_{z0}+417 f_0 f_{2z0}-8 f_0 f_{3z0}+3 f_{z0}^2\\
-6 f_{z0} f_{2z0}+3 f_{2z0}^2)
+12 f_0^2 H_0^2 (967 f_0-26 f_{z0}+26 f_{2z0})+64 H_0^6 (19350 f_0-549 f_{z0}+549 f_{2z0}\\
-23 f_{3z0}-f_{4z0})+3162624 H_0^8]z^5
\biggr]
\end{multline}

\section{Observational Constraints}\label{sec6c}
This section deal with the luminosity distance $d_L$ to constraint $H_0, f_0, f_{z0}, f_{2z0},f_{3z0}$, and $f_{4z0}$,  with the observational data. For this, we have presented three statistical models with different maximum orders of parameters; this method, commonly accepted in the literature, corresponds to a hierarchy of parameters. Now, we are going to constraint the following models:
\begin{equation}
\mathbf{M1}:=\{H_0, \quad f_0, \quad f_{z0}, \quad f_{2z0} \},   \label{c43}
\end{equation}
\begin{equation}
\mathbf{M2}:=\{H_0, \quad f_0,  \quad f_{z0}, \quad f_{2z0}, \quad f_{3z0} \},   \label{c44}
\end{equation}
\begin{equation}
\mathbf{M3}:=\{H_0, \quad f_0, \quad f_{z0}, \quad f_{2z0}, \quad f_{3z0}, \quad f_{4z0} \}. \label{c45}
\end{equation}
The motivation for doing such a hierarchical analysis of the cosmographic functions is that the extension of the sampled distributions by adding more parameters is optimistically expected. The resulting numerical effects on the measured quantities lead to a large distribution error due to the higher orders of Taylor's expansion. This study is concerned with measuring these effects and resolving the limitations of the cosmographic functions. The numerical study is done by the MCMC analysis using SNe Ia data. As we know, SNe Ia is a powerful distance indicator for studying the background evolution of the universe. In this study, to implement the cosmological constraints, we use the largest ``Pantheon'' SNe Ia  sample, which integrates SNe Ia data from the Pan-STARRS1, SNLS, SDSS, low-z, and HST surveys and contains 1049 spectroscopically confirmed data points in the redshift range $z \in [0.01, 2.3]$ \cite{Scolnic/2018}. 

%
%

To perform the standard Bayesian analysis, a Markov Chain Monte Carlo method is employed to obtain the posterior distributions of cosmographic parameters. The best fits of the parameters are maximized by using the probability function
\begin{equation}\label{c46}
\mathcal{L}\propto \exp(-\chi^2/2),
\end{equation}
where $\chi^2$ is the \textit{pseudo chi-squared function} \cite{hobson/2009}.
The marginalized constraining results are displayed in Figs.\ref{f1c}-\ref{f3c} and Table\ref{t1c}. In Table \ref{t1c}, the best fits are shown by the maximum likelihood function of the samples; the cited errors represent the 68\% confidence limits. From Figs. \ref{f1c}-\ref{f3c}, one can see marginalised posteriors lose Gaussianity when we apply additional parameters to model M1. This analysis conclude that considering model M3 over model M2 has the benefit of having more details on the cosmographic $f(Q)$ parameters without expanding the dispersion; however, model M3 is less suitable for post-statistical treatment.

\begin{table*}[!t]
	\renewcommand\arraystretch{1.5}
	\caption{The marginalized constraining results on three cosmographic $f(Q)$ models M1, M2 and M3 are shown by using the Pantheon SNe Ia sample. We quote $1-\,\sigma$ (68$\%$) errors for all the parameters here.
	}\begin{center}
	\begin{tabular} { l |c| c |c }
		\hline
		\hline

		Model              & M1      &M2      &M3        \\
		\hline
		$H_0$ & $79.5\pm2.5$     & $79.2\pm3.1      $    &$79.5\pm2.6$      \\
		\hline
		$f_0$ & $-0.68^{+0.14}_{-0.12}\times 10^5 $     & $-0.66^{+0.23}_{-0.17}\times 10^5 $    &$-0.67^{+0.14}_{-0.12}\times 10^5 $      \\
		\hline
		$f_{z0}$ & $(-0.22\pm0.73)\times 10^5 $     & $-0.04^{+0.86}_{-0.74}\times 10^6 $    &$-0.61^{+0.46}_{-0.56}\times 10^5 $      \\
		\hline
		$f_{2z0}$ & $(-0.30\pm0.74)\times 10^6 $     & $-0.17^{+0.87}_{-0.66}\times 10^6 $    &$-0.87^{+1.7}_{-1.2}\times 10^5 $      \\
		\hline
		$f_{3z0}$ & ---    & $(0.1\pm 1.9)\times10^7        $    &$-4^{+12}_{-11}\times 10^5  $      \\
		\hline
		$f_{4z0}$ & ---     & ---  &$(1\pm3)\times 10^6 $       \\

	    \hline
		\hline
	\end{tabular}
	\end{center}
	\label{t1c}
\end{table*}

\section{Discussions}\label{sec7c}

Cosmography provides a legitimate instrument for investigating cosmic expansion without a cosmological model. The constraints on the cosmographic parameters $(q_0, j_0, s_0, l_0)$ have been obtained by fitting to SNe Ia data. Also, these data completely support the cosmological principle. In certainty, any cosmological model should predict the cosmographic parameter values which align with these values. Such a supposition makes it clear why the study of cosmography allows, as an interpretation of the cosmic speed observed, to verify its viability.

This chapter dealt with the reconstruction of the correct form of $f(Q)$ function in $f(Q)$ gravity using the cosmographic parameters. It uses the cosmographic parameters as a tool to derive $f(z)$ and its derivatives (called functions of the cosmographic set as fCS) in terms of cosmographic parameters. Also, this study estimated the bounds on fCS using statistical analysis with the 1059 points of the Pantheon SNe Ia sample, which includes Pan-STARRS1, SNLS, SDDS, low-z, and HST surveys data points.

Once, we did the expressions for fCS in terms of cosmographic parameters. Then, one can rewrite the expression of Luminosity distance in terms of fCS. Now, one can easily constrain $f(z)$ and its derivatives using numerical analysis. This chapter adopted the MCMC statistical analysis and found the numerical bounds on fCS with the largest Pantheon SNe Ia sample, which are presented in Table \ref{t1c}. This study is able to constraint the fCS for the present cosmographic values.

\chapter{Energy Conditions in Non-minimally Coupled $f(R,T)$ Gravity} 

\label{Chapter6} 

\lhead{Chapter 5. \emph{Energy Conditions in Non-minimally Coupled $f(R,T)$ Gravity}} 

\vspace{10 cm}
* The work, in this chapter, is covered by the following publications: \\
 
\textit{Energy Conditions in Non-minimally Coupled $f(R,T)$ Gravity}, Astronomische Nachrichten, \textbf{342}, 89-95 (2021).

\clearpage
This chapter aims to discuss the simplest non-minimal matter geometry coupling with a perfect fluid distribution in the framework of $f(R,T)$ gravity. The model parameter is constrained by energy conditions(ECs) and a single parameter proposed equation of state (EoS), resulting in the compatibility of the $f(R,T)$ models with the accelerated expansion of the universe. It is seen that the EoS parameter illustrates the quintessence phase in a dominated accelerated phase. Also, the present values of the cosmological constant and the acceleration of the universe are used to check the viability of our linear $f(R,T)$ model of gravity. It is observed that the positive behavior of DEC and WEC  indicates the validation of the model. In contrast, SEC is violating, which causes the accelerated expansion of the universe.

\section{Introduction}\label{sec1}
%

The ECs are the necessary conditions for a good understanding of the singularity theorem, such as black-hole thermodynamics. The well known Raychaudhuri equations \cite{Ray_1955} play a role in describing the attractive nature of gravity and positive energy density. The four basic conditions are the NEC, WEC, DEC, and SEC. All the conditions are derived from the Raychaudhuri equation, which helps us analyze the entire spacetime structure without precise solutions to Einstein's equations playing a vital role in understanding cosmological gravitational interactions. The NEC discusses the second law of black-hole thermodynamics, although its violation corresponds to the universe big rip singularity \cite{Carroll_2004}. On the other SEC is useful for studying the Hawking-Penrose theorem of singularity \cite{Hawking_1973}. SEC is also good at describing the repulsive/attractive nature of gravity under modified theories of gravity. The violation of SEC implies the observed accelerated expansion of the universe. Many works on ECs under modified theories are presented in the literature. Bamba et al.\cite{Bamba_2017} studied the ECs in $f(G)$ gravity. Capozziello et al. present the role of ECs in $f(R)$ cosmology \cite{Capo_2018}. Atazadeh et al. \cite{Atazadeh_2014} considered $f(R, G)$ gravity to study various ECs. Zubair et al. \cite{Zubair_2015} worked on ECs in $f(\mathcal{T})$ gravity with non-minimal torsion-matter coupling etc.\\

In this chapter, the non-minimally coupled $f(R, T)$ gravity i.e. $f(R, T)= R+\alpha R T$ is considered where $\alpha$ is the model parameter, to study various ECs. Also, to check the viability of $f(R, T)$ gravity theory, this study used the present values of the cosmological parameters $q_{0}$ and $H_{0}$. The equation of state parameter $\omega$ provides an acceptable candidate for comparing our models with $\Lambda$CDM. The recent results from Planck Collaboration \cite{Planck/2018} and the $\Lambda$CDM
model indicate that the equation of state parameter $\omega\simeq -1$. This behavior corresponds to the universe's negative pressure framework, which specifies the current accelerated phase. The literature includes several works with certain linear forms of $f(R, T)$ gravity. The linear form of $f(R, T)$ gravity is considered in the last second section, which demonstrates some consistency with $\Lambda$CDM and non-linear form. Using restricted values of cosmological parameters, ECs are examined, which is different from work done earlier.

This chapter is presented and organized as follows: In Section \ref{sec2d}, we briefly describe the formulation of field equations in $f(R, T)$ gravity. In Section \ref{sec3d}, we discuss the solutions of the field equations. The various ECs are studied in section \ref{sec4d}. In Section \ref{sec5d}, we present the comparison with the $\Lambda$CDM model along with the equation of state parameter. Also, the compatibility of model in case of linear form is presented in section \ref{sec6d}. Further discussions and conclusions are presented in section \ref{sec7d}.

\section{Basic Equations of $f(R,T)$ Gravity}\label{sec2d}
%

By varying the action \eqref{57} with respect to the metric $g_{\mu\nu}$ yields
\begin{align}\label{d1}
[f_1'(R)+f_2'(R)f_3(T)]R_{\mu\nu}-\frac{1}{2}f_1(R)g_{\mu\nu}+  \\ \nonumber
(g_{\mu\nu}\Box-\nabla_\mu\nabla_\nu)[f_1'(R)+f_2'(R)f_3(T)]=[8\pi+ \\ \nonumber
f_2(R)f_3'(T)]T_{\mu\nu}+ f_2(R)\left[f_3'(T)p+\frac{1}{2}f_3(T)\right]g_{\mu\nu},
\end{align}
for which it was assumed $f(R,T)=f_1(R)+f_2(R)f_3(T)$ and primes denote derivatives with respect to the argument.

Now, in further study the functional form of $f_1(R)=f_2(R)=R$ and $f_3(T)=\alpha T$ is considered, with $\alpha$ a constant. This is the simplest non-trivial functional form of the function $f(R,T)$ which involves matter-geometry coupling within the $f(R,T)$ formalism. Moreover, it benefits from the fact that GR is retrieved when $\alpha=0$.
%
%

\section{The $f(R,T)=R+\alpha RT$ Cosmology} \label{sec3d}

For a flat Friedmann-Robertson-Walker universe with scale factor $a(t)$ and Hubble parameter $H=\dot{a}/a$, the expressions for energy density and pressure from Eq. \eqref{2} reads the following

\begin{equation}\label{d2}
\rho=\frac{H^2\left[8\pi-27\alpha\left(\dot{H}+2H^2\right)\right]+7\alpha(2\dot{H}+3H^2)\left(\dot{H}+2H^2\right)}{\frac{64\pi^2}{3}-96\pi \alpha \left(\dot{H}+2H^2\right)+18\alpha^2 \left(\dot{H}+2H^2\right)^2},
\end{equation}

\begin{equation}\label{d3}
p=-\frac{9\alpha H^2\left(\dot{H}+2H^2\right)+\left(2\dot{H}+3H^2\right)\left[\frac{8\pi}{3}-3\alpha\left(\dot{H}+2H^2\right)\right]}{\frac{64\pi^2}{3}-96\pi \alpha \left(\dot{H}+2H^2\right)+18\alpha^2 \left(\dot{H}+2H^2\right)^2}.
\end{equation}


Now, using Eq. \eqref{c12}, \eqref{c13} in \eqref{d2} and \eqref{d3}, one can get the following expressions
\begin{equation}\label{d7}
\rho=\frac{3 \alpha  H^4 (q-1) (7 q+10)+12 \pi  H^2}{27 \alpha ^2 H^4 (q-1)^2+144 \pi  \alpha  H^2 (q-1)+32 \pi ^2}
\end{equation}
\begin{equation}\label{d8}
p=\frac{H^2 \left(9 \alpha  H^2 \left(q^2-1\right)+\pi  (8 q-4)\right)}{27 \alpha ^2 H^4 (q-1)^2+144 \pi  \alpha  H^2 (q-1)+32 \pi ^2}
\end{equation}
\section{Energy Conditions}\label{sec4d}
%
%
%
To check the viability of $f(R,T)$ gravity theory, this study uses the present values of the cosmological parameters as $H_0=67.9$ and $q_0=-0.503$ \cite{Planck/2018,Capozziello/2019}. The ECs in section \ref{VIa} reads as follows
\begin{equation}\label{d9}
\text{\textbf{SEC: }} \rho+3p=\frac{3 \alpha  H_0^4 (q_0-1) (16 q_0+19)+24 \pi  H_0^2 q}{27 \alpha ^2 H_0^4 (q_0-1)^2+144 \pi  \alpha  H_0^2 (q_0-1)+32 \pi ^2}\geq 0,
\end{equation}
\begin{equation}\label{d10}
\text{\textbf{NEC: }}\rho+p=\frac{3 \alpha  H_0^4 (q_0-1) (10 q_0+13)+8 \pi  H_0^2 (q_0+1)}{27 \alpha ^2 H_0^4 (q_0-1)^2+144 \pi  \alpha  H_0^2 (q_0-1)+32 \pi ^2}\geq 0
\end{equation}
\begin{multline}\label{d11}
\text{\textbf{WEC: }}\rho= \frac{3 \alpha  H_0^4 (q_0-1) (7 q_0+10)+12 \pi  H_0^2}{27 \alpha ^2 H_0^4 (q_0-1)^2+144 \pi  \alpha  H_0^2 (q_0-1)+32 \pi ^2}\geq 0, \\
 \rho+p=\frac{3 \alpha  H_0^4 (q_0-1) (10 q_0+13)+8 \pi  H_0^2 (q_0+1)}{27 \alpha ^2 H_0^4 (q_0-1)^2+144 \pi  \alpha  H_0^2 (q_0-1)+32 \pi ^2}\geq 0
\end{multline}
\begin{multline}\label{d12}
\text{\textbf{DEC: }}\rho= \frac{3 \alpha  H_0^4 (q_0-1) (7 q_0+10)+12 \pi  H_0^2}{27 \alpha ^2 H_0^4 (q_0-1)^2+144 \pi  \alpha  H_0^2 (q_0-1)+32 \pi ^2}\geq 0,\\
 \rho+p=\frac{3 \alpha  H_0^4 (q_0-1) (10 q_0+13)+8 \pi  H_0^2 (q_0+1)}{27 \alpha ^2 H_0^4 (q_0-1)^2+144 \pi  \alpha  H_0^2 (q_0-1)+32 \pi ^2}\geq 0\\
 \text{or, }\rho-p=\frac{3 \alpha  H_0^4 (q_0-1) (4 q_0+7)-8 \pi  H_0^2 (q_0-2)}{27 \alpha ^2 H_0^4 (q_0-1)^2+144 \pi  \alpha  H_0^2 (q_0-1)+32 \pi ^2}\geq 0
\end{multline}

\begin{figure}[H]
\begin{center}
$%
\begin{array}{c@{\hspace{.1in}}c}
\includegraphics[width=3.0 in, height=2.5 in]{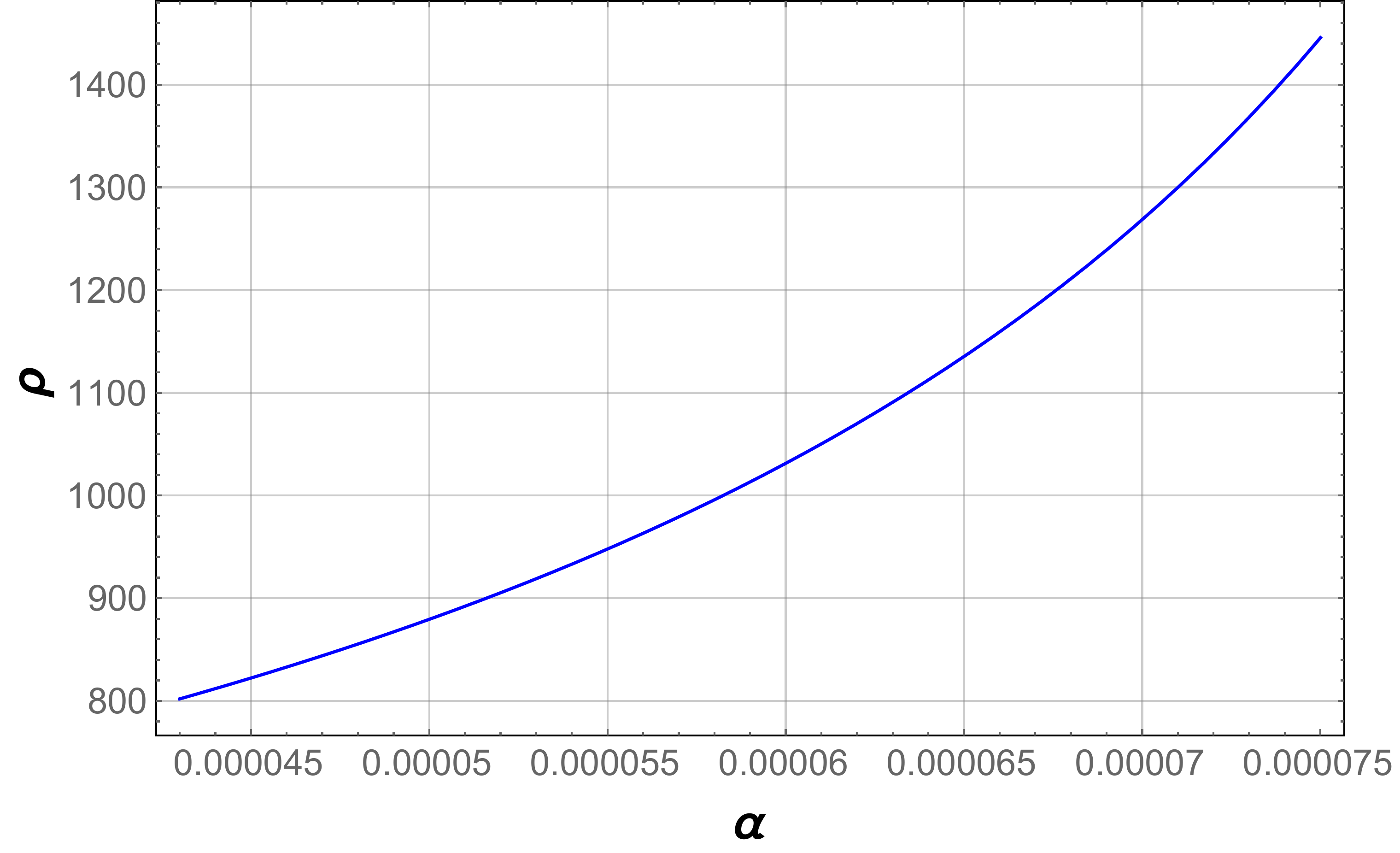} & %
\includegraphics[width=3.0 in, height=2.5 in]{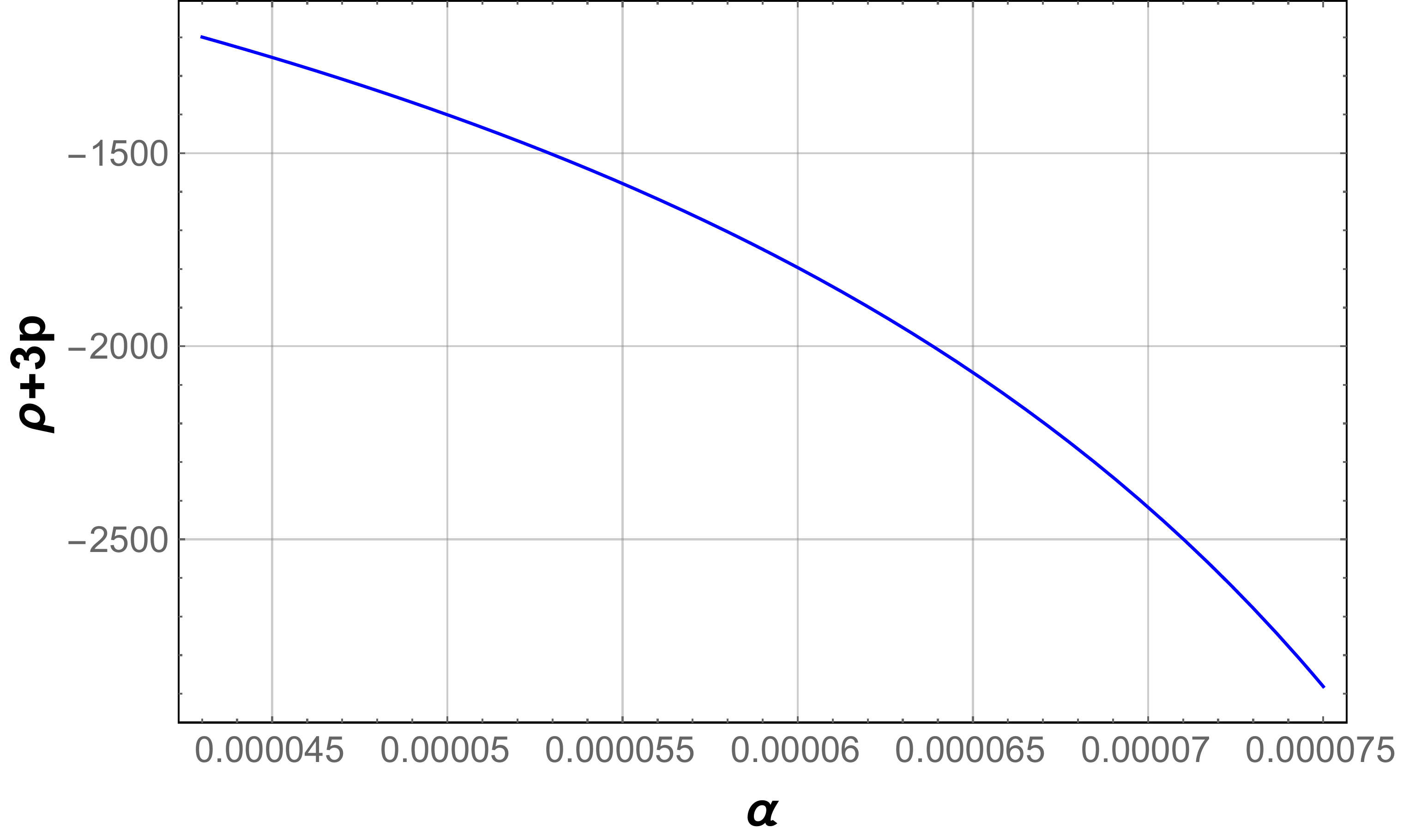}\\
\includegraphics[width=3.0 in, height=2.5 in]{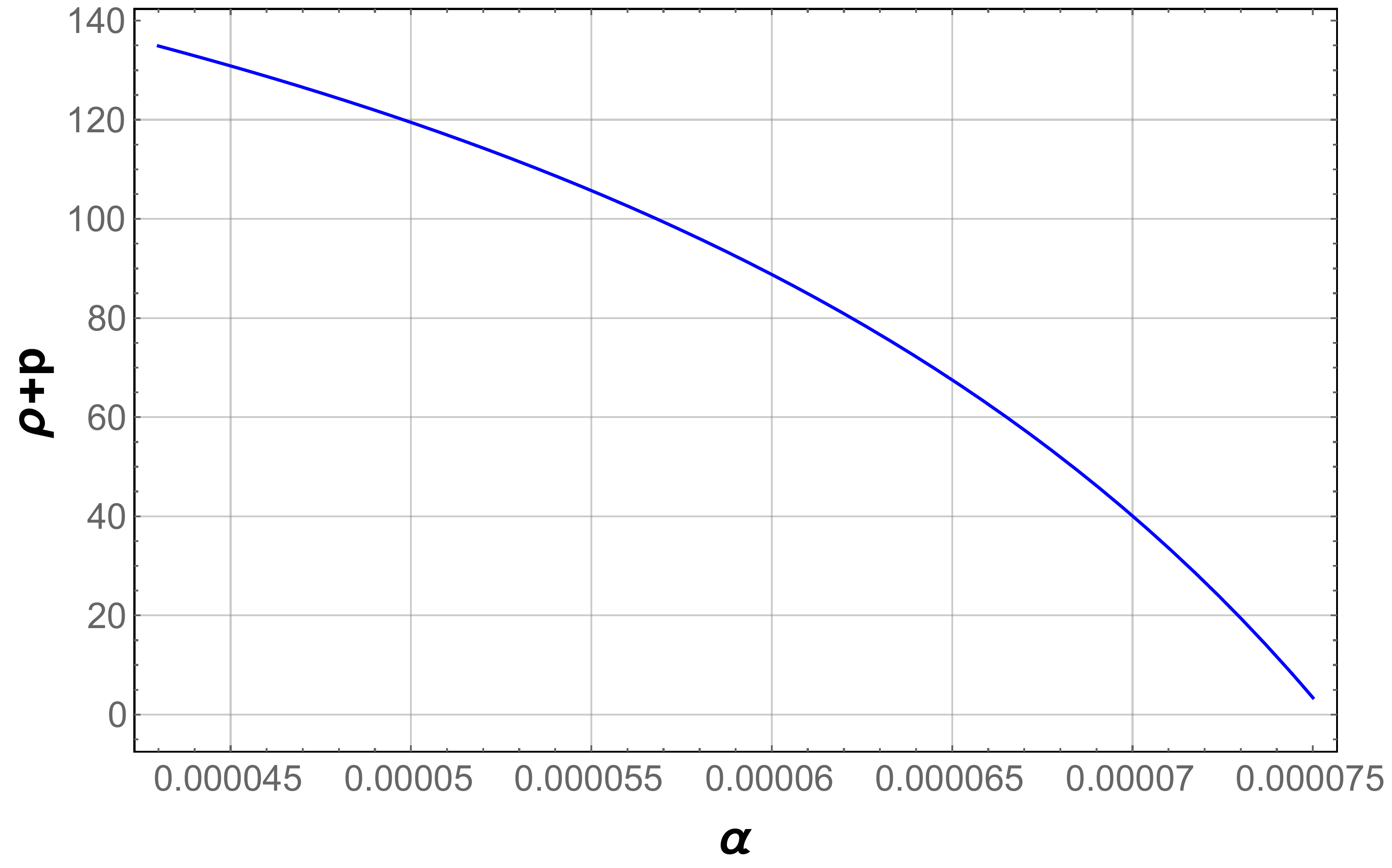} & %
\includegraphics[width=3.0 in, height=2.5 in]{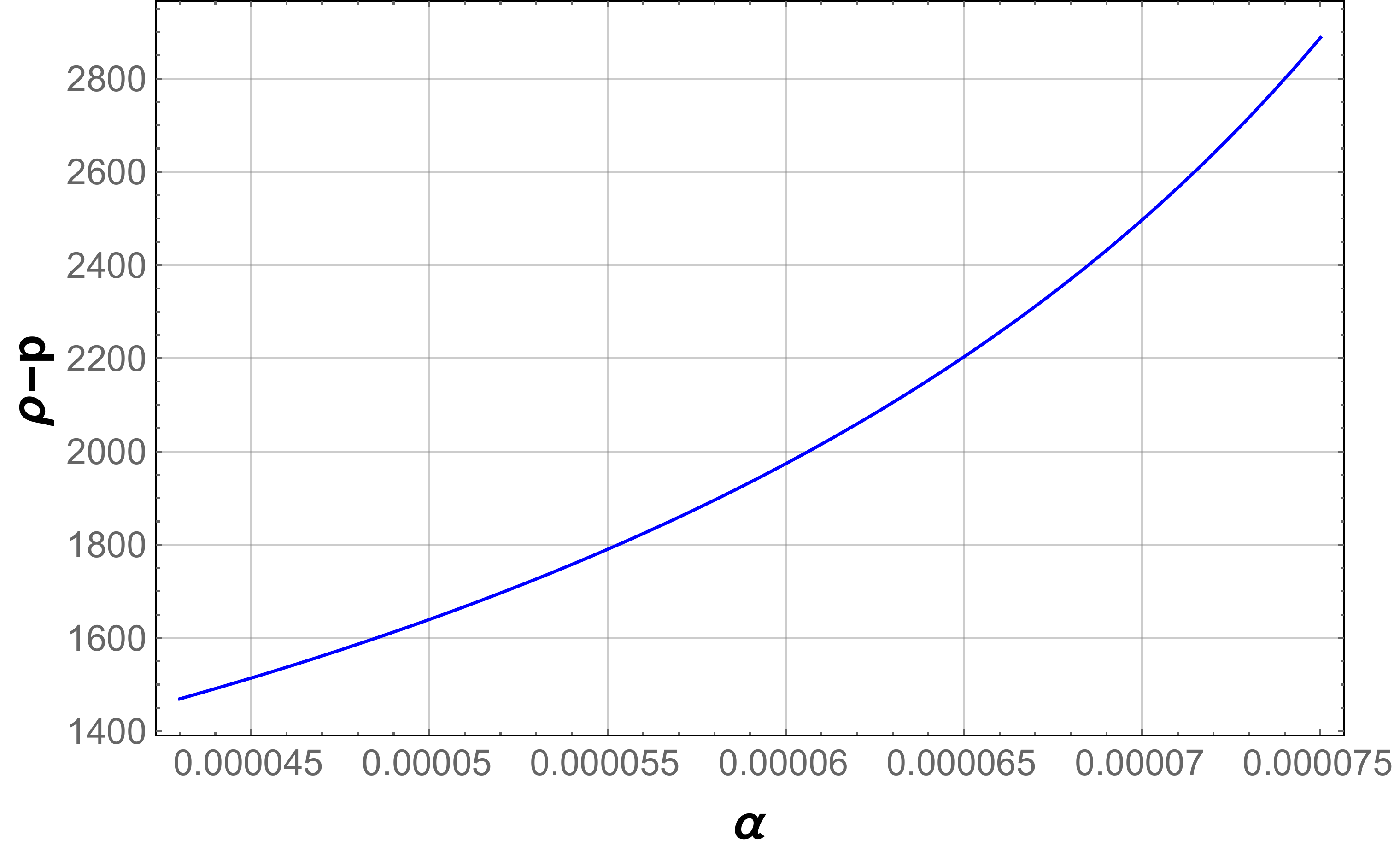}\\
\end{array}%
$%
%
%
%

\caption{ECs for $f(R,T)=R+\alpha RT$ derived with the present values of $H_0$ and $q_0$ parameters.}
\label{f1d}
\end{center}
\end{figure}

From the above expressions \eqref{d9}-\eqref{d12}, one can quickly observe that the ECs depends on the model parameters $\alpha$. However, we can not choose the value of $\alpha$ arbitrarily. If we do that may cause a violation of the current accelerated scenario of the universe. Keeping this thing in mind, we have manipulated the profiles of the ECs in Fig \ref{f1d}. As one can see, WEC, NEC, DEC are satisfied while SEC violated with the present value of the Hubble parameter ($H_0$) and deceleration parameter ($q_0$) in Fig \ref{f1d}. Also, this is an agreement with the current scenario of the universe.

\section{The Standard $\Lambda$CDM Model} \label{sec5d}
\begin{figure}[H]
\begin{center}
\includegraphics[scale=0.25]{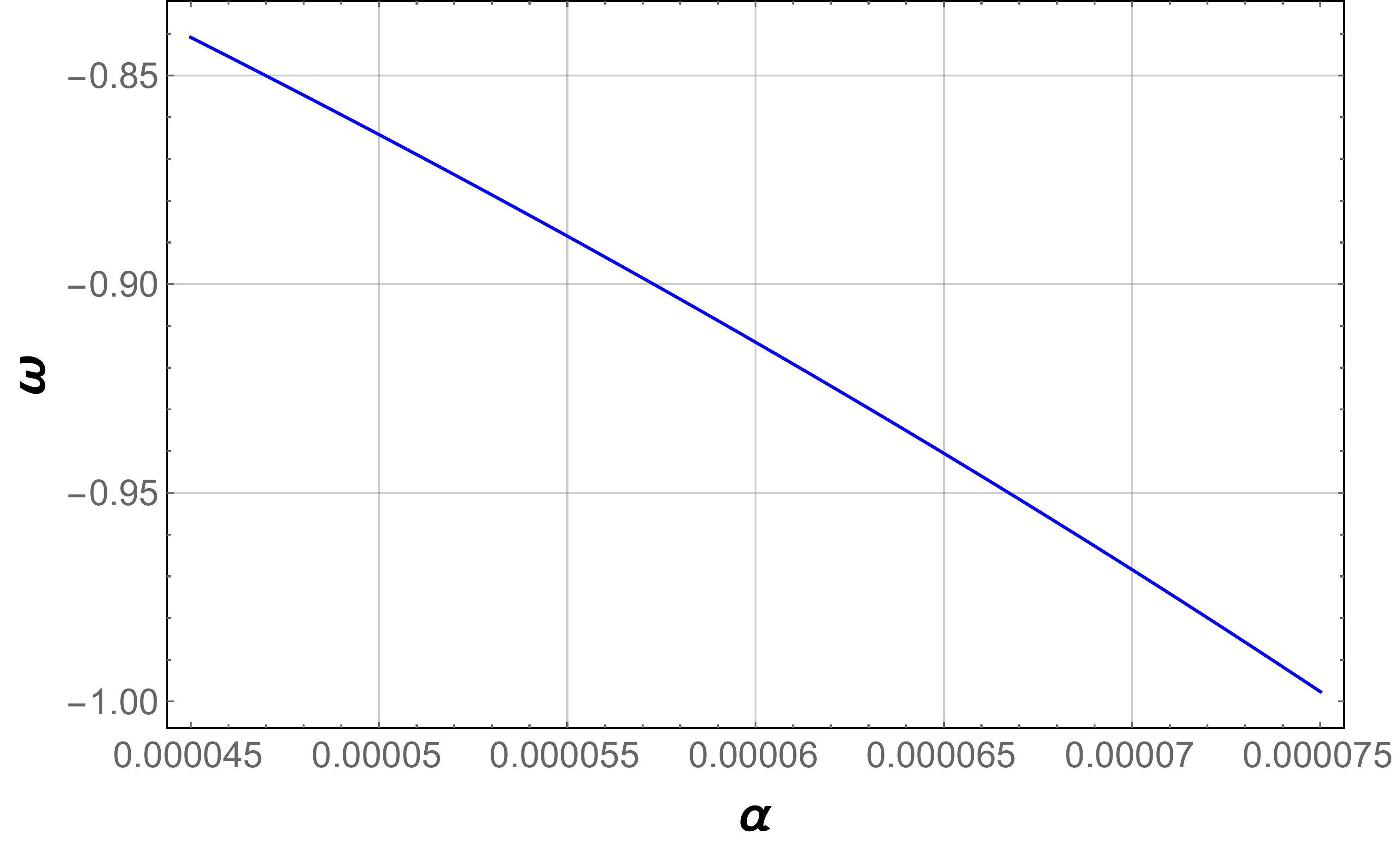}
\caption{EoS parameter for $f(R,T)=R+\alpha RT$ derived with the present values of $H_0$ and $q_0$ parameters.}
\label{f2d}
\end{center}
\end{figure}

The $\Lambda$CDM is a broadly accepted cosmological model that describes the observations of the universe. Several comments and surveys, such as Planck, WAMP, and Dark Energy Survey (DES), tested it in the last two decades. A specific mapping of $f(R, T)$ mimics the $\Lambda$CDM. So, for $\alpha=0$, our model recover the $\Lambda$CDM, and the ECs for this are given by
\begin{itemize}
\item \textbf{SEC : }$6H^2 q \geq 0$,
\item \textbf{NEC : }$2H^2(1+q)\geq 0$,
\item \textbf{WEC : }$3H^2\geq 0$ and  $2H^2(1+q)\geq 0$,
\item \textbf{DEC : }$3H^2\geq 0$ and  $2H^2(1+q)\geq 0$ or, $-2H^2(-2+q)\geq 0$.
\end{itemize}
By applying the present value of $H_0$ and $q_0$, one can check that the NEC, WEC, DEC satisfied whereas SEC violated. And, those profiles of ECs are the proper behavior for a standard model, which shows the accelerated expansion of the universe. Moreover, the constructed model is showing the same profiles for ECs, as demonstrated by $\Lambda$CDM.
Besides, the recent observation, such as the Planck collaboration and the $\Lambda$CDM, confirms the equation of state parameter $\omega $ takes its value as $\omega \simeq -1$ \cite{Planck/2018}. Also, this is an agreement for accelerated expansion. Therefore, $\omega$ is a suitable candidate to compare the constructed model with the $\Lambda$CDM model. The expression of $\omega$ for the constructed model is given by
\begin{equation}
\label{d19}
\omega=\frac{p}{\rho}=\frac{9 \alpha  H^2 \left(q^2-1\right)+\pi  (8 q-4)}{3 \alpha  H^2 (q-1) (7 q+10)+12 \pi }.
\end{equation}
Fig. \ref{f2d} shows the profile of $\omega$ by taking the present value of $H_0$ and $q_0$. One can observe that $\omega$ depends entirely on the model parameter, and its value is very close to $-1$, which is in agreement with the recent observations. Moreover, our model shows the accelerated expansion as good as $\Lambda$CDM.

\section{Compatibility with Linear Case}\label{sec6d}

The present section study the linear case of $f(R, T)$ model i.e., $f(R, T)= R+2 \gamma T$, where $\gamma$ is a constant. Many researchers operated in $f(R, T)$ gravity in the linear case, representing ECs. But, by constrained values of cosmological parameters $H_{0}$ and $q_{0}$, this analysis has attempted to research the ECs. The work includes the present values of $H_{0}=67.9$ and $q_{0}=-0.503$.

From \eqref{d1}, the energy density and pressure can be obtained as,

\begin{equation}
\rho=\frac{-2\dot{H}\gamma+3(1+2\gamma)H^2}{(1+3\gamma)^2-\gamma^2}
\end{equation}
\begin{equation}
p=-\frac{2(1+3\gamma)\dot{H}+3(1+2\gamma)H^2}{(1+3\gamma)^2-\gamma^2}
\end{equation}

The various ECs read as,

\begin{equation}
\rho+3p=\frac{2 H_0^2 [4 \gamma +(10 \gamma +3) q_0]}{8 \gamma ^2+6 \gamma +1}\geq 0
\end{equation}
\begin{equation}
\rho+p=\frac{2 H_0^2 (q_0+1)}{2 \gamma +1}\geq 0
\end{equation}
\begin{equation}
\rho-p=-\frac{2 H_0^2 (q_0-2)}{4 \gamma +1}\geq 0
\end{equation}
\begin{equation}
\omega=\frac{2 H^2 (3 \gamma +1) (q+1)-3 H^2 (2 \gamma +1)}{3 H^2 (2 \gamma +1)+2 H^2 \gamma  (q+1)}
\end{equation}

\begin{figure}[H]
\begin{center}
\includegraphics[scale=0.25]{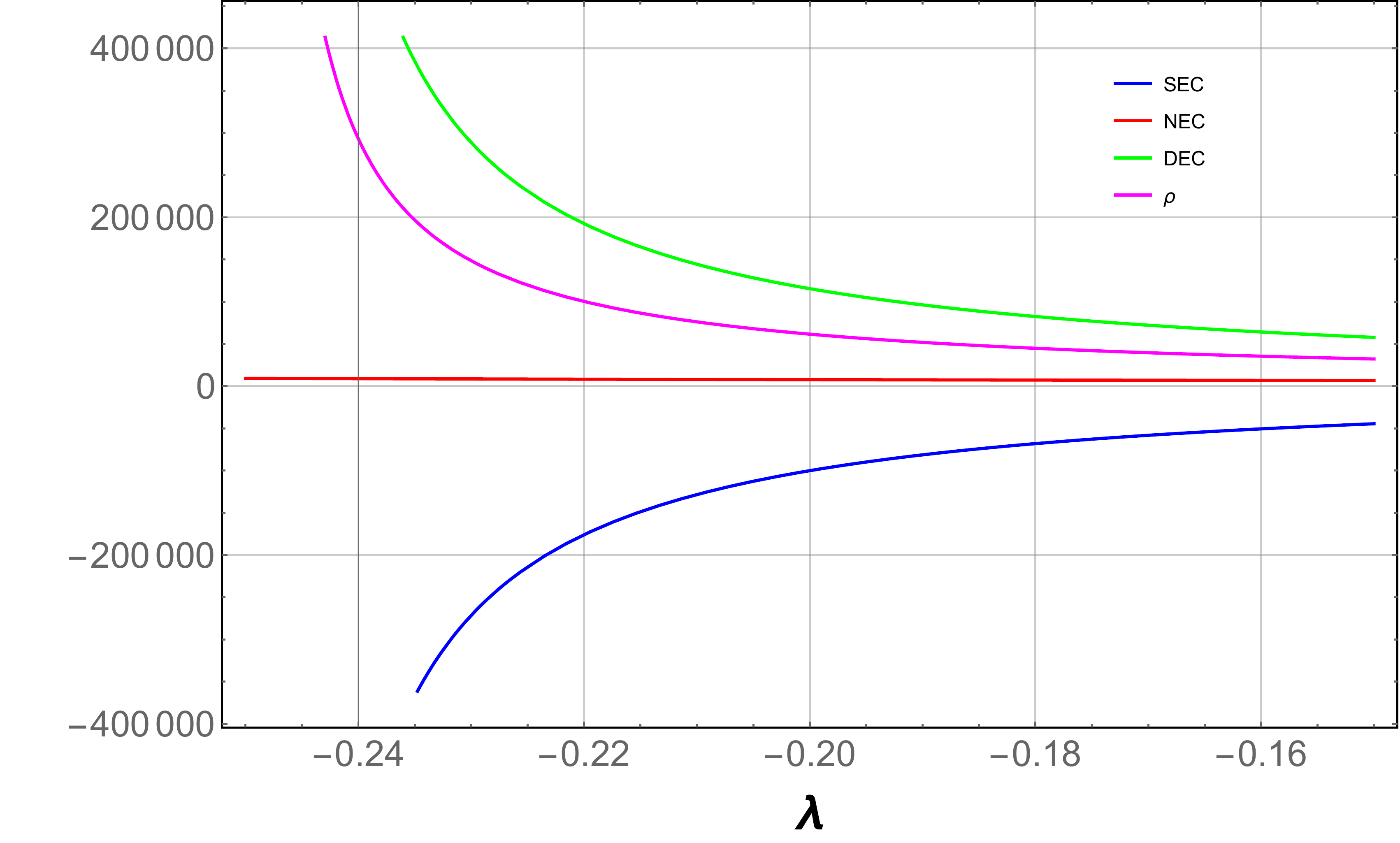}
\caption{ECs for $f(R,T)=R+2\gamma T$ derived with the present values of $H_0$ and $q_0$ parameters.}
\label{f6d}
\end{center}
\end{figure}

\begin{figure}[H]
\begin{center}
\includegraphics[scale=0.25]{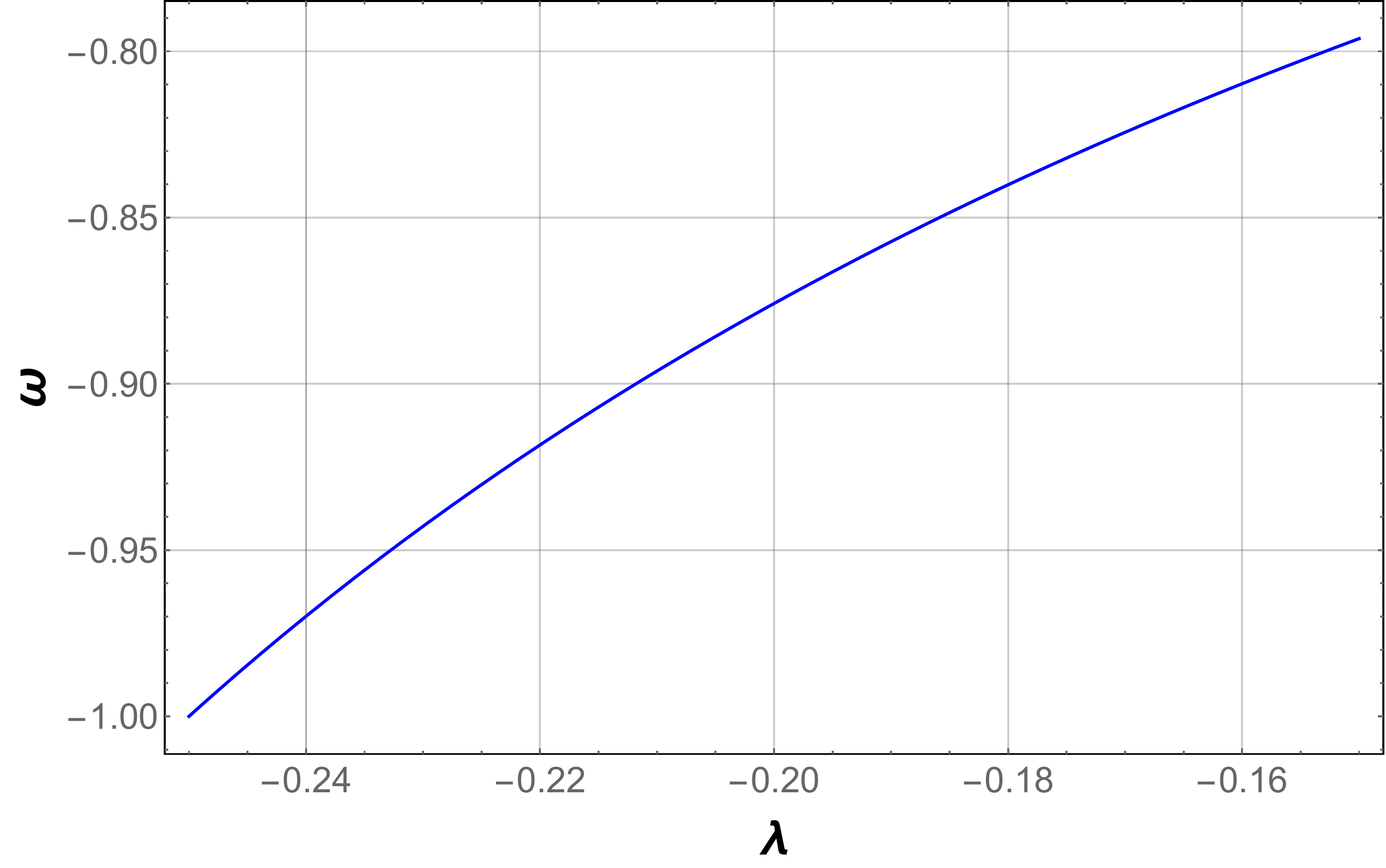}
\caption{EoS parameter for $f(R,T)=R+2\gamma T$ derived with the present values of $H_0$ and $q_0$ parameters.}
\label{f7d}
\end{center}
\end{figure}

From the Fig \ref{f6d}, one can observe that NEC, WEC, and DEC's ECs fulfill their conditions for $-0.25 \leq \gamma \leq -0.15$ while SEC violates its requirement. The SEC violation represents the accelerated expansion of the universe. As we know, the EOS parameter is an acceptable candidate to compare the model with the $\Lambda$CDM, so the $\gamma$ range is selected with the EOS behavior observed.\\
The EoS parameter $\omega$ is approximately equal to -1 according to the Planck observations and $\Lambda$CDM. The Fig. \ref{f7d} also depicts the behavior of $\omega$, taking the current values of $H_{0}$ and $q_{0}$. The values of $\omega$ are very similar to -1, which is in line with the $\Lambda$CDM. So, $f(R, T)$'s linear and non-linear form is in good agreement with $\Lambda$CDM, resulting in an accelerated universe expansion.

\section{Discussions} \label{sec7d}
In this chapter, we built a cosmological paradigm from the simplest non-minimal matter-geometry coupling in the gravitational theory of $f(R, T)$. This study is considered a well-motivated $f(R, T)$ gravity model as $f(R, T)=R+\alpha R T$, where $\alpha$ is the model parameter.\\
The primary motivation for such a theory of gravity is linked to its consistency with spacetime casual and geodesic structure discussed by various ECs. This analysis derived the null, the weak, the dominant, and strong ECs. In particular, the SEC plays an essential role in describing attractive or repulsive nature of the gravity under the modified theory of gravity. Also, the present values of cosmological parameters $H_{0}$ and $q_{0}$ are used to check the viability of $f(R, T)$ gravity theory.\\
According to present values, NEC, WEC, and DEC are observed to satisfy the conditions derived from the Raychaudhuri equations, whereas SEC is violated. The ECs established $0.000045\leq \alpha \leq 0.000075$ constraints to describe an accelerated expansion of the universe.

Further, in this chapter, we study the linear case of $f(R, T)$ model i.e., $f(R, T)= R+2 \gamma T$, where $\gamma$ is a constant and observe that NEC, WEC, and DEC's ECs fulfill their conditions for $-0.25 \leq \gamma \leq -0.15$ while SEC violates its requirement. The SEC violation represents the accelerated expansion of the universe.
This study also compared our energy constraints with those from the $\Lambda$CDM model. If we talk about the $\Lambda$CDM gravity, all ECs are satisfied except SEC. Furthermore, the equation of state parameter, derived from our model, is consistent with a current negative pressure period, showing values close to -1. This activity also confirms the explanation of $\Lambda$CDM for dark energy and recent experimental observations. So, $f(R, T)$ linear and non-linear forms are in good agreement with $\Lambda$CDM, resulting in an accelerated universe expansion. In future study, it will be interesting to constraint the equation of state parameter through cosmological  model.


\chapter{Constraint on the Equation of State Parameter $(\omega)$ in Non-minimally Coupled $f(Q)$ Gravity} 

\label{Chapter2} 

\lhead{Chapter 6. \emph{Constraint on the Equation of State Parameter $(\omega)$ in Non-minimally Coupled $f(Q)$ Gravity}} 

\vspace{10 cm}
* The work, in this chapter, is covered by the following publications: \\
 
\textit{Constraint on the Equation of State Parameter $(\omega)$ in Non-minimally Coupled $f(Q)$ Gravity}, Physics Letters B \textbf{823}, 136786 (2021).

\clearpage
\pagebreak

This chapter aims to study observational constraints on the modified symmetric teleparallel gravity, the non-metricity $f(Q)$ gravity, which reproduces the background expansion of the universe. For this purpose, this study use Hubble measurements, BAO, 1048 Pantheon supernovae type Ia data sample, which integrates SNLS, SDSS, HST survey, and Pan-STARRS1. Then, this chapter confronts the constructed cosmological model against observational measurements to set constraints on the parameters using MCMC methods. This study find the equation of state parameter $\omega=-0.853^{+0.015}_{-0.020}$ and $\omega= -0.796^{+0.049}_{-0.074}$ for Hubble and Pantheon samples, respectively. As a result, the $f(Q)$ model shows the quintessence behavior and deviates from $\Lambda$CDM.


\section{Introduction}

%

This chapter will highlight the cosmological model based on the recently proposed extension of symmetric teleparallel gravity, so-called $f(Q)$ gravity, where the non-metricity $Q$ describes the gravitational interaction \cite{Jimenez/2018}. Investigations on $f(Q)$ gravity have developed rapidly and lead to interesting applications \cite{Lazkoz/2019, Mandal/2020, Mandal/2020a, Harko/2018, Barros/2020, Jimenez/2020, Hasan/2021, Solanki/2021, Frus/2021, Flat/2021, Khyllep/2021,Anag/2021, Ata/2021,Dial/2019,Ayu/2021, Amb/2020}. If we look at the universe's expansion history, one can see that some cosmological parameters play an important role in designating the cosmological model's cosmic evolution. And, it is well known that the equation of state parameter $(\omega)$ predicts various fluid descriptions of the universe. Therefore, it is interesting to constrain this parameter using observational data. For this purpose, we use Hubble measurements, BAO, Pantheon supernovae type Ia integrates SNLS, SDSS, HST survey, Pan-STARRS1. The MCMC methods use to do the numerically analysis.

In this chapter, the ideas are presented in the following sections. In Section \ref{sec3e}, we discuss the cosmological application in FRW space-time. In Section \ref{sec4e}, we discuss the various type of observational datasets, constrain the parameters using the MCMC method, and deliberate our results. Finally, gathering all the information, this chapter conclude in section \ref{sec5e}.

\section{The $f(Q)$ Cosmology}\label{sec3e}

To explore several cosmological applications, we presume the homogeneous, isotropic and spatially flat line element given by
\begin{equation}
\label{e8}
ds^2 = -N^2(t) dt^2 +a^2(t) \delta_{ij} dx^i dx^j,
\end{equation}
where $N(t)$ is the lapse function and for the usual time reparametrization freedom, we can take $N=1$ at any time. $\delta_{ij}$ is the Kronecker delta and $i,\,\, j$ run over spatial components. The expansion rate and dilation rate can be written as
\begin{equation}
\label{e9}
H=\frac{\dot{a}}{a},\,\,\,\,\ T=\frac{\dot{N}}{N},
\end{equation}
respectively. For this line element the non-metricity is read as $Q=6 (H/N)^2$.

This study shall work on the perfect fluid matter distribution, for which the energy momentum tensor \eqref{6} become diagonal. The gravitational equations \eqref{7} in this case generalized to two Friedmann equations:
\begin{equation}
\label{e10}
f_2 \rho =\frac{f_1}{2}-6 F \frac{H^2}{N^2},
\end{equation}
\begin{equation}
\label{e11}
-f_2 p = \frac{f_1}{2}-\frac{2}{N^2}[(\dot{F}-FT)H+F(\dot{H}+3H^2)],
\end{equation}
respectively. Here,
\begin{align}
\label{e12f}
f=f_1(Q)\,+2 f_2(Q)L_M,\\
F=f_1'(Q)+2 f_2'(Q)L_M.
\end{align} 
It is easy to verify that, for $f_1=-Q$ and $f_2=1=-F$, the above Friedmann equations reduce to standard one \cite{Harko/2018}. From the above field equations, the continuity equation for matter field can be derived as
\begin{equation}
\label{e12}
\dot{\rho} +3 H(\rho+p) = -\frac{6 f_2'H}{f_2N^2}(\dot{H}-HT)(L_M+\rho).
\end{equation}
From \eqref{e12}, one can recover the standard continuity equation by imposing $L_M=-\rho$ as
\begin{equation}
\label{e13}
\dot{\rho} +3 H(\rho+p) =0.
\end{equation}
This is compatible with the isotropic and homogeneous background of the universe (see more details about the continuity equation in $f(Q)$ gravity \cite{Harko/2018}).

\subsection{Cosmological Model}\label{sub1}

This subsection shall proceed with $N=1$. Since, this chapter is working on the Friedmann-Robertson-Walker (FRM) framework, the non-metricity $Q$ and dilation rate $T$ reduce to
\begin{equation}
\label{e14}
Q=6H^2,\,\,\,\, T=0.
\end{equation}
Moreover, the modified Friedman equations \eqref{e10} and \eqref{e11} can be rewritten as
\begin{equation}
\label{e15}
3H^2= \frac{f_2}{2 F}\left(-\rho+\frac{f_1}{2f_2}\right),
\end{equation}
\begin{equation}
\label{e16}
\dot{H}+3H^2+\frac{\dot{F}}{F}H = \frac{f_2}{2 F}\left(p+\frac{f_1}{2f_2}\right).
\end{equation}
Now, we have two equations with five unknown such as $H,\,\, \rho,\,\, p,\,\, f_1,\,\, f_2$. To proceed further, this study consider the well-established energy density as
\begin{equation}
\label{e17}
\rho =\rho_0 a(t)^{-3 (1+\omega)},
\end{equation}
where $\rho_0$ is the proportionality constant and $\omega$ is the equation of state parameter.
Without loss of generality, we adopt $\rho_0=1$. In modified theories of gravity the cosmological scenarios can be discussed through the properties of cosmological models. For this purpose, this study introduce two Lagrangian functions $f_1(Q)$ and $f_2(Q)$ as
\begin{align}
\label{e19}
f_1(Q)=\alpha Q^n,\,\,\,\   f_2(Q)= Q,
\end{align}
where $\alpha$ and $n\neq 1$ are the arbitrary constants.

Using \eqref{e17}, \eqref{49}, \eqref{e19} in \eqref{e15}, we find the following expression for $H(z)$
\begin{equation}
\label{e20}
H(z)=\left\lbrace\frac{2\,(1+z)^{3(1+\omega)}}{\alpha\, (2n-1) \,6^{n-1}}\right\rbrace^{\frac{1}{2n-2}}.
\end{equation}
Now, this chapter aim is to put constraint on the parameters $\alpha,\,\, n,\,\, \omega$ using the astronomical observation data.

\section{Data, Methodology and Results}\label{sec4e}

This section deals with the various observational datasets to constrain the parameters $\alpha,\,\, n,\,\, \omega$. For this, we have adopted some statistical analysis to perform the numerical analysis. Specifically, it employs a MCMC method to obtain the posterior distributions of the parameters, using the standard Bayesian technique. This simulation is done by using the Hubble measurements ( i.e., Hubble data) and SNe Ia data. The best fits of the parameetrs are maximized by using the probability function
\begin{equation}
\label{e21}
\mathcal{L} \propto exp(-\chi ^2/2),
\end{equation}
where $\chi^2$ is the \textit{pseudo chi-squared function} \cite{hobson/2009}. The $\chi^2$ functions for various dataset are discussed below.

\subsection{Hubble Datasets}
\begin{figure}[H]
\begin{center}
\includegraphics[scale=0.6]{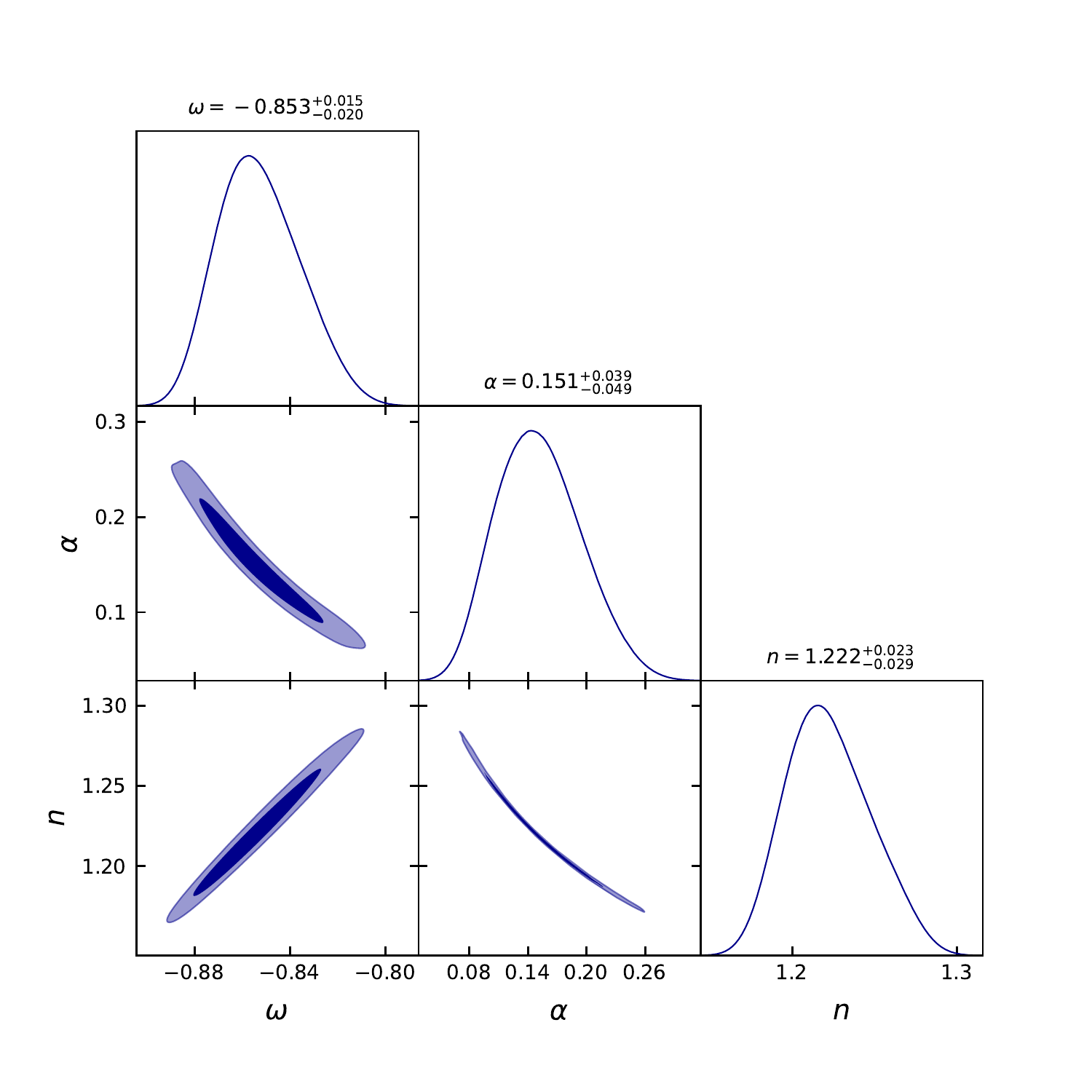}
\caption{The marginalized constraints on the coefficients in the expression of Hubble parameter $H(z)$ in Eqn. \eqref{20} are shown by using the Hubble sample.}
\label{f2e}
\end{center}
\end{figure}

\begin{figure}[H]
\begin{center}
\includegraphics[scale=0.55]{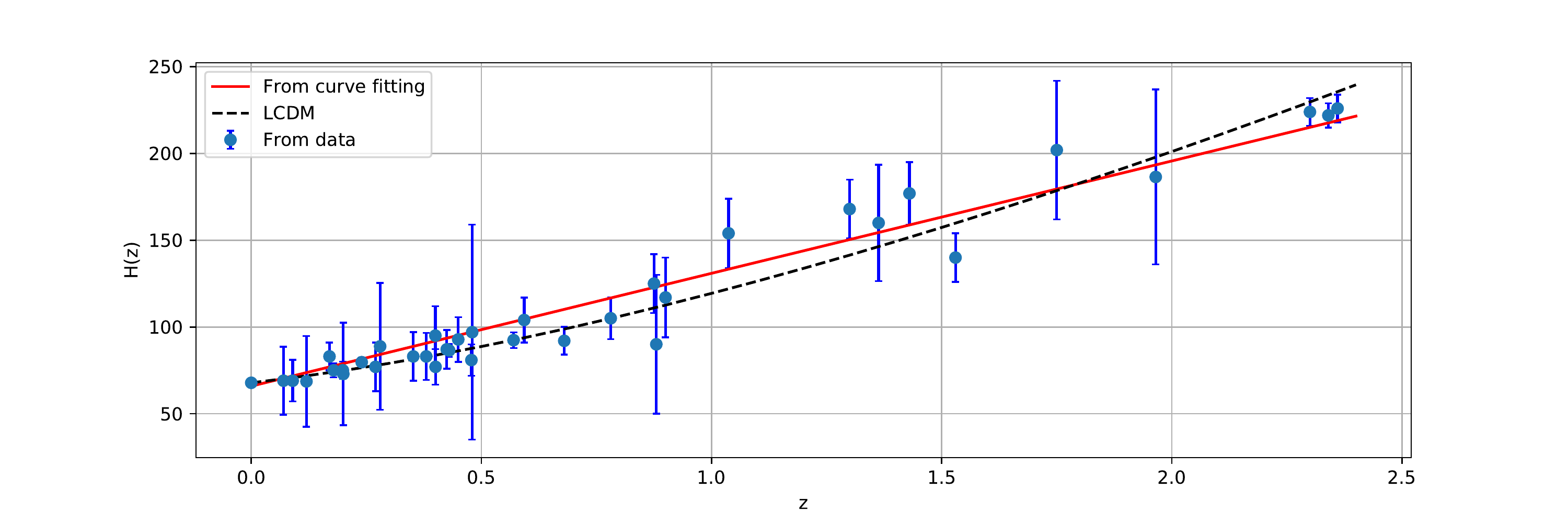}
\caption{The evolution of Hubble parameter $H(z)$ with respect to redshift $z$ is shown here. The red line represents our model and dashed-line indicates the $\Lambda$CDM model with $\Omega_{m0}=0.3$ and $\Omega_{\Lambda 0}=0.7$. The dots are shown the Hubble dataset with error bar.}
\label{f1e}
\end{center}
\end{figure}

Here, this study is considered the Hubble data, which was discussed previously in section \ref{VIIa}. To estimate the model parameters, we use the Chi-square function \eqref{chi}.
In Fig. \ref{f1e}, the profile of our model against Hubble data shown. The marginalized constraining results are displayed in Fig. \ref{f2e}.

\subsection{Pantheon Datasets}

SNe Ia is a powerful distance indicator to explore the background evolution of the universe. Therefore, this study use the latest Pantheon supernovae type Ia sample, which integrates 1048 SNe Ia data points from SNLS, SDSS, Pan-STARRS1, HST surveys, and low-redshift in the redshift-range $z \in [0.01,2.3]$ to constraint the above parameters \cite{Scolnic/2018}. The $\chi^2_{SN}$ function for the Pantheon sample of 1048 SNe Ia \cite{Scolnic/2018} is given by
\begin{equation}
\chi^2_{SN}(p_1,....)=\sum_{i,j=1}^{1048}\bigtriangledown\mu_{i}\left(C^{-1}_{SN}\right)_{ij}\bigtriangledown\mu_{j},
\end{equation}
where $p_j$ represents the free parameters of the presumed model and $C_{SN}$ is the covariance metric \cite{Scolnic/2018}, and $\mu$ represents the distance moduli is given by;
 \begin{align*}
 \mu^{th}(z)& =5\log\frac{D_L(z)}{10pc},\, \quad D_L(z)=(1+z)D_M,\\
 D_M(z)&=c \int_0^{z}\frac{d\tilde{z}}{H(\tilde{z})},\, \quad \bigtriangledown\mu_{i}=\mu^{th}(z_i,p_1,...)-\mu_i^{obs}.
 \end{align*}
 
\begin{figure}[H]
\includegraphics[scale=0.55]{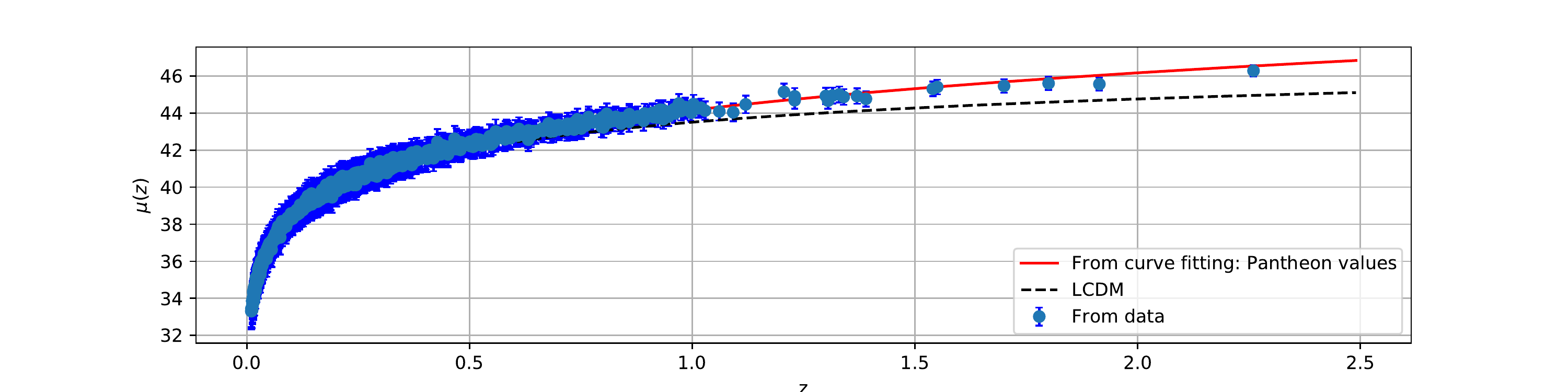}
\caption{The evolution of $\mu(z)$ with respect to redshift $z$ is shown here. The red line represents our model and dashed-line indicates the $\Lambda$CDM model with $\Omega_{m0}=0.3$ and $\Omega_{\Lambda 0}=0.7$. The dots are shown the 1048 Pantheon dataset with error bar. }
\label{f3e}
\end{figure}

\begin{figure}[H]
\begin{center}
\includegraphics[scale=0.6]{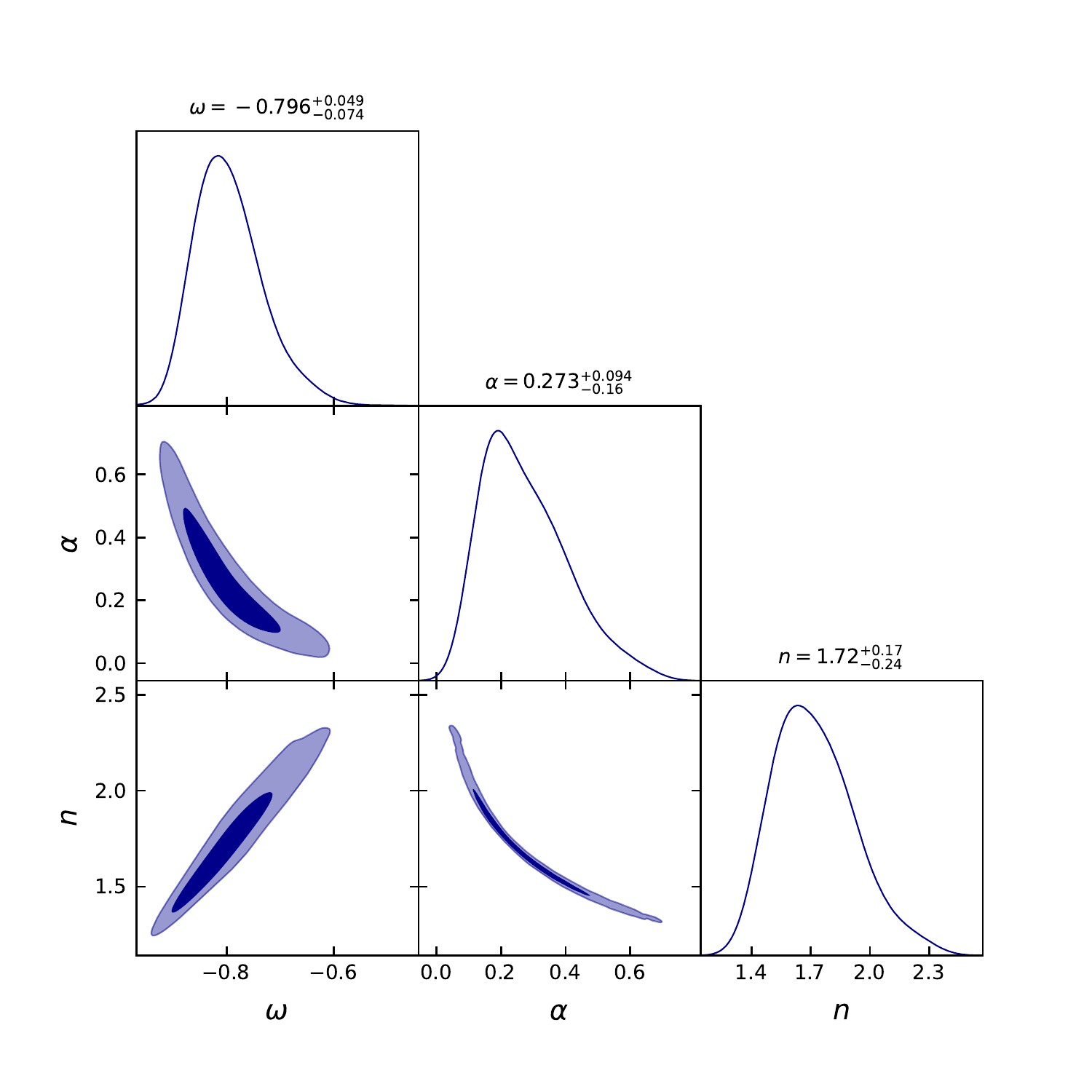}
\caption{The marginalized constraints on the coefficients in the expression of Hubble parameter $H(z)$ in Eqn. \eqref{22} are shown by using the Pantheon sample. }
\label{f4e}
\end{center}
\end{figure}

Figure \ref{f3e} represent the best fit of our model against the Pantheon dataset and the posterior distributions of the parameters are shown in Figure \ref{f4e}.

\subsection{Results}

Table \ref{t1e} show constraints at $68\%$ C.L. of the cosmological parameter $\omega$ and model parameters $\alpha, n$ for the cosmological model. Figs. \ref{f2e}, \ref{f4e} present the contour plots of the parameters at $68\%$ and $95\%$ C.L.  for Hubble and Pantheon samples, respectively. For reference, one can compare the $f(Q)$ model with the $\Lambda$CDM model with constraint values of parameters in Figs. \ref{f1e} and \ref{f3e}. It is observed that the $f(Q)$ model is perfectly fitting with the observational data and deviates slightly from the $\Lambda$CDM. Moreover, it is well-known that the present scenario of the universe, i.e., accelerated expansion, can be discussed with the presence of additional cosmological constant $\Lambda$ in Einstein's field equations or by modifying the fundamental formulation of gravity for the evolution of the universe. Besides this, the equation of state parameter $(\omega)$ plays a vital role for a cosmological model to predict its' different phases of evolution. The present scenario of the universe can be predicted by either quintessence behavior of $\omega\,\, (i.e., -1<\omega<-1/3)$ or phantom behavior of $\omega\,\, (i.e., \omega<-1)$. For our model, we found $\omega=-0.853^{+0.015}_{-0.020}$ for Hubble dataset and $\omega=-0.796^{+0.049}_{-0.074}$ for Pantheon dataset at $1\sigma$ confidence level. One can clearly see that the $f(Q)$ model shows the quintessence behavior, and it is near to $\Lambda$CDM.


\begin{table*}[!t]
	\renewcommand\arraystretch{1.5}
	\caption{The marginalized constraining results on three parameters $\omega,\,\, \alpha,\,\, n$ are shown by using the Hubble and Pantheon SNe Ia sample. We quote $1-\,\sigma$ (68$\%$) errors for all the parameters here.
	}\begin{center}
	\begin{tabular} { l |c| c |c }
		\hline
		\hline

		Dataset    & $\omega$      &$\alpha$      &$n$        \\
		\hline
		Hubble   & $-0.853^{+0.015}_{-0.020}$   & $0.151^{+0.039}_{-0.049}$ &$1.222^{+0.023}_{-0.029}$  \\
		\hline
		Pantheon & $-0.796^{+0.049}_{-0.074}$   & $0.273^{0.094}_{-0.16}$ & $1.72^{+0.17}_{-0.24}$  \\
		
	    \hline
		\hline
	\end{tabular}
	\end{center}
	\label{t1e}
\end{table*}
\section{Conclusion}\label{sec5e}

The rising concern in the current scenarios of the universe compels us to go beyond the standard formulation of gravity. In this context, we have worked on the modified $f(Q)$ gravity to obtain observational constraints for the background candidates of the accelerated expansion of the universe. For this, we used a wide variety of observational samples such as Hubble data and Pantheon data (which includes SNLS, SDSS, Pan-STARRS1, HST surveys, and low-redshift surveys). This analysis also adopted the parametrization technique to obtain the expression for $H(z)$. The best fit ranges of the parameters are obtained by applying the Bayesian method in MCMC simulation. The constraint values of the equation of state parameter $(\omega)$ suggest that the $f(Q)$ model shows quintessence behavior. In addition, this chapter has depicted the profiles of Hubble parameter with the constraint values of parameters for both the datasets in Figs. \ref{f1e} and \ref{f3e}, which helps us to compare the studied model with the $\Lambda$CDM.

In conclusion, this chapter's findings could motivate further research into the $f(Q)$ gravity because it is one of the alternatives to the coherence model that, aside from being preferred by the data. The significance of this model is that it does not face the cosmological constant problem because it does not comprise any additional constant inside the Lagrangian $f(Q)$ form. In a further study, it would be interesting to explore these types of models using weak lensing data, full CMB and LSS spectra, and other datasets. We intend to address some of these tests in the near future and hope to report on them.



\chapter{Concluding Remarks and Future Perspectives} 

\label{Chapter7} 

\lhead{Chapter 7. \emph{Concluding Remarks and Future Perspectives}} 

 \clearpage
 
The objective of this chapter is to summarize the outcomes and future scopes of this thesis in search of an accurate gravitational theory that can explain the evolution process of the universe from early to late-time. In this thesis, we have focused on the accelerated expansion of the universe in the context of modified theories of gravity. Let us discuss the results obtained in five concrete works from the previous chapters \ref{Chapter3}-\ref{Chapter2}. The chapters have not only discussed the theoretical developments of new cosmological models but also confronted the observational measurements to achieve the ultimate goal.

In Chapter-\ref{Chapter1}, we have discussed the history, mathematical notations, basic elements, cosmological applications, fundamental theories of gravity, and cosmological observations. Besides this, it is well-known that the fundamental theory of gravity, like general relativity, fails to address some issues like fine-tuning, flatness problem of the universe, and it seems incomplete. Therefore, its modifications and generalization are more successful in addressing these issues, and the chapter concludes by over-viewing the modified theories of gravity.
 
In Chapter-\ref{Chapter3}, we have discussed the accelerated expansion of the universe in the background of hybrid and logarithmic teleparallel gravity. For this purpose, a well-known deceleration parameter is considered, constraining its free parameter using observational measurements. Next, we have presented a few geometric diagnostics of this parametrization to understand the nature of dark energy and its deviation from the $\Lambda$CDM cosmology. Finally, we have studied the energy conditions to check the consistency of the parameter spaces for both the teleparallel gravity models. We have found that SEC is violated for both models, which is an essential for obtaining an accelerating universe. Chapter-\ref{Chapter4} extended the analysis of Chapter-\ref{Chapter3}, focusing on energy conditions. One promising approach lies in a new class of teleparallel theory of gravity named $f(Q)$ gravity, where the non-metricity $Q$ is responsible for the gravitational interaction, which has been discussed. The important role of these alternative theories is to obey the energy condition constraints. Such constraints establish the compatibility of a given theory with the causal and geodesic structure of space-time. In this chapter, we have presented a complete test of energy conditions for $f(Q)$ gravity models. The energy conditions allowed us to fix our free parameters, restricting the families of $f(Q)$ models compatible with the accelerated expansion of our Universe. Our results have examined the viability of the $f(Q)$ theory, leading us close to the dawn of a complete theory for gravitation. Further extended Chapter-\ref{Chapter4} to Chapter-\ref{Chapter5}, keeping the cosmographic parameters in the front line. In chapter \ref{Chapter5}, we have discussed the cosmography idea and its application to modern cosmology. Cosmography is an ideal tool to investigate the cosmic expansion history of the universe in a model-independent way. The equations of motion in modified theories of gravity are usually very complicated; cosmography may select practical models without imposing arbitrary choices a priori. We have used the model-independent way to derive $f(z)$ and its derivatives up to the fourth order in terms of measurable cosmographic parameters. We have rewritten the luminosity distance in terms of the cosmographic functions. We have performed the MCMC simulation by considering three different sets of cosmographic functions. We have used the recent Supernovae Ia Pantheon data, the constraints on the Hubble parameter $H_0$, and the cosmographic functions estimated. We have found the best fits for the functions of cosmographic sets for three statistical models. The outputs of these studies are aligned with the accelerated expansion of the universe.
 
In Chapter-\ref{Chapter6}, we have discussed the simplest non-minimal matter geometry coupling with a perfect fluid distribution in the $f(R,T)$ gravity framework. The model parameter is constrained by energy conditions and a single parameter proposed equation of state, resulting in the compatibility of the $f(R,T)$ models with the accelerated expansion of the universe. It is seen that the EoS parameter illustrates the quintessence like expansion. Also, the present values of the cosmological constant and the acceleration of the universe are used to check the viability of our linear $f(R,T)$ model of gravity. It has been observed that the positive behavior of DEC and WEC  indicates the validation of the model. In contrast, SEC is violated, which indicates the accelerated expansion of the universe. Chapter-\ref{Chapter6} is extended to Chapter-\ref{Chapter2}, focusing on the EoS parameter. In Chapter-\ref{Chapter2}, we have studied observational constraints on the modified symmetric teleparallel gravity. For this purpose, we have used Hubble measurements, BAO, 1048 Pantheon supernovae type Ia data sample, which spans SNLS, SDSS, HST survey, Pan-STARRS1. We have confronted our cosmological model against observational samples to set constraints on the parameters using MCMC methods. We have found the equation of state parameter $\omega=-0.853^{+0.015}_{-0.020}$ and $\omega= -0.796^{+0.049}_{-0.074}$ for Hubble and Pantheon samples, respectively. As a result, the $f(Q)$ model is shown the quintessence behavior and deviates from $\Lambda$CDM.
 
In the future, it would be interesting to extend the above analysis for a better understanding of the evolution of the universe. There are a lot of scopes in extending the work done in this thesis. For example, in these studies, there is wide range of freedom for our free parameters, enabling several testable scenarios for $f(Q)$ gravity. Such tests could include the absence of ghost modes, gravitational wave constraints, and cosmological parameters derived from the Cosmic Microwave Background. Besides, it would be interesting to study carefully the coupling of $f(Q)$ with inflaton fields, looking for possible analytic models or cosmological parameters constraints. Besides this, one can extend the present work using the perturbation analysis and dynamical system analysis. We intend to address some of these investigations in the near future and hope to report on them.






\addtocontents{toc}{\vspace{2em}} 

\backmatter


\label{References}
\lhead{\emph{References}}

\cleardoublepage
\pagestyle{fancy}

\label{Publications}
\lhead{\emph{List of Publications}}

\chapter{List of Publications}
\section*{Thesis Publications}
\begin{enumerate}

\item \textbf{Sanjay Mandal}, P.K. Sahoo, \textit{Constraint on the equation of state parameter ($\omega$) in non-minimally coupled $f(Q)$ gravity}, \textcolor{blue}{Physics Letters B} \textbf{823}, 136786 (2021).

\item P. K. Sahoo, \textbf{Sanjay Mandal}, Simran Arora \textit{Energy Condition in Non-minimally $f(R,T)$ Gravity}, \textcolor{blue}{Astronomische Nachrichten} \textbf{342}, 89 (2021).

\item \textbf{Sanjay Mandal}, Deng Wang, P.K. Sahoo, \textit{Cosmography in $f(Q)$ gravity}, \textcolor{blue}{Physical Review D} \textbf{102}, 124029 (2020).

\item \textbf{Sanjay Mandal}, P.K. Sahoo, J.R.L. Santos \textit{Energy Conditions in $f(Q)$ gravity}, \textcolor{blue}{Physical Review D} \textbf{102}, 024057 (2020).

\item \textbf{Sanjay Mandal}, S. Bhattacharjee, SKJ Pacif, P.K. Sahoo, \textit{Accelerating universe in hybrid and logarithmic teleparallel gravity}, \textcolor{blue}{Physics of the Dark Universe} \textbf{28},  100551 (2020).
\end{enumerate}
\section*{Other Publications}
\begin{enumerate}

\item G. N. Gadbail, \textbf{Sanjay Mandal}, P.K. Sahoo, \textit{Reconstruction of $\Lambda$CDM universe in $f(Q)$ gravity,} \textcolor{blue}{Physics Letters B} \textbf{ 835}, 137509 (2022).

\item N. S. Kavya, V. Venkatesha, \textbf{Sanjay Mandal}, P.K. Sahoo, \textit{Constraining Anisotropic Cosmological Model in $f(R,L_m)$ Gravity,}  \textcolor{blue}{Physics of the Dark Universe} \textbf{38}, 101126 (2022).

\item S. Arora, \textbf{Sanjay Mandal}, S. Chakraborty, G. Leon, P.K. Sahoo, \textit{Can f (R) gravity isotropize a pre-bounce contracting universe ?,} \textcolor{blue}{Journal of Cosmology and Astroparticle Physics} \textbf{09}, 042 (2022).

\item \textbf{Sanjay Mandal}, P.K. Sahoo, J.R.L. Santos, \textit{`Reply to "Comment on Energy Conditions in $f(Q)$ Gravity"'},  \textcolor{blue}{Physical Review D} \textbf{106}, 048502 (2022).

\item L. V. Jaybhaye, R. Solanki, \textbf{Sanjay Mandal}, P.K. Sahoo, \textit{Cosmology in $f(R,L_m)$ gravity}, \textcolor{blue}{Physics Letters B} \textbf{831}, 137148 (2022).

\item R. Solanki, A. De, \textbf{Sanjay Mandal}, P.K. Sahoo, \textit{Accelerating expansion of the universe in modified symmetric telaparallel gravity}, \textcolor{blue}{Physics of the Dark Universe}  \textbf{36}, 101053 (2022).

\item O. Sokoluik, \textbf{Sanjay Mandal}, P.K. Sahoo, A. Baransky, \textit{Generalised Ellis-Bronnikov Wormholes in $f(R)$ Gravity}, \textcolor{blue}{European Physical Journal C} \textbf{82}, 280 (2022).

\item \textbf{Sanjay Mandal}, Abhishek Parida, P.K. Sahoo, \textit{Observational constraints and some toy models in $f(Q)$ gravity with bulk viscous fluid}. \textcolor{blue}{Universe} \textbf{8}, 240 (2022).

\item A. De, \textbf{Sanjay Mandal}, J.T. Beh, T.H. Loo, I\textit{sotropization of locally rotationally symmetric Bianchi-I Universe in $f(Q)$ garvity}. \textcolor{blue}{European Physical Journal C}  \textbf{82}, 71 (2022).

\item \textbf{Sanjay Mandal}, G. Mustafa, Z. Hassan, P.K. Sahoo, A\textit{ study of anisotropic spheres in $f(Q)$ gravity with quintessence field},  \textcolor{blue}{Physics of the Dark Universe} \textbf{35}, 100934 (2021).

\item Tee-How Loo, Avik De, \textbf{Sanjay Mandal}, P.K. Sahoo, \textit{How a projectively flat geometry regulates $F(R)$-gravity theory?}, \textcolor{blue}{Physica Scripta} \textbf{96}, 125034 (2021).

\item L. V. Jaybhaye, \textbf{Sanjay Mandal}, P. K. Sahoo, \textit{Constraints on Energy Conditions in $f(R,L_m)$ Gravity}, \textcolor{blue}{International Journal of Geometric Methods in Modern Physics}  \textbf{19}, 2250050 (2021)

\item \textbf{Sanjay Mandal}, N. Myrzakulov, P. K. Sahoo, R. Myrzakulov, \textit{Cosmological bouncing scenarios in symmetric teleparallel gravity}. \textcolor{blue}{European Physical Journal Plus} \textbf{136}, 760 (2021). 

\item \textbf{Sanjay Mandal}, Avik De, Tee How Loo, P.K. Sahoo, \textit{Almost-Pseudo-Ricci Symmetric FRW Universe with a Dynamic Cosmological Term and Equation of State}, \textcolor{blue}{Universe} \textbf{7}, 205 (2021).

\item Zinnat Hassan, \textbf{Sanjay Mandal}, P. K. Sahoo, \textit{Traversable Wormhole Geometries in $f(Q)$ gravity}, \textcolor{blue}{Fortschritte der Physik} \textbf{69}, 2100023 (2021).

\item \textbf{Sanjay Mandal}, P.K. Sahoo, \textit{On the temporal evolution of particle production in $f(T)$ gravity}, \textcolor{blue}{Mod. Phys. Lett. A} \textbf{35} 2050328 (2020).

\item \textbf{Sanjay Mandal}, P.K. Sahoo, \textit{A Complete Cosmological Scenario in Teleparallel Gravity}, \textcolor{blue}{European Physical Journal Plus} \textbf{135}, 706 (2020).

\item  Parbati Sahoo, \textbf{Sanjay Mandal}, P.K. Sahoo, \textit{Wormhole model with a hybrid shape function in $f(R,T)$ gravity}, \textcolor{blue}{New Astronomy} \textbf{80}, 101421 (2020).

\item K. M. Singh, \textbf{Sanjay Mandal}, L. P. Devi, P.K. Sahoo, \textit{Dark Energy and Modified Scale Covariant Theory of Gravitation}, \textcolor{blue} {New Astronomy} \textbf{77}, 101353 (2020).

\item Y Aditya, \textbf{Sanjay Mandal}, P.K. Sahoo, D.R.K. Reddy, \textit{Observational constraint on interacting Tsallis holographic dark energy in logarithmic Brans-Dicke theory}, \textcolor{blue}{European Physical Journal C} \textbf{79}, 1020 (2019).

\end{enumerate}
\cleardoublepage
\pagestyle{fancy}
\lhead{\emph{Biography}}

\chapter{Biography}

\section*{Brief Biography of the Candidate:}
\textbf{Mr. Sanjay Mandal} obtained his Bachelor's degree in Mathematics from Utkal University, Odisha, in 2016 and Master's degree in Mathematics from Berhampur University, Odisha, in 2018. His academic credentials are excellent, his research activities are impressive, and his performance as a research scholar is outstanding. He received many academic awards like two-time gold medals in Mathematics (B.Sc, and M.Sc.), Institute of Mathematics and Applications (IMA) scholarship in 2016, University Rank Holder (URH-UGC) fellowship from 2016 to 2018 for his Master's degree, DST Inspire Fellowship, Govt. of India from 2019-2024 for Ph.D., Best Paper Award (in International conference) in 2022. In his Ph.D. research career, he has published 26 research articles in various renowned international journals. He has presented research papers at several national and international conferences (such as Brazil, South Africa, and Taiwan).

\section*{Brief Biography of the Supervisor:}
\textbf{Prof. P.K. Sahoo} is currently serving as Professor and Head in the department of Mathematics of Birla Institute of Technology and Science-Pilani, Hyderabad Campus. He received his Ph.D. degree from Sambalpur University, Odisha, India in 2004. In his research career, he has published more than 150 research articles in various renowned national and international journals. In 2020 and 2021, he has been placed among the top $2\%$ scientists of the world according to the survey by researchers from Stanford University, USA in Nuclear and Particle Physics. He has presented several research papers and delivered invited talks in these areas in national and international conferences held in India and abroad. He has conducted several academic and scientific events in the department. He serves as an expert reviewer and editorial member for a number of reputed scientific journals, and also Ph.D. examiner in several universities. He is also recipient of Science Academics Summer Research Fellowship, UGC Visiting fellow, Fellow of Institute of Mathematics and its Applications (FIMA), London, Fellow of Royal Astronomical Society (FRAS), London, elected as a foreign member of the Russian Gravitational Society, Expert Reviewer, Physical Science Projects, SERB-DST, Govt. of India and lifetime membership of many scientific societies. He has received scientific projects from UGC, Govt. of India in 2004, DAAD-RISE Worldwide in 2018, 2019, CSIR, Govt. of India in 2019, NBHM. Govt. of India in 2022. He has also visited countries like Canada, South Korea, Germany, Belgium, Japan, Poland, Switzerland, UK, and presented his research work and delivered invited talks in different scientific events.

\end{document}